\documentclass[superscriptaddress,prx,nofootinbib,notitlepage,aps,10pt]{revtex4-2}
\usepackage{graphicx}
\usepackage{amssymb,amsmath,amsthm,amsfonts}
\usepackage{thmtools}
\usepackage{mathtools}
\mathtoolsset{centercolon}
\usepackage{thm-restate} 
\usepackage{placeins}
\usepackage{multirow}
\usepackage{ulem}
\usepackage[colorlinks]{hyperref}
\hypersetup{
	pdfstartview={FitH},
	pdfnewwindow=true,
	colorlinks=true,
	linkcolor=blue,
	citecolor=blue,
	filecolor=blue,
	urlcolor=blue}

\usepackage{siunitx}
\usepackage[shortlabels]{enumitem}

\allowdisplaybreaks

\usepackage{url}
\usepackage{hypcap}
\usepackage{dsfont}
\usepackage{algorithm}
\usepackage{dcolumn}
\newcolumntype{d}[1]{D{.}{.}{#1}}

\newcommand{\eq}[1]{Eq.~\hyperref[eq:#1]{(\ref*{eq:#1})}}
\renewcommand{\sec}[1]{\hyperref[sec:#1]{Section~\ref*{sec:#1}}}
\newcommand{\app}[1]{\hyperref[app:#1]{Appendix~\ref*{app:#1}}}
\newcommand{\tab}[1]{\hyperref[tab:#1]{Table~\ref*{tab:#1}}}
\newcommand{\fig}[1]{\hyperref[fig:#1]{Figure~\ref*{fig:#1}}}
\newcommand{\figa}[2]{\hyperref[fig:#1]{Figure~\ref*{fig:#1}#2}}
\newcommand{\figx}[2]{\hyperref[fig:#1]{Figure~\ref*{fig:#1}(#2)}}
\newcommand{\thm}[1]{\hyperref[thm:#1]{Theorem~\ref*{thm:#1}}}
\newcommand{\lem}[1]{\hyperref[lem:#1]{Lemma~\ref*{lem:#1}}}
\newcommand{\cor}[1]{\hyperref[cor:#1]{Corollary~\ref*{cor:#1}}}
\newcommand{\defn}[1]{\hyperref[def:#1]{Definition~\ref*{def:#1}}}
\newcommand{\alg}[1]{\hyperref[alg:#1]{Algorithm~\ref*{alg:#1}}}

\newcommand{\norm}[1]{\left\lVert#1\right\rVert}
\newcommand{\upper}{\aleph}
\newcommand{\evals}{\mathcal{M}}

\def\bra#1{\mathinner{\langle{#1}|}}
\def\ket#1{\mathinner{|{#1}\rangle}}

\newcommand{\nn}{\nonumber \\}

\newcommand{\locl}{u_{\rm loc}^{\alpha_\ell}}

\newcommand{\nonlocl}{u_{\rm non}^{\alpha_\ell}}
\newcommand{\loc}{u_{\rm loc}^{\alpha}}
\newcommand{\nonloc}{u_{\rm non}^{\alpha}}
\newcommand{\eli}{\iota}
\newcommand{\elj}{\xi}

\usepackage{color}

\makeatletter
\renewcommand*\env@matrix[1][\arraystretch]{%
	\edef\arraystretch{#1}%
	\hskip -\arraycolsep
	\let\@ifnextchar\new@ifnextchar
	\array{*\c@MaxMatrixCols c}}
\makeatother

\newcommand{\MQ}{\affiliation{%
School of Mathematical and Physical Sciences,
Macquarie University, Sydney, NSW, Australia}}

\newcommand{\Google}{\affiliation{%
Google Quantum AI, Venice, CA, United States}}

\newcommand{\covestro}{\affiliation{%
Covestro Deutschland AG, 51373 Leverkusen, Germany}}

\newcommand{\Harvard}{\affiliation{Department of Chemistry and Chemical Biology, Harvard University, Cambridge, MA, United States}}

\newcommand{\Florida}{\affiliation{Department of Chemistry and Biochemistry, Florida State University, Tallahassee, FL, United States}}

\begin{document}

\title{Quantum Simulation of Realistic Materials in First Quantization\\ Using Non-local Pseudopotentials}

\date{\today}

\author{Dominic W.~Berry}
\email[Corresponding author: ]{dominic.berry@mq.edu.au}
\MQ

\author{Nicholas C.~Rubin}
\email[Corresponding author: ]{nickrubin@google.com}
\Google

\author{Ahmed O. Elnabawy}
\covestro

\author{Gabriele Ahlers}
\covestro

\author{A.~Eugene DePrince~III}
\Google
\Florida

\author{Joonho Lee}
\Google
\Harvard

\author{Christian Gogolin}
\covestro

\author{Ryan Babbush}
\email[Corresponding author: ]{babbush@google.com}
\Google

\begin{abstract}
This paper improves and demonstrates the usefulness of the first quantized plane-wave algorithms for the quantum simulation of electronic structure, developed by Babbush \emph{et al.}~\cite{BabbushContinuum} and Su \emph{et al.}~\cite{SuPRXQuantum21}. We describe our quantum algorithm for first quantized simulation that accurately includes pseudopotentials. We focus on the Goedecker-Tetter-Hutter (GTH) pseudopotential, which is among the most accurate and widely used norm-conserving pseudopotentials enabling the removal of core electrons from the simulation. The resultant screened nuclear potential regularizes cusps in the electronic wavefunction so that orders of magnitude fewer plane waves are required for a chemically accurate basis. Despite the complicated form of the GTH pseudopotential, we are able to block encode the associated operator without significantly increasing the overall cost of quantum simulation.
This is surprising since simulating the nuclear potential is much simpler without pseudopotentials, yet is still the bottleneck.
We also generalize prior methods to enable the simulation of materials with non-cubic unit cells, which requires nontrivial modifications. Finally, we combine these techniques to estimate the block-encoding costs for commercially relevant instances of heterogeneous catalysis (e.g.\ carbon monoxide adsorption on transition metals) and compare to the quantum resources needed to simulate materials in second quantization. We conclude that for computational cells with many particles, first quantization often requires meaningfully less spacetime volume.
\end{abstract}

\maketitle

\section{Introduction}
Simulation of quantum systems is an area where quantum computers could provide major speedups over the best classical methods, and quantum simulation of chemistry or materials is an area likely to yield considerable practical benefit.
To realize the potential of quantum computing it is crucial to carefully optimize the representation of the simulation problem in order to have the lowest complexity implementation on the quantum computer and to make most efficient use of quantum resources such as ancilla qubits and Toffoli/T-gates.
Crucial here is the choice of basis in which the electronic degrees of freedom of a system are discretized, a selection of the most relevant such degrees of freedom and an effective treatment of the remaining ones (e.g., core electrons), and methods through which the resulting Hamiltonian can be represented in the most compact way possible on a quantum computer. 

Recent work has focused on efficient block encodings of chemical and materials Hamiltonians within second quantization by optimizing the construction of the linear combination of unitaries (LCU) representation of the Hamiltonian and estimating phase estimation costs~\cite{BabbushSparse1, BabbushSparse2, PhysRevX.8.041015, vonBurg2020, Lee2020, goings2022reliably, PRXQuantum.4.040303, PhysRevResearch.5.013200, Oumarou2023}. While second quantized representations of the Hamiltonian in a localized basis provide a number of advantages, such as efficient representation of low-density systems and compact basis sets, plane-wave basis sets provide complimentary advantages which have been explored within the context of efficient quantum algorithm representations using a first quantized approach~\cite{BabbushContinuum, SuPRXQuantum21, PhysRevA.106.032428}.

Plane wave based methods are routinely applied in the classical simulation of the electronic structure of periodic systems. According to Bloch's theorem, the wave function for an electron in a periodic potential is a product of a plane wave and a periodic function. These Bloch functions form an ideal single-particle basis for calculations on periodic many-electron systems. This basis naturally reflects the translational symmetry of the crystal lattice, captures the extended nature of the wave functions for these systems, and is systematically improvable via the inclusion of higher-energy plane waves. Importantly, plane-wave-based representations of the operators encountered in non-relativistic quantum chemistry are trivial to evaluate; matrix elements of the kinetic energy and Coulomb operators in the momentum basis are simple functions of the momenta associated with the plane-wave functions. These nice properties have contributed to the success of plane wave density functional theory (DFT) in materials science applications \cite{Hickel13_438}.

The primary disadvantage of plane waves is that a large number of functions is required to accurately resolve tightly localized or rapidly oscillating features, such as the electronic wave function in close proximity to nuclei. This challenge has motivated the development of a variety of approaches (e.g., norm-conserving pseudopotentials \cite{Chiang79_1494}, ultra-soft pseudopotentials \cite{Vanderbilt90_7892}, the projector augmented wave (PAW) method \cite{Blochl94_17953,Joubert99_1758}, {etc.}) that avoid explicitly considering the core electrons and orbitals. Pseudopotential approaches achieve this aim by replacing the true ionic potential for the valence electrons with a smoothly-varying effective potential (with a different functional form) that reproduces the behavior of the valence electrons in the presence of the core orbitals. Pseudopotentials thus reduce the computational cost of plane wave calculations in two ways: (i) one no longer needs to include high-energy plane waves to resolve core orbitals, and (ii) the total number of electrons in the wave function is limited to those that reside within the valence space. Pseudopotentials can also be parameterized to capture physics beyond that contained in the Coulomb Hamiltonian, such as scalar relativistic effects \cite{Cao12_403}. In this work, we focus on the Goedecker-Teter-Hutter (GTH) \cite{GTH1996} and Hartwigsen-Goedecker-Hutter (HGH) \cite{HGH1998} parameterizations of the norm-conserving pseudopotential, the latter of which accounts for scalar relativistic effects.  This parameterization of the norm-conserving pseudopotentials has an analytical form for the necessary matrix elements that is simple enough to implement coherently with a quantum circuit.

Reference~\cite{Shokrian} proposed the idea of using pseudopotentials to improve efficiency of first-quantized plane-wave basis simulations and proposed a method for simulating non-cubic unit cells. 
Their proposal leverages the fact that the QROM \cite{PhysRevX.8.041015} primitive can be used for state preparation, and the pseudopotential becomes substantially simpler if angular momentum projectors are ignored (see Eq.~(29) of that work). There are two significant drawbacks to their approach that render them of questionable utility in practice: 1) the QROM-based state preparation for the pseudopotential (based on Grover-Rudolph \cite{Grovrud}) requires QROM to output as much data as the entire plane-wave basis leading to very high costs and even more importantly, 2) ignoring off-diagonal angular momentum projectors of the pseudopotential results in energy errors that are hundreds to thousands of times chemical accuracy \textit{per atom}. Errors of this magnitude simply mean one is simulating a completely different Hamiltonian than the intended system. In Section~\ref{sec:omiterr}, we confirm this error numerically for large-core pseudopotentials by demonstrating that LDA DFT energies differ by as much as Hartrees when ignoring off-diagonal angular momentum projectors. While it is possible to extend the QROM heavy approach of Ref.~\cite{Shokrian} to include higher angular momentum terms this would incur a substantial increase in the amount of data output by QROM and could give a further order of magnitude increase in Toffoli costs. A further difficulty with QROM in the form used in that work (with many ancilla qubits), is that there is a large overhead of Clifford gates and thus a Toffoli-only cost model is less appropriate for estimating quantum resources.  Finally, their non-cubic unit cell treatment uses an even distribution of grid points regardless of the simulation cell dimensions, resulting in an uneven discretization of momentum space.  Thus, while the use of pseudopotentials in first-quantized plane-wave calculations has been discussed in prior literature, a scalable and practical quantum simulation strategy incorporating these features has not been articulated.

Our approach provides a scalable block encoding cost while appropriately generalizing to non-cubic unit cells. In order to avoid the large ancilla and Toffoli complexity associated with QROM, we only use the QROM primitive for data that scales linearly with the number of atoms, or logarithmically in the basis size. We primarily use arithmetic to implement the SELECT oracle in order to
to yield (poly)logarithmic scaling in the total basis set size.
That is an \emph{exponential} improvement over the approach of Ref.~\cite{Shokrian}.

In our approach, the highest complexity step in SELECT is (like the approach without pseudopotentials \cite{BabbushContinuum, SuPRXQuantum21}) still $\widetilde{\mathcal{O}}(\eta)$ for $\eta$ electrons. Specifically, we use a QROM to output position data for the nuclei, as well as for function approximation by interpolation in order to efficiently compute the primitive function composing the GTH-pseudopotential -- the negative exponential.
The local and nonlocal components of the pseudopotential use the same functional form and thus we minimize costs by using the same set of arithmetic primitives to compute the core common functions. Further leveraging coherent arithmetic, we are able to use a plane-wave basis in a non-cubic unit cell and variable grid resolutions in each Miller direction. Accounting for different numbers of bits in each Miller direction requires modification to the block encoding primitives. While the omission of large components of the nonlocal pseudopotential and the different handling of non-cubic unit cells make it difficult to directly compare our block encoding costs with those in Ref.~\cite{Shokrian}, our arithmetic approach for the entire Hamiltonian is approximately two orders of magnitude more efficient than the previous strategy that implemented the incomplete operator.

To demonstrate the cost savings in block encoding and to frame the total quantum resources needed to resolve an open materials question we perform resource estimation on the heterogeneous catalysis systems for carbon-monoxide adsorption on transition metal catalysts. Accurate calculations of the energetics of CO adsorption is important in a number of applications such as steam reforming of fossil fuels~\cite{jones2008first}, methanol synthesis~\cite{grabow2011mechanism, behrens2012active}, water-gas-shift
and its reverse~\cite{grabow2008mechanism}. In order to perform resource estimates at grid resolutions and system sizes that are realistic for resolving the CO adsorption question we perform DFT calculations on a variety metal-CO systems to resolve the number of plane waves and grid resolution needed to accurately resolve binding energies. Using the grid resolution estimates we estimate the total Toffoli complexity to perform phase estimation to chemical precision ($1.6 \times 10^{-3}$ Hartree). We estimate that while the block encoding costs are on the order of $10^{4}$ in Toffoli gates the total costs for phase estimation are substantial (on the order of $10^{14}$ Toffoli gates) due to the large $\lambda$-value associated with the weight in the block encoding of the Hamiltonian.

A full block encoding of the pseudopotential now allows us, for the first time, to accurately compare first quantized simulation methods to second quantized simulations in order to identify which encoding strategy offers the current best approach for simulating materials. We compared the second quantized simulation costs for the Lithium-Nickel-Oxide (LNO) cathode simulation, recently presented in Ref.~\cite{PRXQuantum.4.040303}, to first quantized simulation costs. We select a grid resolution by DFT calculations of LNO system with increasing cutoff energies. We find that while first quantized simulation, costing about $10^{14}$ Toffoli gates, are more expensive than second quantized simulations, costing about $5.0\times 10^{12}$ Toffoli gates, the space resources for first quantization are substantially lower.  For the LNO system the dominant space complexity is the system register using about 1,400 qubits while for second quantized calculations the dominant space complexity is QROAM costing 75,000 qubits.

In Section~\ref{sec:Bravais} we describe the method of including non-cubic unit cells in the analysis.
Then in Section~\ref{sec:GTH} we describe the GTH pseudopotential and the principle of the method used to block encode it.
In Section~\ref{sec:omiterr} we analyze the error incurred by omitting terms in the pseudopotential.
We give an overview of the method of block encoding the GTH pseudopotential in Section~\ref{sec:blockenc}, then go into details of the cost of the block encoding in Section~\ref{sec:GTHcosting}.
The cost of block encoding other parts of the Hamiltonian is explained in Section~\ref{sec:othercosts}.
The overall cost needs calculation of the $\lambda$-value of the block encoding, which is addressed in Section~\ref{sec:lambdas}.
These results are used to determine cost estimates for examples of interesting materials in Section~\ref{sec:examples}, then we conclude in Section~\ref{sec:conclude}.

\section{The plane-wave basis from non-orthogonal Bravais vectors}\label{sec:Bravais}
Most work simulating first-quantized plane-wave basis electronic structure on a quantum computer has either been confined to non-periodic systems, or systems that have the simplest possible periodicity; e.g.\ cubic periodicity. Cubic periodicity results in a simple expression for the reciprocal vectors (as scaled Euler vectors) and thus the computational basis states of a quantum register are used as Miller indices leading to an exponential space savings. Furthermore, the arithmetic needed for block encoding is simplified by the fact that the Gramian of reciprocal lattice vectors for a cubic unit cell is proportional to the identity.
For non-cubic unit cells the reciprocal lattice vectors no longer have a simple form (most generally computed by scaled cross products). We still use the computational basis of a quantum register to represent Miller indices, but we require more complicated arithmetic to block encode using a plane-wave basis generated from reciprocal lattice vectors whose Gramian does not form a scaled identity matrix. Computing basis state norms in the generalized basis can involve up to six fixed-point multiplications of the squared Miller indices with elements of the reciprocal lattice Gramian. 

In what follows, we introduce the details of the basis and the representation we use in the quantum calculation. The three-dimensional simulation cell is described with three vectors for the sides of the cell arranged as rows in a matrix denoted $\boldsymbol{a}$. Given the geometry of the simulation cell the reciprocal lattice vectors are described as
\begin{align}
\boldsymbol{g}^{(1)}= \frac{2\pi}{\Omega}\boldsymbol{a}_{2} \times \boldsymbol{a}_{3} \, , \qquad \boldsymbol{g}^{(2)}= \frac{2\pi}{\Omega}\boldsymbol{a}_{3} \times \boldsymbol{a}_{1}  \, , \qquad \boldsymbol{g}^{(3)} = \frac{2\pi}{\Omega}\boldsymbol{a}_{1} \times \boldsymbol{a}_{2} \, ,
\end{align}
where $\Omega$ is the volume of the simulation cell computed by taking the determinant of $\boldsymbol{a}$. We will not be using $k$-point sampling and thus commonly the simulation cell will be multiples of the primitive cell (unit cell containing one lattice point).
Examples of vectors for different classes of materials are given in Table \ref{TAB:LATTICE_VECTORS} in Appendix \ref{app:vectors}.
For simulation cells that are multiples of the primitive cells, the vectors can be easily scaled.
For example, using two primitive cells in the first direction doubles $\boldsymbol{a}_{1}$ and halves $\boldsymbol{g}^{(1)}$.

The reciprocal lattice vectors can be efficiently computed by $[\boldsymbol{g}^{(1)},\boldsymbol{g}^{(2)},\boldsymbol{g}^{(3)}] = 2\pi (\boldsymbol{a}^{-1})^{T}$. For cubic simulation cells the reciprocal lattice vector, now only shown for $\boldsymbol{g}^{(1)}$, is simply
\begin{align}
\boldsymbol{g}^{(1)} = 2\pi \boldsymbol{a}_{1}/|\boldsymbol{a}_{1}| = \frac{2\pi}{\Omega^{1/3}}\boldsymbol{e}_{1} \, ,
\end{align}
where $\boldsymbol{e}_{1}$ is the unit vector along the primitive cell direction $\boldsymbol{a}_{1}$.

The basis is indexed by the Miller indices $(p_x,p_y,p_z)$, a 3-vector of integers describing reciprocal cell shifts.
For a cubic simulation cell the momentum vector is proportional to the vector of Miller indices, so the inner product would be computed as $\langle k_{p}, k_{q}\rangle=\frac{4\pi^{2}}{\Omega^{2/3}}\left(p_{x}q_{x}+p_{y}q_{y}+p_{z}q_{z}\right)$.
In contrast, for the general reciprocal lattice vectors we would compute the momentum vector as
\begin{equation}
\label{eq:non_orthogonal}
k_p = \left(p_x \boldsymbol{g}^{(1)} + p_y \boldsymbol{g}^{(2)} + p_z \boldsymbol{g}^{(3)}\right) \, , \qquad p_{a} \in \left[-\frac{N_{a}-1}{2},\frac{N_{a}-1}{2}\right] \subset \mathbb{Z} \, .
\end{equation}
Then the inner product of basis vectors is obtained by treating $k_{p}, k_{q}$ as Cartesian vectors to give
\begin{equation}
\label{eq:innerprod_nonorthog}
    \langle k_{p},k_{q}\rangle=
    \begin{bmatrix}
        p_x & p_y & p_z
    \end{bmatrix}
    \begin{bmatrix}
        g_{11} & g_{12} & g_{13}\\
        g_{21} & g_{22} & g_{23}\\
        g_{31} & g_{32} & g_{33}\\
    \end{bmatrix}
    \begin{bmatrix}
        q_x \\ q_y \\ q_z
    \end{bmatrix} ,
\end{equation}
where $g_{ij} = (\boldsymbol{g}^{(i)})^{T}\boldsymbol{g}^{(j)}$.
Here $[g_{ij}]$ are elements of the Gramian matrix that is real, symmetric, and positive definite.
The squared norm of basis function, $\|k\|^2$, is now computed as
\begin{equation}
    \|k\|^2 = g_{11} p_x^2 + g_{22} p_y^2 + g_{33} p_z^2 + 2 g_{12} p_x p_y + 2 g_{23} p_y p_z + 2 g_{13} p_x p_z \, .
\end{equation}

When considering sets of Miller indices with different numbers of points in each direction we would use three registers to store signed integer sets
\begin{equation}
    G = \left[-\frac{N_x-1}{2},\frac{N_x-1}{2}\right]\otimes \left[-\frac{N_y-1}{2},\frac{N_y-1}{2}\right] \otimes \left[-\frac{N_z-1}{2},\frac{N_z-1}{2}\right]  \subset \mathbb{Z}^3 \, ,
\end{equation}
which will use $n_x,n_y,n_z$ bits in each direction
\begin{equation}\label{eq:bits_to_Nx}
    2^{n_x-1}-1 = \frac{N_x-1}{2} \quad \implies \quad  N_x = 2^{n_x}-1
\end{equation}
and similarly for $y$ and $z$.
We will also consider a set for differences in momenta
\begin{equation}
    G_d = \left[-(N_x-1),N_x-1\right]\otimes \left[-(N_y-1),N_y-1\right] \otimes \left[-(N_z-1),N_z-1\right]  ,
\end{equation}
where the width in each direction is doubled.
We also consider $G_0$, which is $G_d$ excluding the origin.
That is used where $1/\|k_\nu\|^2$ weights are required, but we can use $G_d$ for pseudopotentials because they do not diverge at $\nu=0$.

For indexing $G_d$ or $G_0$, we have maximum absolute values of $N_x-1=2^{n_x}-2$, and similarly for $y$ and $z$.
That means that we can use $n_x,n_y,n_z$ qubits for magnitudes of components of vectors $\nu$ in $G_d$ or $G_0$, as well as sign qubits.
Note also that the number of points in $G$ is $N=N_x N_y N_z$, whereas the number of points in $G_d$ is $(2N_x-1)(2N_y-1)(2N_z-1)<8N$.

To compute $\|k\|^2$ using coherent arithmetic, there are up to three squares of integers $p_x,p_y,p_z$, three products of integers, and six multiplications of integers by the real numbers $g_{ij}$.
In practice, we will always be multiplying $\|k\|^2$ by a real number in another part of the algorithm, so we can scale $g_{ij}$ and one of those multiplications is not required.
Moreover, for real examples there are often simplifications where values of $g_{ij}$ are zero or equal to each other.
That enables us to eliminate or combine multiplications by real numbers, as well as to eliminate the need for some of the products of integers.
The simplifications for specific examples are detailed in Appendix \ref{app:bravaisnorm}.  We note that while the aforementioned strategy involves some additional arithmetic it does easily allow for differing resolution in each Miller index direction.  This allows one to maintain a uniform sampling in reciprocal space no matter how non-cubic the simulation cell. In Ref.~\cite{Shokrian} a uniform distribution of the total number of plane waves is taken in each Miller index direction which can lead to non-uniform simulation grids and ultimately skew results. Modifying the grid size in each direction requires modification of the block encoding oracles discussed later.

\section{The Goedecker-Teter-Hutter separable dual-space Gaussian pseudopotential}
\label{sec:GTH}
\subsection{Local and nonlocal pseudopotentials}
Nonlocal pseudopotentials are perhaps the most popular type of pseudopotential. This family of pseudopotentials includes norm-conserving pseudopotentials. As an example, Kleinman-Bylander pseudopotentials can be written as
\begin{equation}
v_{\rm pp} =
	\sum_{\ell=1}^L \left( v^{\alpha_\ell,R_{\ell}}_{\rm loc} + v^{\alpha_\ell, R_{\ell}}_{\rm nonloc}\right) \, ,
\end{equation}
where $\ell$ indexes the $L$ nuclei in the simulation.
In this equation $v^{\alpha_\ell,R_{\ell}}_{\rm loc}$ is the local Coulomb potential (e.g.\ this might be expressed as $Z_{\alpha_\ell} / (r-R_{\ell})$ in position space), where we use $\alpha_\ell$ to indicate the species (type of nuclei) associated with the $\ell$ index, $R_{\ell}$ is used to indicate the coordinates of the nuclei, and $Z_{\alpha}$ is the charge of a nucleus of species $\alpha$ (including the charge associated with the core). We can write the form of a Kleinman-Bylander nonlocal pseudopotential component as
\begin{equation}
v^{ \alpha,R}_{\rm nonloc} =   \sum_{l,m} \, \ket{\chi^{ \alpha,R}_{lm} } E_{l} \bra{ \chi^{\alpha, R}_{lm}} ,
\end{equation}
where $l$ and $m$ are the quantum numbers corresponding to angular momentum and the projection of angular momentum. The nonlocal pseudopotentials are written this way because the purpose of the nonlocal part is to capture the angular momentum dependent interactions of the core electrons that we remove using the pseudopotential. 

In this paper we will focus on implementing a specific type of nonlocal pseudopotential known as the Goedecker-Teter-Hutter (GTH) pseudopotential \cite{GTH1996}, which has been popularized by CP2K developers over many years. Specifically, we focus on the relativistic parameterization of this pseudopotential introduced by Hartwigsen, Goedecker and Hutter (HGH) \cite{HGH1998}. These two papers have been cited over ten thousand times in total, and are among the most commonly used pseudopotentials.
A comprehensive review of the pseudopotential literature \cite{PP2016} found that the HGH parameterization of the GTH pseudopotential is comparable in accuracy to other accurate norm-conserving, ultrasoft, and projector-augmented wave pseudopotentials. This study also demonstrated that for consistency of equations-of-state (EOS) calculations, using the HGH parameterization yields a less consistent root-mean-square error than PAW, when comparing against all-electron calculations, by a factor of 2-3. The consistency of EOS does not benchmark absolute accuracy though and norm-conserving pseudopotentials like those of HGH are considered accurate~\cite{PP2016}. Here we focus on HGH parameterizations as a step towards high-accuracy quantum computations of real materials.

The local component of the GTH pseudopotential can be expressed as follows in real space:
\begin{align}
    {}^{\rm GTH} v_{\rm loc}^{\alpha,R}\left(r\right) &= \left[-\frac{Z_\alpha}{r-R} \textrm{erf}\left(\frac{r-R}{\sqrt{2}\, r_{\rm loc}^\alpha}\right) \right. \nn
    & \quad \left. + \exp\left(-\frac{1}{2}\left(\frac{r-R}{r_{\rm loc}^\alpha}\right)^2\right)\!\left(C_1^\alpha + C_2^\alpha \left(\frac{r-R}{r_{\rm loc}^\alpha}\right)^2 + C_3^\alpha \left(\frac{r-R}{r_{\rm loc}^\alpha}\right)^4 + C_4^\alpha \left(\frac{r-R}{r_{\rm loc}^\alpha}\right)^6 \right)\right]  .
\end{align}
We note that the $C_n^\alpha$ constants as well as $r_{\rm loc}^\alpha$ are all $\alpha$ (nuclear species type) dependent parameters.
Parameters that were optimized for a density functional calculation using the local density approximation (LDA) as an exchange-correlation function and $\alpha$ corresponding to several atoms are given in Appendix~\ref{app:lda_pp_params} Table \ref{TAB:GTH_LOCAL_LDA_PARAMETERS} \cite{HGH1998}. It is a special property of the GTH pseudopotential (beneficial for our purposes) that the local part also has an analytical form if expressed in the reciprocal space. This form of the pseudopotential would be useful if our intention was to implement the pseudopotential in a real space grid parameterization (for example, in conjunction with the algorithms of Kassal \emph{et al.}~\cite{Kassal2008} or with the algorithm described in Appendix K of \cite{SuPRXQuantum21}).  However, here we will focus on implementing the operator in momentum space. The Fourier transform of the local component of the GTH pseudopotential is
\begin{align}
\label{eq:locpp}
^{\rm GTH}v_{\rm loc}^{\alpha, R}\left(k\right) & =  - \frac{4 \pi Z_\alpha}{\Omega}e^{-ik \cdot R-(r_{\rm loc}^\alpha\left \|k \right\|)^2/2}\frac{1}{\|k\|^2} + \sqrt{8 \pi^3}\frac{(r_{\rm loc}^\alpha)^3}{\Omega} e^{-ik \cdot R-(r_{\rm loc}^\alpha\left \|k \right\|)^2/2}  \nonumber\\
& \quad \times
\left\{C_1^\alpha + C_2^\alpha\left[3-\left(r_{\rm loc}^\alpha\left \|k \right\|\right)^2\right] + C_3^\alpha\left[15 - 10 \left(r_{\rm loc}^\alpha\left \|k \right\|\right)^2 + \left(r_{\rm loc}^\alpha\left \|k \right\|\right)^4\right] \right. \nonumber\\
 & \quad \left. + \, C_4^\alpha\left[105 - 105 \left(r_{\rm loc}^\alpha\left \|k \right\|\right)^2 + 21 \left(r_{\rm loc}^\alpha\left \|k \right\|\right)^4 - \left(r_{\rm loc}^\alpha\left \|k \right\|\right)^6\right]\right\} \, .
\end{align}
The matrix elements of the local part of the pseudopotential are determined in the same way as for any other periodic one-body operator, i.e., $\langle G' | ^{\rm GTH}v_{\rm loc}^{\alpha, R}\left(\hat r\right) | G'' \rangle$ = $^{\rm GTH}v_{\rm loc}^{\alpha, R}\left(G'-G''\right)$.
Note that Eq.~\eqref{eq:locpp} diverges for $k = G'-G'' = 0$.
Since the case with $k=0$ would correspond to the identity we do not include it in the sum, and there is no problem with divergent values.

The matrix elements for the nonlocal contribution to the GTH pseudopotential are 
\begin{align}
^{\rm GTH}v^{\alpha,R}_\text{nonloc}(k_p,k_q)
& =
\frac{1}{\Omega}e^{-i(k_p-k_q) \cdot R}\sum_{l = 0}^{l_{\text{max}}^\alpha}
(-1)^{l}
\sum_{m = -l}^l
Y^{l}_m\left(\hat{k}_p\right) Y^{l}_m\left(\hat{k}_q\right) 
\sum_{i = 1}^3
\sum_{j = 1}^3
E_{l\alpha}^{ij}
F_{l\alpha}^i\left(\left\|k_p\right\|\right) F_{l\alpha}^j\left(\left\|k_q\right\|\right)\nn
& =
\frac{1}{\Omega}e^{-i(k_p-k_q) \cdot R}\sum_{l = 0}^{l_{\text{max}}^\alpha}(-1)^{l}
\frac{\left(2 l + 1\right)}{4\pi} P_l \left(\frac{k_p \cdot k_q}{\left\|k_p\right\| \left\|k_q\right\|}\right)
\sum_{i = 1}^3
\sum_{j = 1}^3
E_{l\alpha}^{ij}
F_{l\alpha}^i\left(\left\|k_p\right\|\right) F_{l\alpha}^j\left(\left\|k_q\right\|\right)
\end{align}
where we have taken the sum over the projection $m$ of the angular momentum $l$ so that
	\begin{equation}
	\sum_m Y^{l}_m\left(\hat{k}_p\right) Y^{l}_m\left(\hat{k}_q\right) = \frac{2 l + 1}{4 \pi} P_l \left(\frac{k_p \cdot k_q}{\left\|k_p\right\| \left\|k_q\right\|}\right) \, ,
	\end{equation}
	where $P_l(x)$ is the $l^{\rm th}$ Legendre polynomial of $x$.
In these expressions $l_{\text{max}}^\alpha$ depends on $\alpha$, and $E_{l\alpha}^{ij}$ is a constant tabulated for each atom type $\alpha$ and angular momentum.
The function
$F_{l\alpha}^j\left(\left\|k\right\|\right)$
takes the form
\begin{equation}
\label{eqn:projectors}
F^i_{l\alpha}(\|k\|) = \|k\|^l C_{li}^\alpha \left ( \sum_{x=0}^{i-1} c_{x, li}(r_l^\alpha\|k\|)^{2x}\right ) e^{-\frac{1}{2}(r_l^\alpha\|k\|)^2} ,
\end{equation}
where $C_{l i}^\alpha$, $\{c_{x,l i}\}$, and $r_l^\alpha$ are all tabulated constants available for each atom, angular momentum, and a projector index $i$.
The $C_{li}^\alpha$ and $c_{x,li}$ parameters that enter Eq.~\eqref{eqn:projectors} are given in Appendix~\ref{app:lda_pp_params} Table \ref{TAB:PROJECTORS}.
Note that $C_{li}^\alpha$ depends on $\alpha$ only through $r_l^\alpha$,
and the ratio $C_{li}^\alpha/(r_l^\alpha)^{l+3/2}$ is independent of $\alpha$.
Examples of $r_l^\alpha$ and $E_{l\alpha}^{ij}$ parameters that were optimized for LDA with $\alpha$ corresponding to several atoms are given in Appendix~\ref{app:lda_pp_params} Table \ref{TAB:GTH_NONLOCAL_LDA_PARAMETERS} \cite{HGH1998}.

\subsection{The complete Hamiltonian for the GTH pseudopotential}

For the complete Hamiltonian we need to sum $^{\rm GTH}v_{\rm loc}^{\alpha, R}\left(k\right)$ and $^{\rm GTH}v^{\alpha,R}_\text{nonloc}(k_p,k_q)$ over all $L$ nuclei, \emph{and} include the kinetic and two-electron parts of the Hamiltonian, which are unchanged from what we used in \cite{SuPRXQuantum21}.
We also need to sum over the electrons, so we have $k$ for each of the electron registers.
For each of the nuclei there is a distinct position of the nucleus $R_\ell$ (which is a vector).
The quantity $\alpha$ indexes the different types of nuclei.
For each type of nucleus, there are values tabulated for $Z_\alpha$, $r_{\rm loc}^\alpha$, $C_n^\alpha$, $l_{\max}^\alpha$, $r_{l}^\alpha$, $E_{l\alpha}^{ij}$, and $C_{li}^\alpha$, with $c_{x,li}$ independent of the nucleus.
The simplest case is that where there is only one type of nucleus, with the amount of data needed here growing as the number of types of nuclei.
To be more specific, we have
\begin{align}
H&= T + U_{\rm loc} + U_{\rm nonloc} +V \\
 T &=  \sum_{\elj=1}^{\eta}  \sum_{p\in G} \frac{\left \| k_p\right\|^2}{2} \ket{p}\!\!\bra{p}_{\elj} \\
U_{\rm loc} &= \sum_{\ell=1}^L \sum_{\nu\in G_0} \locl\!\left(k_\nu\right) \left( e^{-\mathrm{i}k_{\nu}\cdot R_\ell}\sum_{\elj=1}^\eta \sum_{\substack{q\in G\\ (q-\nu) \in G}} \ket{q-\nu}\!\!\bra{q}_{\elj} \right) \\
U_{\rm nonloc} &= \sum_{\ell=1}^L\sum_{\nu\in G_d} e^{-\mathrm{i}k_{\nu}\cdot R_\ell} \sum_{\elj=1}^{\eta}\sum_{\substack{q\in G\\ (q-\nu) \in G}} \nonlocl(k_q,k_{q-\nu})  \ket{q-\nu}\!\!\bra{q}_{\elj} \\
    	V &=\frac{2 \pi}{\Omega} \sum_{\nu\in G_0}\frac{1}{\left\| k_{\nu}\right\|^2} \left( \sum_{\eli\neq {\elj}=1}^{\eta} \sum_{\substack{p,q\in G\\(p+\nu)\in G\\(q-\nu)\in G}} \ket{p + \nu}\!\!\bra{p}_{\eli} \ket{q-\nu}\!\!\bra{q}_{\elj} \right)
\end{align}
where
\begin{align}
    \loc\!\left(k_\nu\right) & \coloneqq e^{\mathrm{i}k_{\nu}\cdot R_\ell}\times  {}^{\rm GTH}v_{\rm loc}^{\alpha, R_\ell}\!\left(k_\nu\right) \\
    \nonloc(k_q,k_{q-\nu})  & \coloneqq e^{\mathrm{i}k_{\nu}\cdot R_\ell}\times  {}^{\rm GTH}v^{\alpha,R_\ell}_\text{nonloc}(k_q,k_{q-\nu}) 
\end{align}
are real coefficients where we have removed the phase factor.
Note that these now no longer depend on the nuclear coordinate $R_\ell$.
We have modified the description of the nuclear part of the Hamiltonian $U_{\rm loc}$ and $U_{\rm nonloc}$ so that it is a sum over $q$ and the difference $\nu=q-p$.
The complete Hamiltonian is similar to that in prior work \cite{SuPRXQuantum21}, except there the potential $U$ was
\begin{equation}
    	U = -\frac{4\pi}{\Omega}\sum_{\ell=1}^{L} Z_\ell \sum_{\nu\in G_0}\frac{1}{\norm{k_{\nu}}^2}\left(e^{-\mathrm{i}k_{\nu}\cdot R_\ell} \sum_{\elj=1}^{\eta}\sum_{\substack{q\in G\\ (q-\nu) \in G}}\ket{q-\nu}\!\bra{q}_{\elj} \right).
\end{equation}
Therefore, for the simulation we can apply many of the techniques from before, except that now we need to account for the much more complicated form of $U$.

\subsection{Method for block encoding of \texorpdfstring{$U$}{U}}

The way that this work differs from Ref.~\cite{SuPRXQuantum21} is that we have a different form of $U$, composed of functions involving sums where each term involves an exponential with a different argument.
If we were to attempt to perform the block encoding using coherent arithmetic for the entire sum with repeated calculation of the exponential, then the complexity would be large due to the repeated coherent arithmetic.
Instead, we introduce a method to block encode these functions using coherent arithmetic with only a single evaluation of the exponential.
We will explain this in more detail for just the case of $U_{\rm nonloc}$, and it is straightforward to combine this method with the block encoding of $U_{\rm loc}$, which also requires an exponential.

\begin{figure}[tbh]
\includegraphics[width=12.4cm]{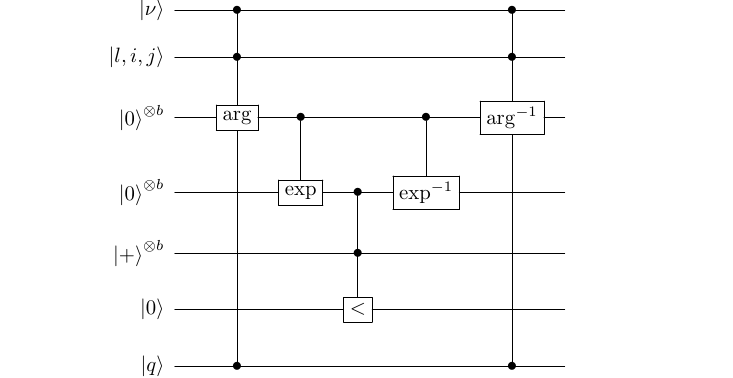}
\caption{\label{fig:combinedencoding}A circuit showing the block encoding of a simplified form of $U_{\rm nonloc}$.
For simplicity we are ignoring the sums over nuclei and electrons.}
\end{figure}

To illustrate, a simplified form of the circuit is shown in Figure~\ref{fig:combinedencoding}, which is omitting details such as the sums over the nucleus and electron number for simplicity.
The register $\ket{\nu}$ is an ancilla prepared for the block encoding, and $\ket{q}$ is the target system.
We also include another register with indices $l,i,j$ corresponding to indices in the sum for $\nonlocl(k_q,k_{q-\nu})$.
This register is also prepared in a superposition state before using this circuit.
The values of $q,\nu,l,i,j$ are all used to calculate the argument of the exponential function in an ancilla register with $b$ bits (indicated by $\ket{0}^{\otimes b}$), shown as `arg'.
That is then used to calculate the exponential function once.
The result is then used in an inequality test with a register in an equal superposition state to provide a result in a flag register that is used to flag `success', thereby achieving the correct amplitude in the block encoding.

The calculations then need to be inverted to reset the ancilla qubits, shown as $\exp^{-1}$ and arg$^{-1}$.
This inversion normally doubles the cost of the calculation, but it is also possible to invert the calculations using Clifford gates.
That can be achieved if ancilla qubits are preserved from the initial calculation, so that the Toffolis for the inverse calculation can be performed with measurements and Clifford gates.
In many cases the overall number of qubits needed just to store the system state is large and the number of Toffolis for the calculation is small, so it is preferable to perform the inversion in this way.

This is the key method to modify the block encoding to enable efficient encoding of the pseudopotentials.
Below in Section \ref{sec:blockenc} we give a detailed explanation of how the block encodings of the different parts are combined, the methods for the state preparation for this block encoding, and the methods for performing the coherent arithmetic including efficient calculation of the exponential.

\section{Errors from omitting angular momentum projectors}
\label{sec:omiterr}

Before describing the block encoding of the pseudopotential Hamiltonian and the generalized plane-wave basis, we first motivate the need to simulate the entire nonlocal pseudopotential by quantifying the error incurred by ignoring 
all but one projector for each angular momentum function, 
as is done in Ref.~\cite{Shokrian}. Energetically, this approximation appears to be a crude one. We have computed LDA energies for three systems containing metals for which the HGH nonlocal pseudopotential has up to three projectors for at least one angular momentum function: Al, Mn, and Ni. Table \ref{TAB:LDA_ENERGIES} provides LDA energies computed using full HGH-type pseudopotentials for varying numbers of plane waves ($N_{\rm pw}$). Also provided are errors incurred from the use of approximate HGH-type pseudopotentials (in parentheses). The error is evaluated as the difference in the LDA energy computed using the approximate pseudopotential and the full one, both using the same orbitals obtained from self-consistent calculations with the full pseudopotential (in the DFT calculation, a determinant constructed from these orbitals approximates the eigenstate).
Approximating the nonlocal part of the pseudopotential can lead to errors on the order of hundreds of milliHartree (mE$_{\rm h}$). 

Consider first AlN; the HGH pseudopotential for Al has up to two projectors per angular momentum function. For the primitive cell of AlN, calculations with large numbers of plane waves demonstrate that large errors ($\approx$ 0.1658 Hartree (E$_{\rm h}$) with 15,525 plane waves) persist  with reasonably converged calculations. These errors are extensive -- i.e.\ scale with simulation cell size. For the $1\times 1\times 2$ supercell, the error grows to $\approx$ 0.4058 E$_{\rm h}$ for a calculation performed with 16,965 plane waves. Comparable errors are observed in LDA calculations using Gaussian-type basis functions performed in PySCF. For example, for the primitive cell of AlN, described by the GTH-cc-pVQZ basis set~\cite{ye2022correlation}, energies evaluated using full and approximate pseudopotentials differ by 0.1653 E$_{\rm h}$. Consider now NiOF; the HGH pseudopotential for nickel has up to three projectors per angular momentum function. For the primitive cell with 14,981 plane-wave basis functions, LDA energies evaluated with full versus approximate pseudopotentials differ by 0.0686 E$_{\rm h}$. The approximate pseudopotential also leads to large energy errors for LiMnO. For the primitive cell, described by 12,673 plane-wave basis functions, this error is nearly one-quarter of a Hartree. 

\begin{table*}[tbh]
    \begin{tabular}{|c|c|c|c|c|}
    \hline
    \hline
    system & super cell size &  $N_{\rm pw}$ & energy (E$_{\rm h}$) & band gap (eV) \\
    \hline
AlN (wurzite) & $1\times 1 \times 1$ & 1087    &   -22.4279  ( -0.1656 )  &  3.51  ( -3.28 ) \\
              &                      & 2929    &   -22.8956  ( -0.1655 )  &  3.57  ( -3.28 ) \\
              &                      & 5479    &   -22.9905  ( -0.1657 )  &  3.59  ( -3.28 ) \\
              &                      & 8497    &   -23.0084  ( -0.1657 )  &  3.60  ( -3.28 ) \\
              &                      & 11837   &   -23.0120  ( -0.1658 )  &  3.60  ( -3.28 ) \\
              &                      & 15525   &   -23.0128  ( -0.1658 )  &  3.60  ( -3.28 ) \\

              &                      & GTH-cc-pVQZ &  -23.0060 ( -0.1653 )     &  3.59 ( -3.22 )     \\ \cline{2-5}
              & $1\times 1 \times 2$ & 2159    &   -45.4762  ( -0.4111 )  &  3.83  ( -3.44 ) \\
              &                      & 5939    &   -46.3985  ( -0.4062 )  &  3.89  ( -3.43 ) \\
              &                      & 10995   &   -46.5768  ( -0.4059 )  &  3.93  ( -3.43 ) \\
              &                      & 16965   &   -46.6109  ( -0.4058 )  &  3.93  ( -3.43 ) \\ \hline
NiOF          & $1\times 1 \times 1$ & 2331    &   -73.7002  ( -0.0785 )  &  0.65  ( -0.60 ) \\
              &                      & 4229    &   -75.3878  ( -0.0694 )  &  0.71  ( -0.71 ) \\
              &                      & 6529    &   -75.6967  ( -0.0688 )  &  0.73  ( -0.73 ) \\
              &                      & 11879   &  -75.7986  ( -0.0687 )   &  0.73  ( -0.73 ) \\
              &                      & 14981   &  -75.8087  ( -0.0686 )   &  0.73  ( -0.73 ) \\  \cline{2-5}
              & $2\times 1 \times 1$ & 4585    & -147.5385  ( -0.1774 )   &  0.26  ( -0.12 ) \\
              &                      & 8443    & -151.1023  ( -0.1956 )   &  0.65  ( -0.51 ) \\
              &                      & 12983   & -151.7137  ( -0.1929 )   &  0.61  ( -0.47 ) \\ \hline
LiMnO         & $1\times 1 \times 1$ & 3227    &  -29.6997  ( -0.8034 )   &  1.07  ( -0.16 ) \\
              &                      & 5931    &  -31.1207  ( -0.5136 )   &  0.79  ( -0.57 ) \\
              &                      & 9149    &  -32.5721  ( -0.3470 )   &  1.36  (  0.02 ) \\
              &                      & 12673   &  -33.8616  ( -0.2495 )   &  1.58  ( -0.50 ) \\
    \hline
    \hline
        \end{tabular}
    \caption{LDA energies in Hartree (E$_{\rm h}$) computed using full HGH pseudopotentials; the numbers in parentheses are errors in energies from approximate pseudopotentials that consider only one projector per angular momentum function.
    Band gaps in eV computed with full HGH pseudopotentials are also provided, with errors in band gaps from approximate pseudopotentials in parentheses. 
    AlN and LiMnO are treated as having singlet spin states. For NiOF, the primitive cell is treated as doublet, and the supercell is treated as a singlet. Energies and band gaps computed using the approximate pseudopotentials are evaluated using orbitals obtained from self-consistent LDA calculations using the full pseudopotentials.}
    \label{TAB:LDA_ENERGIES}
\end{table*}

In Table \ref{TAB:LDA_ENERGIES}, errors incurred from the use of the approximate pseudopotentials were evaluated with the state (the determinant of Kohn-Sham orbitals) optimized with the full pseudopotentials. These errors can be substantially larger when the state is optimized using the approximate pseudopotential. For example, for the primitive cell of AlN and 2,929 plane waves, the LDA energy with the approximate pseudopotential is -23.3658 E$_{\rm h}$, which differs from the LDA energy using the full pseudopotential and the same number of plane-wave basis functions by 0.3753 E$_{\rm h}$. These large energy differences suggest that the orbitals / state obtained from fully self-consistent LDA simulations using the approximate pseudopotential differ significantly from those obtained from calculations using the full pseudopotential. In other words, one could, in principle, obtain qualitatively different solutions depending on the treatment of the pseudopotential. These data suggest that similarly large errors would be observed in the quantum algorithm, where the energies correspond to eigenstates of different Hamiltonians when using either the full or approximate pseudopotential. The eigenstates simulated in these cases could themselves differ significantly.
Note also that LDA calculations take far longer to converge with the approximate pseudopotential than with the full pseudopotential. With our implementation in qcpanop~\cite{qcpanop}, the LDA calculations with 5,749 plane waves required 297 and 18 iterations, using the approximate and full pseudopotentials, respectively, to converge the energy to $10^{-8}$~E$_{\rm h}$ and the orbital gradient to $10^{-6}$ E$_{\rm h}$.

Aside the impact on absolute energies, approximating the nonlocal pseudopotential terms as is done in Ref.~\cite{Shokrian} leads to substantial errors in energy differences, as well. Table~\ref{TAB:LDA_ENERGIES} provides band gaps from self-consistent LDA calculations carried out using full HGH-type pseudopotentials, as well band gaps derived from Kohn-Sham matrices constructed using approximate pseudopotentials and orbitals obtained from the self-consistent calculations carried out with the full pseudopotentials. Clearly, errors in band gaps can be substantial. For the primitive cell of AlN, the LDA bandgap computed using the full HGH pseudopotential and 15,525 plane-wave basis functions is 3.60 eV, and this gap reduces to only 0.32 eV when using the approximate pseudopotential. As in the case of absolute energies, we observe similar errors from calculations using Gaussian-type basis functions. Calculations on NiOF and LiMnO confirm that band gaps change dramatically when approximating the non-local pseudopotential. In particular, LDA with the full pseudopotential predicts NiOF to be a semiconductor, with a band gap of $\approx$ 0.6 eV -- 0.7 eV depending on the size of the simulation cell. On the other hand, the approximate pseudopotential would lead us to conclude that NiOF is gapless. 

These data confirm the substantial error incurred by limiting the number of projectors per angular momentum function to one in the nonlocal pseudopotential. We have observed not only large absolute energy errors due to this approximation, but also larger errors in energy differences, which could lead to qualitatively incorrect predictions. Thus, when using pseudopotentials, it is important to simulate the \textit{entire} Hamiltonian or numerically confirm that, for the particular system of interest, the truncation incurs minimal energy error. 

\section{Block encoding the GTH pseudopotential}
\label{sec:blockenc}

First we will give the general overview of how the block encoding of the Hamiltonian is performed in prior work, then we explain how we amend the block encoding for the pseudopotential.
The general principles for the block encoding from \cite{SuPRXQuantum21} are as follows.
\begin{itemize}
    \item Two qubits are used to select between the $T$, $U$, and $V$ parts of the Hamiltonian.
    \item A special superposition state is produced with weighting as $1/\|\nu\|$ that is in common between the block encodings of $U$ and $V$.\label{second}
    \item Equal superposition states are used for $\eli$ and $\elj$ (selecting the electron registers), and a weighted superposition is used for $\ell$.
    \item The registers encoding $\eli$ and $\elj$, which encode the numbers for electron registers, are used to control the swap of the values in these registers into ancilla registers, the select operation is performed on the ancilla, and it is swapped back.
    This avoids an overhead of $\eta$ (the number of electrons) from performing the select operation on all $\eta$ electron registers.
    \item The value of $\nu$ is subtracted from one ancilla for both $U$ and $V$.
    \item The value of $\nu$ is added to the other ancilla only for $V$.
    \item For $U$ alone the phase factor $e^{-\mathrm{i}k_{\nu}\cdot R_\ell}$ is applied, where $R_\ell$ was output via a QROM on $\ell$.
    \item The complexity of performing arithmetic for the kinetic part is completely avoided by expressing the sum of squares as a linear combination of bitwise products.
\end{itemize}

In this work we are changing the potential $U$, so the block encoding of this component needs to be changed.
Previously, the preparation over $\nu$ with amplitudes $1/\|\nu\|$ was in common with that needed for $V$, so did not require additional complexity.
Here we allow non-orthogonal Bravais vectors, so need to replace $\|\nu\|$ with $\|k_\nu\|$.
Moreover, we have a more complicated function of $\nu$ required for the local part of the pseudopotential.
The nonlocal part of the pseudopotential is even more complicated, because the factor that is applied depends on the momentum state $\ket{q}$.
This means that an amplitude factor needs to be applied according to $\ket{q}$ as well as having a preparation over $\nu$.
This is no longer unitary, but can be block encoded by standard techniques of using an inequality test with an equal superposition state.
For consistency with the terminology used for linear combinations of unitaries, we will separate the block encoding into a `prepare' step and a `select' step.
Unlike the case of a linear combination of unitaries the `select' is not unitary, but also involves a state-dependent amplitude; nevertheless, the principle is similar.

Other than these amendments to the block encoding, the other parts can still be performed relatively unchanged.
We still perform controlled swaps of momentum registers into working registers,
we still perform subtraction and addition of $\nu$ from the ancilla registers,
and we still apply the phase factor $e^{-\mathrm{i}k_{\nu}\cdot R_\ell}$.

There are a number of choices in how we can block encode the local and nonlocal parts of the pseudopotential.
There is a tradeoff between the complexity of the block encoding and the value of $\lambda$.
In order to minimize $\lambda$, we would perform a tight initial preparation over $\nu$ using amplitude amplification similar to the preparation of $1/\|k_\nu\|$ for $V$.
Then we would apply an amplitude factor for the nonlocal part of the pseudopotential.
The functions of $\|k_\nu\|$ required could be applied using QROM interpolation, but the amplitude amplification requires computation of the function three times.
That is because there is a computation, reflection on an ancilla, inversion of the computation, reflection about zero, then computation again.
Some improvement in non-Clifford cost can be obtained by retaining working qubits to invert computations, but this is still a high-complexity step.
Moreover we need separate functions for each species of nucleus, further increasing the complexity.

Then the application of the amplitude factor for the nonlocal part of the pseudopotential has a high complexity because it depends on both the momentum $q$ and $\nu$.
The QROM approach to function interpolation is for functions of a single variable, and interpolation of a function of two would be unreasonably high cost.
For example, if there were $200$ points needed to interpolate a function of one variable, there would then be 40,000 needed to interpolate a function of two.
For this reason, we would aim to interpolate the individual functions $F_{l\alpha}^i$, and use those to determine the overall pseudopotential.
However, that would also be high cost because we would need to interpolate $9$ functions of both $q$ and $q-\nu$, compute three Legendre polynomials, and perform many multiplications and additions to determine the complete pseudopotential.

To avoid the overhead from separately interpolating and computing all these functions, we can instead break up the sum over $l,i,j$ for the nonlocal pseudopotential.
We can prepare a superposition over these three indices, and perform controlled operations based on them.
This substantially reduces the complexity, but at the cost of increasing $\lambda$, because we need to account for the sum of the absolute values over these three indices, rather than summing first then taking the absolute value.

The drawback to separating out the sum over $l,i,j$ in the preparation stage is that the function of $\nu$ now needs to depend on $l,i,j$.
Moreover these functions are complicated, and would need to all be individually interpolated in the preparation further increasing the complexity.
However, it is not necessary for the preparation to be a tight function of $\nu$.
That is, in the preparation we are preparing a function over $\nu$, then we are applying an amplitude that is a function of $q$ and $\nu$ that is no greater than $1$.
The overall amplitude (corresponding to the square root of the function for the nonlocal pseudopotential) is a product of that in the two stages.
This means that the initial function of $\nu$ needs to be an upper bound on the function required, but it does not need to be a tight upper bound.

Recall that in the standard scheme for preparation of $1/\|k_\nu\|$ as described for example in Ref.~\cite{SuPRXQuantum21}, one initially prepares a superposition over a set of nested boxes.
The number of nested boxes corresponds to the number of bits in a single component of $\nu$, so is quite small, and there is low cost for the preparation.
The amplitudes on the individual boxes correspond to the maximum of the desired function within that box.
We can use a similar principle here, but instead of initially performing the full preparation over $\nu$ we just prepare the appropriate amplitudes over the boxes.
That substantially reduces the complexity of the preparation, but also increases the value of $\lambda$, because we are using upper bounds on each box instead of a tight bound on $\|k_\nu\|$.
The details of how the nested boxes are constructed are explained in Appendix \ref{app:nested}.

Next, we aim to reduce the complexity of the block encoding by combining operations between the local part of the pseudopotential and the nonlocal part as much as possible.
With the simplifications in the implementation of the nonlocal part above, all the arithmetic and QROM function interpolation is relegated to the select part of the block encoding, with a very simple preparation.
In order to reduce the overall complexity we apply a similar principle to the local part of the pseudopotential.
That is, we perform a simple preparation over $\nu$, then apply the more complicated function interpolation as part of the select operation.

One advantage of this approach is that the interpolation needed for the nonlocal part of the Hamiltonian is for the negative exponential function, and this is also the interpolation needed for the local part of the pseudopotential.
This means that we can perform this function interpolation in common between the two parts of the Hamiltonian, and just need to perform the further arithmetic in order to compute the functions.
This arithmetic can also be combined between the local and nonlocal parts of the pseudopotential.
Note that the functions given to be multiplied by $C_1^\alpha$, $C_2^\alpha$, and $C_3^\alpha$ correspond to the polynomials used for computing $F_{0\alpha}^j$ for $j=0,1,2$; that is, with $l=0$.
For $l=0$ the Legendre polynomial is just equal to $1$ as well, as is the polynomial for $F_{0\alpha}^i$ for $i=1$.
That means the implementation of the terms in the local pseudopotential with $C_1^\alpha,C_2^\alpha,C_3^\alpha$ can be completely combined with those for the nonlocal pseudopotential with $l=0$ and $i=1$.

One could also use $j=1$ and select between the terms with $i$, but it is convenient to use $i=1$ to simplify the management of the input to the function.
For the nonlocal pseudopotential we use $q$ and $q-\nu$, but for the local pseudopotential we just use $\nu$.
Therefore, if we use zero in place of $q$ for the local pseudopotential, then we will obtain the desired result (note that $\|k_{-\nu}\|=\|k_\nu\|$).
This will also be appropriate for the input to the negative exponential for the local pseudopotential, where we want $(r_{\rm loc}^\alpha\|k_\nu\|)^2/2$ rather than $(r_0^\alpha\|k_q-k_\nu\|)^2/2+(r_0^\alpha\|k_q\|)^2/2$ as for the nonlocal pseudopotential (so we obtain the correct result with $q=0$).

The remaining part of the local pseudopotential that needs to be accounted for separately is that with $1/\|k_\nu\|$.
In that case one simply needs to account for the $1/\|k_\nu\|$ factor by rearranging the inequality test in the usual way.
That is, the standard way to apply an amplitude factor in the block encoding is to calculate a function, prepare an ancilla register in an equal superposition state over integers, then perform an inequality test and flag on success for the block encoding \cite{SandersPRL18,SuPRXQuantum21}.
When dividing by $\|k_\nu\|^2$ in the amplitude, instead of performing arithmetic for the division, simply perform a multiplication of $\|k_\nu\|^2$ by the integer in the equal superposition.
This multiplication will need to be controlled by a qubit selecting this component of the pseudopotential.

Note that, in this procedure, we have separated the local pseudopotential into parts as well, which will further increase the $\lambda$.
Nevertheless, we find that the local pseudopotential has a relatively small contribution to $\lambda$, so this increase is unimportant.
In order to block encode the complete pseudopotential, the explicit steps needed are as follows.
These steps are in addition to the standard steps listed above for the block encoding without pseudopotentials, except we are no longer using the state prepared for $\nu$ with weights $1/\|\nu\|$ for implementing $U$.

\begin{enumerate}
    \item \label{step1} First we need to prepare an appropriately weighted superposition state over the nuclear species type, a register $\mu$ selecting the boxes for $\nu$, a register selecting between the local and nonlocal pseudopotential, registers for $l,i,j$, and a further register to select the first term for the local pseudopotential.
    We initially perform state preparation of a single contiguous register.
    \item \label{step2} Then QROM on this register may be used to output all of the desired registers ($\mu,j,i,j$, etc.), as well as further data needed for later steps.
    \item \label{step3} Based on the output value of $\mu$ we prepare a superposition over $\nu$ for the box $B_\mu$ (defined in Appendix \ref{app:nested}).
    \item \label{step4} Based on data output from the QROM in Step \ref{step2}, we prepare a superposition over nucleus number $\ell$.
    \item \label{step5} We apply QROM on $\ell$ in order to determine $R_\ell$ to use in the phase factor.
    \item \label{step6} We will copy the value of $q$ in the working ancilla into another working ancilla, with the copy being controlled by the register selecting between the local and nonlocal components of the pseudopotential.
    This new working register will now have zero for the local pseudopotential.
    \item \label{step7} We determine $q-\nu$ in a new register based on the value in the previous step.
    For the local part of the pseudopotential this will be $-\nu$.
    \item \label{step8} Compute the following with $p=q-\nu$,
   and $r_l^\alpha$ replaced with $r_{\rm loc}^\alpha$ for the local component of the pseudopotential.
    \begin{enumerate}
        \item $\|k_q\|^2$ \& $\|k_p\|^2$
        \item $(r_l^\alpha\|k_q\|)^2$ \& $(r_l^\alpha\|k_p\|)^2$
        \item $(r_l^\alpha\|k_q\|)^4$ \& $(r_l^\alpha\|k_p\|)^4$
        \item $(r_l^\alpha\|k_q\|)^2+(r_l^\alpha\|k_p\|)^2$
        \item $k_p \cdot k_q$
        \item $[3(k_p \cdot k_q)^2-\|k_p\|^2\|k_q\|^2]/2$
    \end{enumerate}
    \item \label{step9} Perform QROM interpolation of the negative exponential using the value of $(r_l^\alpha\|k_q\|)^2+(r_l^\alpha\|k_p\|)^2$.
    \item \label{step10} Compute $c_{0,li}+c_{1,li}(r_l^\alpha\|k_q\|)^2+c_{2,li}(r_l^\alpha\|k_q\|)^4$ and $c_{0,lj}+c_{1,lj}(r_l^\alpha\|k_p\|)^2+c_{2,lj}(r_l^\alpha\|k_p\|)^4$ using the values determined in (b) and (c) above, and using the values of $c_{x,li},c_{x,lj}$ output by the QROM.
    \item \label{step11} Use the value of $l$ to copy either $1$, or the computed items (e) or (f) into a working register for the Legendre polynomial. This is because $l$ is the degree of the Legendre polynomial needed.
    \item \label{step12} Multiply the exponential from the QROM interpolation in Step \ref{step9}, the $q$ and $p$ polynomials from Step \ref{step10}, and the Legendre polynomial from Step \ref{step11}.
    \item \label{step13} Multiply $\Psi$, a weighting factor used in the preparation described in Section \ref{sec:calcform} below, by the value in an equal superposition register.
    The value of $\Psi$ has previously been output via QROM.
    \item \label{step14} Controlled by the register selecting the first term for the local pseudopotential, multiply the value in the equal superposition register by $\|k_\nu\|^2$.
    \item \label{step15} Determine the absolute value of the result of the product and perform an inequality with the result from the multiplications with the equal superposition register.
    The success flag for the inequality test applies the appropriate amplitude for the block encoding.
    \item \label{step16} Use the sign from the product in Step \ref{step12} to control a $Z$ operation.
\end{enumerate}

The key parts of these steps are that Steps \ref{step1} to \ref{step4} are the state preparation (omitted from Figure~\ref{fig:combinedencoding}), Step \ref{step8} is computing the argument needed for the QROM interpolation (`arg' in Figure~\ref{fig:combinedencoding}), and Step \ref{step9} is computing the exponential (`exp' in Figure~\ref{fig:combinedencoding}).
Steps \ref{step10} to \ref{step13} are further arithmetic needed (omitted for simplicity in Figure~\ref{fig:combinedencoding}), then Step \ref{step15} is the inequality test in Figure~\ref{fig:combinedencoding}.
The value of $r_l^\alpha$ or $r_{\rm loc}^\alpha$ is output in Step \ref{step2} so that the arithmetic may be performed in common between the local and nonlocal parts of the pseudopotential.
Steps \ref{step6} and \ref{step14} are also used to take account of how the block encoding of the local pseudopotential is combined with the nonlocal pseudopotential, then Step \ref{step16} is just ensuring that the terms are added together with the correct signs.

We will go into further detail for these steps and their costs below.
A further subtlety that needs to be accounted for in this block encoding is how $\nu$ is prepared between $U$ and $V$.
To avoid making all operations controlled, it is convenient to prepare separate registers for $\nu$ with the weightings for $U$ and $V$.
Then we may perform a controlled swap into a working register.
A further observation is that the value of $\|k_\nu\|^2$ needs to be computed in the preparation for $V$ multiple times.
However, we are also computing this quantity for the block encoding of the local part of the pseudopotential.
We can combine the two to avoid computing this quantity twice.

This is possible even though we have included it as part of the preparation for $V$, but part of the selection for $U$.
The reason is that, for $V$, we complete what we have described as the `prepare' step by multiplying $\|k_\nu\|^2$ by a register in an equal superposition and performing an inequality test.
That is exactly what is done for the first term of the local part of the pseudopotential, so the implementation of these two parts can be combined, with the only distinction being the small extra cost of making the multiplication by the exponential controlled.

\section{Details of the costing}
\label{sec:GTHcosting}
\subsection{Preparation of registers for selecting between terms}
\label{sec:stateprep}

Throughout this work, we give the costing of the complexity in terms of Toffoli gates, as that is the non-Clifford gate used in these algorithms, and much more costly than Clifford gates to implement in an error-corrected architecture.
We also use the convention that logarithms are base 2.
Here we describe in more detail the preparation of the indices for selecting between the components of $U_{\rm loc}$ and $U_{\rm nonloc}$.
These are as follows.
\begin{itemize}
    \item The nuclear species $\alpha$.
    \item The box number for $\nu$, which we will denote $\mu$.
    \item Selection between the term in $U_{\rm loc}$ with $1/\|k_\nu\|$, the rest of $U_{\rm loc}$, and $U_{\rm nonloc}$.
    We will denote the variable for selection $\varsigma$, and it can be stored on two qubits, with one selecting between $U_{\rm loc}$ and $U_{\rm nonloc}$ and the other between the components of $U_{\rm loc}$.
    \item The value of $l$, which is selecting between terms for $U_{\rm nonloc}$ but just taken to be equal to $0$ for $U_{\rm loc}$.
    \item The values of $i,j$. We will prepare $i\le j$ for $U_{\rm nonloc}$, but for $U_{\rm loc}$ we will take $i=1$ and just use $j$ to select between terms.
\end{itemize}
In order to produce the full range of $i,j$ a swap is performed, controlled by a $\ket{+}$ state.
This controlled swap will also need to be controlled by the qubit selecting between $U_{\rm loc}$, and $U_{\rm nonloc}$.
These controlled swaps and their inversion can be performed with only 5 Toffolis, which can be ignored in comparison to other parts of the procedure.

This preparation is achieved by initially preparing a contiguous register, Step \ref{step1} above, then using QROM on that register to output the indices, Step \ref{step2} above.
To determine the complexity of preparing the contiguous register, we need to determine the number of values this contiguous register must take.
To explain how these values are counted,
we will initially consider just for a fixed $\alpha$ and $\mu$ (box number).
Nominally there would appear to be 5 for $U_{\rm loc}$, with the first term with $1/\|k_\nu\|$ and the following terms with $C_1^\alpha$, $C_2^\alpha$, etc.
However, for particular nuclear species as for example listed in Table \ref{TAB:GTH_LOCAL_LDA_PARAMETERS} many of these will be zero, and in fact there are no more than $C_1^\alpha$ and $C_2^\alpha$ given in this table.
For $U_{\rm nonloc}$, the number of values corresponds to the number of independent values of $E_{l\alpha}^{ij}$ (accounting for symmetry between $i,j$).
There are four different numbers for the different species of nuclei as given in Table \ref{TAB:GTH_NONLOCAL_LDA_PARAMETERS}.
The total values for the listed nuclei are as follows.
\begin{itemize}
    \item $3 + 1 = 4$ -- For nuclei like B, C, N, and O, there are three parts to $U_{\rm loc}$ (with $C_1$ and $C_2$ nonzero), but only one part to $U_{\rm nonloc}$ with $l=0$.
    That gives a total of four parts.
    \item $3+2=5$ -- For Li there are still three parts for $U_{\rm loc}$, but there are now two for $E_{l\alpha}^{ij}$ for a total of $5$.
    \item $2 + 4=6$ --  For Al, Si and Cl there are now only two parts for $U_{\rm loc}$ since $C_2=0$, but there are four $E_{l\alpha}^{ij}$ (with $l=0,1$) for a total of $6$.
    \item $1+10=11$ -- For examples like Mn and Ni there is only one part for $U_{\rm loc}$, but now for $U_{\rm nonloc}$ there is a maximum $l$ of 2, for 10 independent values of $E_{l\alpha}^{ij}$, and a total of $11$.
\end{itemize}
We need to sum these values over all the nuclear species required for the simulation.

We would also multiply by the number of boxes for $\nu$.
A subtlety here is that the sum for $U_{\rm loc}$ excludes $\nu=0$ (since it is just a term in the Hamiltonian proportional to the identity), whereas the sum for $U_{\rm nonloc}$ includes $\nu=0$.
In that case the momentum is unchanged, but the amplitude is modified according to $q$, so this is not just proportional to the identity.
In practice we find that the difference in values needed for $\nu=0$ and the next box are trivial, so we can just use the same box number for $\nu=0$ as for the next smallest box.
This means that the number of values needed is multiplied by $\mu_{\max}-1$, where $\mu_{\max}$ is given as in Eq.~\eqref{eq:maxmu} as
\begin{equation}
    \mu_{\max}=\max(n_x+\delta_x,n_y+\delta_y,n_z+\delta_z)+1.
\end{equation}
Here $\delta_x,\delta_y,\delta_z$ are shifts to account for different lengths of Bravais vectors, which are zero or small integers.
See Appendix \ref{app:nested} for further explanation.
In the simple case where $\delta_x,\delta_y,\delta_z$ are all zero and the numbers of bits for the components of the vectors are all equal to $n$, the number of boxes is just $n$.
We will use this in the following explanation for simplicity.

If we give the total number of values as per the list above summed over all nuclei as $\evals$, and the number of boxes is $n$, then the total number is $\evals n$.
Then costings for preparing the initial superposition over the contiguous register (Step \ref{step1}) are as follows.
\begin{itemize}
    \item $4\lceil \log (\evals n)\rceil +1$ for preparing the equal superposition over $\evals n$ basis states \cite{Sanders2020}.
    \item $\evals n$ for the QROM iterating over the index. (This can also be made smaller by using the advanced QROM.)
    \item A cost of the number of bits of precision for the inequality test with the keep register.
    This step does not need to be very precise, and the complexity can be taken to be the same as the number of bits, $b$, used in later arithmetic.
    That is more than enough, and has a complexity much smaller than other steps.
    \item There is a cost of $\lceil \log (\evals n)\rceil$ for a controlled swap on the index being prepared and the alt register.
\end{itemize}
After this preparation, Step \ref{step2} is QROM on this contiguous register with cost $\evals n$.
This QROM outputs $\varsigma,\alpha,\mu,l,i,j$, the values of $c_{x,li},c_{x,lj},r_l^\alpha,r_{\rm loc}^\alpha$,
and a quantity $\Psi_{\varsigma,\alpha,\mu,l,i,j}$ to use in inequality testing.
We will also use it to output rotation angles to be used in the preparation of $\nu$ and $\ell$, and the number of nuclei of each species $L_\alpha$.

Next, Step \ref{step3} is to prepare $\nu$ based on the box number $\mu$.
For the pseudopotentials it is preferable to prepare a superposition eliminating the inner boxes and the negative zeros.
To test if the value of $\nu$ is in an inner box, we can perform a triply-controlled NOT for a given value of $\mu$ to test if there are no leading ones in any of the three components.
Because $\mu$ is in a superposition, we need a quadruply-controlled NOT (controlling on a qubit for $\mu$ as well), and need to do it for each possible value of $\mu$.
That gives a cost of $3n$.
To test for a negative zero, we have cost $n$ to tell if each component is zero, and a unit cost to determine if the sign qubit is negative.
That gives a $3n$ cost to test a negative zero on any of the components.
There is also a $3n$ cost for controlled Hadamards on each of the components, as always when preparing superpositions with nested boxes.
That gives a cost of $9n$, which is performed three times in the amplitude amplification.
There is also a $3n$ cost for reflection on the $\nu$ registers in the amplitude amplification for a total of $30n$.
We also need a $\mu$-dependent rotation (to be performed twice) in the amplitude amplification to give near-unit amplitude for success.
That angle is given by the QROM in the preceding step, and can be taken to be $b$ bits for cost $2b$ for the two rotations.

Step \ref{step4} is then to prepare a superposition over $\ell$.
For given species, the nucleus number $\ell$ does not affect the amplitude, only the phase through the factor of $e^{-ik_{\nu}\cdot R_\ell}$.
Therefore, for each $\alpha$ an equal superposition state over $\ell$ needs to be prepared, over a number of basis states $L_\alpha$.
That number is given in a quantum register output by the QROM in Step \ref{step2}.
There is a standard procedure to prepare that equal superposition with logarithmic complexity \cite{Lee2020}.

The total complexity is then $(2\evals +30)n$ plus $3b$ and a logarithmic complexity in $\evals$ and $n$.
This complexity is doubled when accounting for inverse preparation in the block encoding.

\subsection{The form of the calculation}
\label{sec:calcform}
Now we give more detail on the form of the expressions that need to be evaluated.
We first explain how this is done for the case of the nonlocal pseudopotential, as this is more complicated, and the calculation of the local pseudopotential is combined with it.
The nonlocal part of the pseudopotential is given by
\begin{equation}
U_{\rm nonloc} = \sum_{\ell=1}^L\sum_{\nu\in G_d} e^{-ik_{\nu}\cdot R_\ell} \sum_{j=1}^{\eta}\sum_{\substack{q\in G\\ (q-\nu) \in G}} \nonlocl(k_q,k_{q-\nu})  \ket{q-\nu}\!\!\bra{q}_j    \, ,
\end{equation}
where
\begin{align}
\nonloc(k_p,k_{q}) \coloneqq
\frac 1{\Omega} \sum_{l = 0}^{l_{\text{max}}^\alpha}(-1)^{l}
\frac{\left(2 l + 1\right)}{4\pi} P_l \left(\frac{k_p \cdot k_q}{\left\|k_p\right\| \left\|k_q\right\|}\right)
\sum_{i = 1}^3
\sum_{j = 1}^3
E_{l\alpha}^{ij}
F_{l\alpha}^i\left(\left\|k_p\right\|\right) F_{l\alpha}^j\left(\left\|k_q\right\|\right) \, ,
\end{align}
and
\begin{equation}
F^i_{l\alpha}(\|k\|) = \|k\|^l C_{li}^\alpha \left ( \sum_{x=0}^{i-1} c_{x, li}(r_l^\alpha\|k\|)^{2x}\right ) e^{-(r_l^\alpha\|k\|)^2/2} \, .
\end{equation}
In this expression $E_{l\alpha}^{ij}$ and $r_l^\alpha$ are dependent on the nuclear species $\alpha$, but $c_{x, li}$ is independent of the nuclear species.

We now define a scaled $\widetilde F_l^i$ 
\begin{equation}
\widetilde F^i_{l}(\|k\|) :=
\|k\|^l \left ( \sum_{x=0}^{i-1} c_{x, li}\|k\|^{2x}\right ) e^{-\|k\|^2/2} \, .
\end{equation}
This quantity is effectively $\widetilde F_l^i$ without $C_{li}^\alpha$, but including the factor of $(r_l^\alpha)^l$ from $C_{li}^\alpha$, so
\begin{equation}
\widetilde F^i_{l}(r_l^\alpha\|k\|) = \frac{(r_l^\alpha)^l F^i_{l\alpha}(\|k\|)}{C_{li}^\alpha} \, .
\end{equation}
We also define a quantity
    \begin{equation}
        \upper_{\alpha,\nu,l,i,j} \coloneqq   \frac 1{(r_l^\alpha)^{2l}} \max_{q}
\left| \widetilde F_{l}^i\!\left(r_l^\alpha\|k_{q+\nu}\|\right) \widetilde F_{l}^j\!\left(r_l^\alpha\|k_q\|\right) \right| \, .
    \end{equation}
Note that this expression is dependent on the nuclear species through $r_l^\alpha$ as input to $\widetilde F^i_{l}$.
That only gives a scaling in its variation, and it can be calculated independently of the nuclear species if $\nu$ is scaled by $r_l^\alpha$.
We have also included a factor of $1/{(r_l^\alpha)^{2l}}$, because we will not be including it explicitly in the calculation of $\widetilde F_{l}^i$.

Then we can rewrite $\nonloc(k_p,k_{q})$ as
\begin{align}\label{eq:supform}
\nonloc(k_p,k_{q}) =
\sum_{l = 0}^{l_{\text{max}}^\alpha}
\sum_{i = 1}^3
\sum_{j = 1}^3
\frac{\left(2 l + 1\right)}{4\pi\Omega} E_{l\alpha}^{ij} C_{li}^\alpha C_{lj}^\alpha  \upper_{\alpha,\nu,l,i,j}
\left[ \frac {(-1)^{l}}{(r_l^\alpha)^{2l} \upper_{\alpha,\nu,l,i,j}} P_l \left(\frac{k_p \cdot k_q}{\left\|k_p\right\| \left\|k_q\right\|}\right)
\widetilde F_{l}^i\!\left(r_l^\alpha\|k_p\|\right) \widetilde F_{l}^j\!\left(r_l^\alpha\|k_q\|\right) \right] \, .
\end{align}
The expression in the square brackets is constructed such that its absolute value is no larger than 1 (recall that the absolute value of the Legendre polynomial is no larger than 1).

The expression for $\upper_{\alpha,\nu,l,i,j}$ can, in general, be a complicated function of $\nu$, meaning that it would have high complexity to approximate.
As discussed above, in the state preparation it is convenient to just approximate the functions within the individual boxes.
Then the upper bound we will use is
\begin{equation}
    \Psi_{2,\alpha,\mu,l,i,j} =  \max_{\nu \in B_{\mu}\backslash B_{\mu-1}}\upper_{\alpha,\nu,l,i,j} \, .
\end{equation}
Here the maximum is over values of $\nu$ within the box $B_\mu$ but excluding the smaller inner box $B_{\mu-1}$.
We are now also including $2$ for $\varsigma=2$ to select the nonlocal part of the pseudopotential.

For the implementation we therefore can prepare a state over $\alpha,\nu,l,i,j$ with (squared) amplitudes proportional to
\begin{equation}
    L_\alpha \frac{\left(2 l + 1\right)}{4\pi} E_{l\alpha}^{ij} C_{li}^\alpha C_{lj}^\alpha \Psi_{2,\alpha,\mu,l,i,j} \, ,
\end{equation}
where $L_\alpha$ is the number of nuclei of type $\alpha$.
Note that this is \emph{independent} of $q$, so does not depend on the system state.
Recall that we prepare a superposition over nuclear species $\alpha$ initially, then prepare the superposition over $L_\alpha$ values of $\ell$ in Step \ref{step4}.

Then, dependent on the state momentum, we apply an amplitude corresponding to the expression in square brackets in Eq.~\eqref{eq:supform}.
This amplitude can be applied in a block encoding by first computing
\begin{equation}\label{eq:ampfac}
    \frac 1{(r_l^\alpha)^{2l}} P_l \left(\frac{k_p \cdot k_q}{\left\|k_p\right\| \left\|k_q\right\|}\right)
\widetilde F_{l}^i\!\left(r_l^\alpha\|k_p\|\right) \widetilde F_{l}^j\!\left(r_l^\alpha \|k_q\|\right) \, ,
\end{equation}
where the factor of $1/{(r_l^\alpha)^{2l}}$ is to simplify the computation.
The computation of this value is Step \ref{step12}.
In the usual way for applying an amplitude factor in block encoding, we would prepare another register in an equal superposition state, and perform an inequality test; this is Step \ref{step15}.
However, the register in an equal superposition would first be multiplied by $\Psi_{2,\alpha,\mu,l,i,j}$ before the inequality test to account for the maximum value; that multiplication is Step \ref{step13}.
We would also need to compute the absolute value of the computed quantity before the inequality test, and apply a phase factor according to its sign.
We would also need to apply a sign according to $(-1)^l$ in the square brackets in Eq.~\eqref{eq:supform}.
Application of the sign is in Step \ref{step16}.

For the computation of the expression in Eq.~\eqref{eq:ampfac}, we group the factor of $(\left\|k_p\right\| \left\|k_q\right\|)^l$ from $\widetilde F^i_{l}$ and $\widetilde F^j_{l}$ together with the Legendre polynomial in order to avoid needing to compute the ratio.
Since $r_l^\alpha$ appears in $\widetilde F_{l}^i\!\left(r_l^\alpha\|k_p\|\right)$ but \emph{not} the Legendre polynomial, this computation gives the factor of $1/{(r_l^\alpha)^{2l}}$ above.
For the nuclei we are considering, $l\le 2$, and the expressions obtained for the product of the Legendre polynomial and $(\left\|k_p\right\| \left\|k_q\right\|)^l$ are as follows.
\begin{enumerate}
    \item For the case $l=0$, the Legendre polynomial is just equal to 1 and there is no cancellation.
    \item For $l=1$, the Legendre polynomial is $x$, so we have $k_p \cdot k_q$ and the denominator cancels.
    \item For $l=2$, the Legendre polynomial is $(3x^2-1)/2$, so multiplying by $(\left\|k_p\right\| \left\|k_q\right\|)^l$ gives
    \begin{equation}
        \frac 12 \left[ 3(k_p \cdot k_q)^2-\left\|k_p\right\|^2 \left\|k_q\right\|^2 \right] .
    \end{equation}
\end{enumerate}
The parts for degree $l>0$ of the Legendre polynomial are nontrivial, so are calculated in Step \ref{step8} parts (e) and (f).
The correct order of the Legendre polynomial is then selected in Step \ref{step11}.

The remaining expression we need to compute is then
\begin{equation}\label{eq:nonlocexp}
    \left ( \sum_{x=0}^{i-1} c_{x, li}(r_l^\alpha\|k_p\|)^{2x}\right )
    \left ( \sum_{x=0}^{i-1} c_{x, lj}(r_l^\alpha\|k_q\|)^{2x}\right )
    e^{-\frac{1}{2}[(r_l^\alpha\|k_q\|)^2+(r_l^\alpha\|k_p\|)^2]} \, .
\end{equation}
This expression can be computed using just additions and multiplications, except for the exponential, which is the most demanding part of the expression to accurately approximate.
However, this exponential is in common between the parts of the sum, so need only be evaluated once.
This evaluation is Step \ref{step9} in the list above, and from Eq.~\eqref{eq:nonlocexp} it can be seen that the argument is proportional to $(r_l^\alpha\|k_q\|)^2+(r_l^\alpha\|k_p\|)^2$ as described in Step \ref{step9}.
The approximation of the negative exponential function is analysed in Ref.~\cite{Sanders2020}, and we provide further details below in Section \ref{sec:interpolate}.

Step \ref{step8} involves calculation of many of the quantities needed for the calculation, including the argument of the exponential in part (d), and powers of $r_l^\alpha\|k_q\|$ and $r_l^\alpha\|k_p\|$ in parts (b) and (c).
The squares are used for the argument of the exponential, as well as the polynomials in Eq.~\eqref{eq:nonlocexp}, which are calculated in Step \ref{step10}.
Note that the squares $\left\|k_q\right\|^2$, $\left\|k_p\right\|^2$ are computed separately, because they are needed for the Legendre polynomials.

For the local part of the pseudopotential, recall that we have
\begin{align}
U_{\rm loc} &= \sum_{\ell=1}^L \sum_{\nu\in G_0} \locl\!\left(k_\nu\right) \left( e^{-\mathrm{i}k_{\nu}\cdot R_\ell}\sum_{\elj=1}^\eta \sum_{\substack{q\in G\\ (q-\nu) \in G}} \ket{q-\nu}\!\!\bra{q}_{\elj} \right) \\
u_{\rm loc}^{\alpha, R}\!\left(k\right) & =  - \frac{4 \pi Z_\alpha}{\Omega\|k\|^2}e^{-(r_{\rm loc}^\alpha \|k \|)^2/2} + \sqrt{8 \pi^3}\frac{(r_{\rm loc}^\alpha)^3}{\Omega} e^{-(r_{\rm loc}^\alpha \|k \|)^2/2} 
\sum_{j=1}^4 C_j^\alpha \sum_{x=0}^{j-1} c_{x,0j} (r_{\rm loc}^\alpha \|k \|)^{2x} \, .
\end{align}
Here, we have written the sum in terms of $c_{x,0j}$ which is in common with the nonlocal pseudopotential.
We can also write it as
\begin{align}\label{eq:simplocal}
u_{\rm loc}^{\alpha}\!\left(k\right) & =  - \frac{4 \pi Z_\alpha}{\Omega\|k\|^2}e^{-(r_{\rm loc}^\alpha \|k \|)^2/2} + \sqrt{8 \pi^3}\frac{r_{\rm loc}^3}{\Omega} 
\sum_{j=1}^4 C_j^\alpha \widetilde F_0^j \! \left( r_{\rm loc}^\alpha \|k \| \right) \, .
\end{align}
For the second term here we would prepare a state with squared amplitudes proportional to
\begin{align}
    & L_\alpha \sqrt{8\pi^3} \frac{r_{\rm loc}^3}{\Omega} C_j^\alpha \Psi_{1,\alpha,\mu,0,1,j} \, , \\
    &\Psi_{1,\alpha,\mu,0,1,j} :=  \max_{\nu \in B_{\mu}\backslash B_{\mu-1}} \left| \widetilde F_0^j \! \left( r_{\rm loc}^\alpha \|k_\nu \| \right) \right| .
\end{align}
Again the maximum is so we only need to prepare a state with a superposition over boxes, rather than an exact $\nu$-dependent superposition.
The quantity $\Psi_{1,\alpha,\mu,0,1,j}$ here has $\varsigma=1$ for selecting the second term.
Then, for the select operation, we can compute the expression as in Eq.~\eqref{eq:nonlocexp}, except with $q=0$, which gives us $\widetilde F_0^j \! \left( r_{\rm loc}^\alpha \|k_\nu \| \right)$.
That is why this computation can be performed in common with the nonlocal part of the pseudopotential.
It is accounted for by copying zero into the working register rather than $q$ in Step \ref{step6}.
To apply the required amplitude factor we would then prepare an equal superposition state and multiply it by $\Psi_{1,\alpha,\mu,0,1,j}$.
The sign of the expression is again applied by taking the absolute value before the inequality test, and applying a $Z$ gate according to the sign in Step \ref{step16}.

For the first term in Eq.~\eqref{eq:simplocal} we prepare squared amplitudes proportional to
\begin{align}
    & L_\alpha \frac{4 \pi Z_\alpha}{\Omega} \Psi_{0,\alpha,\mu,0,1,1} \, ,\\
    & \Psi_{0,\alpha,\mu,0,1,1} := \max_{\nu \in B_{\mu}\backslash B_{\mu-1}} \frac{e^{-(r_{\rm loc}^\alpha \|k_\nu \|)^2/2}}{\|k_\nu\|^2} .
\end{align}
Here, $\Psi_{0,\alpha,\mu,0,1,1}$ has $\varsigma=0$ to select this part of the local pseudopotential.
In this case we take $l=0$, $i=j=1$, so computation of the expression in Eq.~\eqref{eq:nonlocexp} with $q=0$ gives $e^{-(r_{\rm loc}^\alpha \|k_\nu \|)^2/2}$.
That enables this part of the Hamiltonian to be block encoded in a common way with the nonlocal pseudopotential as well.
To apply the desired amplitude factor in the select operation, we again prepare an equal superposition state and multiply by $\Psi_{0,\alpha,\mu,0,1,1}$.
In this case, however, we multiply the equal superposition state by $\|k_\nu \|^2$ to account for the division by $\|k_\nu \|^2$.
That is Step \ref{step14} above.
The value of $\|k_\nu\|^2$ is computed in the preparation of the superposition state for $V$ and can be used here.

\subsection{QROM interpolation}\label{sec:interpolate}
For the QROM interpolation, since we are approximating a negative exponential function, it is convenient to apply the following procedure.
\begin{itemize}
    \item First multiply the argument by $1/\ln 2$.
    This is so that we can interpolate $2^{-z}$ rather than $e^{-z}$.
    \item Apply QROM interpolation on the fractional part of $z$ (i.e., after multiplying by $1/\ln 2$).
    \item Use the integer part of $z$ to control a bit shift of the output.
\end{itemize}

Since about half the bits in the binary expansion of $1/\ln 2$ are zero, the multiplication by $1/\ln 2$ has a complexity of approximately $b^2/4$ when performing calculations with $b$ bits.
In Section II E of \cite{Sanders2020} the error for the case of the linear interpolation is estimated in terms of the second derivative (see Eqs.~(92) to (94)) as
\begin{equation}
    \frac{(\delta z)^2}{8} |f''(z)| \, .
\end{equation}
When the function is $2^{-z}$, the \emph{relative} error is
\begin{equation}
    \frac{(\delta z \, \ln 2)^2}{8} \, .
\end{equation}
For example, if $\delta z=1/256$, then the relative error is less than $10^{-6}$.
For interpolation on the fractional part of $z$, this would correspond to a Toffoli complexity of about $256$.
The interpolation requires a multiplication with complexity $b^2$ (we're ignoring the complexity of additions).
Then the complexity of the controlled bit shift is about $b^2/2$.

In cases where higher precision is required, a higher-order polynomial can be used.
In general, for an order $n-1$ polynomial, one can use interpolation on $n$ Chebyshev nodes, which gives a maximum error
\begin{equation}
    \frac{(\delta z)^n}{2^{2n-1} n!} \max_{z\in [z_0-\delta z/2,z_0+\delta z/2]} \left| f^{(n)}(z) \right| \, ,
\end{equation}
for the interval $[z_0-\delta z/2,z_0+\delta z/2]$.
For example, quadratic interpolation with $n=3$ gives
\begin{equation}
    \frac{(\delta z)^3}{192} \max_{z\in [z_0-\delta z/2,z_0+\delta z/2]} \left| f^{(3)}(z) \right| \, .
\end{equation}
Then $128$ interpolation points would give an accuracy better than one part in $10^9$.
The quadratic can be arranged so only one more multiplication is needed, and the square does not need to be computed explicitly.
This extra multiplication has cost approximately $b^2$ for $b$ digits.

This means that, for linear interpolation, the complexity is approximately $(3/4)b^2+256$ when aiming for a relative error of one part in $10^6$.
More generally it would be $(7/4)b^2+\ln 2/\sqrt{8\, \delta\! f}$ when aiming for relative error $\delta\! f$.
With quadratic interpolation the complexity can be given as $(7/4)b^2+128$ (for error of one part in $10^9$), or more generally $(11/4)b^2+\ln 2/(192\, \delta\! f)^{1/3}$.
The QROM cost scales only as the $1/3$ power of the allowed error, meaning that it increases only slowly with the required error.
For example, precision of one part in $10^{12}$ could be achieved with QROM cost about 1200.
That is a small contribution to the total block encoding costs, which are on the order of 20,000 to 50,000 in Table \ref{TAB:BE_Costs}.
As the required precision is increased, the arithmetic is a much more significant contribution to the increase in complexity.

\begin{figure}
    \centering
    \includegraphics[width=10.79cm]{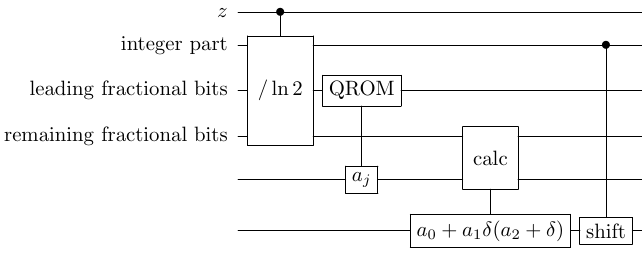}
    \caption{The method for the interpolation.
    First the input is divided by $\ln 2$ so we are interpolating $2^{-z}$.
    QROM outputs coefficients for a polynomial, then those are used for computing the interpolated values on the fractional part of the input.
    The integer part of the input is used to control a bit shift on the output.}
    \label{fig:Interpolate}
\end{figure}

To explain the method in a little more detail, the circuit diagram is shown in Figure~\ref{fig:Interpolate}.
After multiplying by $1/\ln 2$, the output is divided into the integer part, the leading fractional bits and the remaining fractional bits.
For $128=2^7$ interpolation points, the QROM for the interpolation is applied to the leading 7 bits of the fractional part.
Each of those can be used to output parameters for the interpolation $a,b,c$.
The interpolated value can be calculated on the remaining fractional bits as $a_0+a_1\delta(a_2+\delta)$.
That is, if the value given by the leading fractional bits is $x_0$, then the interpolated value is calculated relative to that as $a_0+a_1(x-x_0)(a_2+x-x_0)$, so it is only the shift $\delta$ from $x_0$ that is used, which is given by the remaining fractional bits.
Lastly the integer part is used to control a shift of the output.

\subsection{Completing the pseudopotential costing}
\label{ref:completepseudo}
Using the explanation in the preceding subsections, we now go on to provide the costing of the steps as presented in Section~\ref{sec:blockenc}, using the same step numbers.
We ignore $\mathcal{O}(1)$ complexities, and linear complexities in the number of bits for the arithmetic involving multiplications.
\begin{enumerate}
    \item As detailed above in Section~\ref{sec:stateprep}, the cost of the state preparation can be taken to be $\evals n + 5 \log(\evals n)+ b$, where $\evals$ is the number of independent values needed for each nucleus summed over the nuclei.
    \item There is a further cost of $\evals n$ to output all the required parameters using QROM.
    \item Again as detailed above in Section~\ref{sec:stateprep}, the cost of preparing $\nu$ and updated box numbers is $30n+2b$.
    \item The nuclear species $\alpha$ is used to prepare an equal superposition over a register $\ell$ indexing between the different nuclei of that type.
    Given that there are $\lceil \log L \rceil$ qubits used for the number of nuclei, the Toffoli cost is $7\lceil \log L \rceil + 2b_r - 6$ according to Appendix C of \cite{Lee2020} (step 3(a) on page 42).
    Here $b_r$ is the number of bits of accuracy for the rotation, which can be taken to be 7 which would give complexity $7\lceil \log L \rceil + 8$.
    For $\ell$ we do not need to make it index over all nuclei, it can just index over nuclei for each type, so there are the same $\ell$ values for different $\alpha$.
    \item The QROM on $\ell$ to output $R_{\ell}$ has cost $L$.
    This QROM can use both the register with $\alpha$ and $\ell$.
    This is the only stage where $\ell$ is used, which is why $\ell$ can just index over nuclei of each type individually, rather than all nuclei.
    \item The controlled copy of $q$ into a new register has complexity $n_x+n_y+n_z$, or $3n$ if all three are equal.
    \item Computing $p=q-\nu$ also has complexity $n_x+n_y+n_z$.
    \item The complexities of computing the main quantities needed are as follows.
    Here we give the expressions in terms of $r_l^\alpha$, but this is also $r_{\rm loc}^\alpha$ for the local pseudopotential.
    \begin{enumerate}
        \item $\|k_q\|^2$ \& $\|k_p\|^2$ -- As described in Appendix \ref{app:bravaisnorm}, the leading order complexity of the computation of this expression using a Gramian is as given in Eq.~\eqref{eq:normcomplex} for $b$ bits for the arithmetic.
        \item $(r_l^\alpha\|k_q\|)^2$ \& $(r_l^\alpha\|k_p\|)^2$ -- Each of these multiplications has a leading-order complexity of $b^2$, using the result in Eq.~(D18) of \cite{Sanders2020}.
        \item $(r_l^\alpha\|k_q\|)^4$ \& $(r_l^\alpha\|k_p\|)^4$ -- Each squaring of a real number has a leading-order complexity of $b^2/2$, as per Eq.~(D37) of \cite{Sanders2020}.
        We do not need to perform these calculations in the case that $E_{l\alpha}^{ij}$ does not go up to $i,j=2$, though (so it is only needed for Mn, Ni, Ca, and Ti out of the listed nuclei).
        \item $(r_l^\alpha\|k_q\|)^2+(r_l^\alpha\|k_p\|)^2$ -- The cost of the addition can be ignored in this costing.
        \item $k_p \cdot k_q$ -- The complexity of the dot product using a Gramian is given in Eq.~\eqref{eq:dotcomplex}.
        \item $[3(k_p \cdot k_q)^2-\|k_p\|^2\|k_q\|^2]/2$ -- For this expression a square (complexity $b^2/2$) and product (complexity $b^2$) are required for a total of $3b^2/2$.
        The multiplication by 3/2 can be applied with a single addition with complexity $\mathcal{O}(b)$, and similarly the subtraction is $\mathcal{O}(b)$; we ignore these complexities linear in $b$.
        This computation is only needed in the case that $l$ goes up to $2$.
    \end{enumerate}
    \item The complexities given above for the QROM are $(7/4)b^2+256$ if the goal is relative error below $10^{-6}$, or $(11/4)b^2+128$ to give accuracy of one part in $10^9$.
    These are using linear or quadratic interpolation, respectively.
    \item The complexity of computing $c_{0,lj}+c_{1,lj}(r_l^\alpha\|k_q\|)^2+c_{2,lj}(r_l^\alpha\|k_q\|)^4$ (and similarly for $i$) depends on the maximum values of $i,j$.
    \begin{enumerate}
        \item First is the case $i=j=1$, which corresponds to $l_{\max}^\alpha=0$ in most cases, though Li has $l_{\max}^\alpha=1$ with the same maximum values of $i,j$.
        In that case we just want $c_{0,l1}$, since the sums over $x$ are now just giving $x=0$.
        There is no extra Toffoli cost then, since $c_{0,l1}$ is output by QROM at an earlier stage.
        \item In the case where the maximum of $i,j$ is $2$, we need $c_{0,lj}+c_{1,lj}(r_l^\alpha\|k_q\|)^2$ and similarly for $i$.
        Then $c_{1,lj}$ can only be $0$ or $-1$, so the multiplication by $c_{1,lj}$ is just a controlled copy with cost $b$.
        The addition has a cost of $b$ for a total of $2b$, then there is a factor of 2 to perform this calculation for both $i$ and $j$ for a total of $4b$.
        \item In the case where the maximum of $i,j$ is $3$, we need to 
        compute $c_{0,lj}+c_{1,lj}(r_l^\alpha\|k_q\|)^2+c_{2,lj}(r_l^\alpha\|k_q\|)^4$, and
        also allow values of $c_{1,lj}$ of $-10$ and $-14$.
        We do not have $c_{1,lj}=-18$ for the materials listed because we do not have a combination of $l=2$ and $j=3$.
        The bits selecting $1$ and $2$ (in the value of $c_{1,lj}$) cannot both be $1$, so we may simply perform a controlled copy into an ancilla register for both with cost $2b$.
        There is a controlled addition of $8$ times the argument with cost approximately $2b$.
        Next, $c_{2,lj}$ has just 1 bit, so the multiplication by $c_{2,lj}$ is just a controlled copy with cost $b$.
        Together with the two additions this is a cost of approximately $7b$, then the factor of 2 for the calculation with both $i$ and $j$ gives a total of $14b$.
    \end{enumerate}
    \item Use the value of $l$ to copy either $1$, or the computed items (e) or (f) from part 8 into a working register for the Legendre polynomial.
    The complexity is $2b$ for two controlled copies of $b$ bits.
    The controlled copy of 1 has complexity $\mathcal{O}(1)$.
    \item Multiply the exponential from the QROM interpolation in Step \ref{step9}, $p$ and $q$ polynomials, and the Legendre polynomial.
    The three multiplications have complexity $3b^2$.
    \item Multiply the value of $\Psi_{\varsigma,\alpha,\mu,l,i,j}$ that has previously been output via QROM by the value in an equal superposition register.
    This yields another complexity of $b^2$.
    \item The controlled multiplication of the value in the superposition register by $\|k_\nu\|^2$ has complexity $b^2$.
    The control just introduces complexity linear in $b$, which we ignore here.
    \item Determine the absolute value of the result of the product from Step \ref{step12} and perform an inequality with the result from Step \ref{step14}.
    The success flag for the inequality test applies the appropriate amplitude for the block encoding.
    This has complexity linear in $b$.
    \item Use the sign from the product from Step \ref{step12} to control a $Z$ operation.
    This has $\mathcal{O}(1)$ complexity.
\end{enumerate}

The primary costs here are those of the multiplications and squarings.
Assuming these are $b$ bits, there is a maximum cost of $11\tfrac 14 b^2+14b$ for all these operations
(including the cost linear in $b$ from computing the polynomials with $c_{x,lj}$, and assuming quadratic interpolation, but omitting the Gramian-dependent costs).
There are various savings in other cases.
\begin{itemize}
    \item There is a saving of $b^2$ if linear interpolation is used instead of quadratic.
    \item There is a saving of $3b^2/2$ if $l_{\max}^\alpha<2$.
    \item There is a saving of $b^2+10b$ if the maximum of $i,j$ is 2 instead of 3.
    \item There is a further saving of $4b$ if the the maximum of $i,j$ is 1.
\end{itemize}

The cost of the QROM for the interpolation is considerably smaller.
For accuracy of about one part in $10^6$ we can use linear interpolation with $256$ interpolation points, as well as $b=20$.
Then the cost is about $5000$ Toffolis (the cost of the pseudopotential excluding other parts of the block encoding).
If one were aiming for a higher precision, closer to one part in $10^8$, one could choose $b=27$ and use quadratic interpolation with $64$ interpolation points.
That would increase the cost to around 10,000 Toffolis.
This means that the precision can be improved by two orders of magnitude with about a factor of 2 increase in the complexity.

A choice we have made in the implementation is to interpolate just the exponential, instead of separately interpolating $F_{l\alpha}^i$ and $F_{l\alpha}^j$.
Separately interpolating would save the multiplications by $c_{x,li},c_{x,lj}$, but we would have much higher QROM cost.
We would need to interpolate $9$ combinations of each of $l,i$ and $l,j$.
Even if the interpolations only use $100$ points each, the extra cost over just interpolating the exponential is nearly $2000$.
It would also be a considerably more complicated error analysis because we would not be interpolating just an exponential.
In comparison, the multiplication cost we have here is about $14b$; this is not large because $c_{x,li}$ has a small number of bits.

There is also a cost of determining two squares of vectors and a dot product using the Gramian, but the cost will be highly dependent on the form of the Gramian.
Because $p=q-\nu$, and $\|k_\nu\|^2$ is computed in a different part of the algorithm, we can compute $k_q \cdot k_\nu$, which will give $\|k_p\|^2$ and $k_p \cdot k_q$ with only additions.
This means that there is an additional complexity at this step of one norm and one dot product, rather than \emph{two} norms.

In this section we are only considering the overhead in the block encoding for the pseudopotentials.
The costs for other parts of the procedure are discussed further below in Section \ref{sec:othercosts}.

\section{Other costs in the block encoding}
\label{sec:othercosts}
\subsection{Preparing the superposition over \texorpdfstring{$\nu$}{nu}}
\label{sec:nuprep}
The primary cost for the other part of the superposition is in preparing the state with amplitudes $1/\|k_\nu\|$ for $V$.
This is the same state as in prior work, except here we consider the amendment that we use a Gramian to compute the norm, so $\|k_\nu\|$ is not just proportional to $\|\nu\|$.
Also, this superposition is used only for $V$, not $U$.

The general principle used to prepare the state is to first prepare a set of nested boxes indexed by $\mu$, then use that to prepare a superposition of over $\nu$, then use inequality testing in order to apply the correct amplitude factors.
The details of how the nested boxes are applied is described in detail in Appendix \ref{app:nested}.
As explained in Section \ref{sec:stateprep}, there is a complexity linear in $n$ (the number of bits for components of the momentum) in order to construct the superposition over $\nu$ given the superposition over $\mu$.

The superposition over $\mu$ is prepared using controlled Hadamard gates in Ref.~\cite{SuPRXQuantum21}.
Here we are considering the case of non-orthogonal Bravais vectors, so the weightings are not the same as in Ref.~\cite{SuPRXQuantum21} and more general preparation is needed.
We can use a similar preparation over $\mu$ here as is used in Section \ref{sec:stateprep}, by using coherent alias sampling.
The cost is just linear in $n$, plus a logarithmic (in $n$) cost for preparing an equal superposition and another logarithmic (in the precision) cost for performing the inequality test.

In the simple case discussed in Ref.~\cite{SuPRXQuantum21}, the inequality is written in the form (see Eq.~(84) of that work)
\begin{equation}
    (2^{\mu-2})^2 M > m \|\nu\|^2 \, ,
\end{equation}
where $m$ is the value prepared in an equal superposition state, $M$ is the number of values in the superposition for $m$, and $(2^{\mu-2})^2$ is a value resulting from the simple form of nested boxes in that work.
As discussed there, $(2^{\mu-2})^2M$ can be given with no extra Toffoli gates.
For the more general set of nested boxes considered here, the value will not be so simple, but the appropriate $\mu$-dependent value can be obtained from a QROM on $\mu$ with complexity linear in $n$.
This cost, and the other costs for preparing the superposition over $\mu$, are very small compared the other costs in the procedure, and will be ignored in our approximate costing.

The main cost in the procedure is from determining the value of $\|k_\nu\|^2$.
That is no longer proportional to $\|\nu\|$ here, and the details of the complexity are given in Appendix \ref{app:bravaisnorm}.
In preparing the state via inequality testing and amplitude amplification, we use the following steps.
\begin{itemize}
\item Prepare the superposition over $\mu$ for the nested boxes.
    \item Prepare the simple superposition over $\nu$ using the value of $\mu$ as discussed above.
    \item Compute $\|k_\nu\|^2$ with complexity described in Appendix \ref{app:bravaisnorm}.
    \item Multiply the result by a register in an equal superposition state, with complexity $b^2$ when performing arithmetic with $b$ bits.
    \item Perform an inequality test between the result and a $\mu$-dependent value given by QROM.
    \item Reflect on the result of the inequality test.
    \item Invert the inequality test, multiplication and computation of $\|k_\nu\|^2$.
    This may be performed with Clifford gates, provided working ancillas are retained.
    \item Invert the preparation of $\mu,\nu$ from the first two steps, reflect about zero on the ancillas, and reprepare.
    \item Compute $\|k_\nu\|^2$ and the multiplication again.
    \item Perform the inequality test to prepare the state.
\end{itemize}

For the total cost we also need to account for inverting this state preparation.
If we retain qubits from the calculation we can invert the last multiplication and computation of $\|k_\nu\|^2$ with Clifford gates, then need to pay the cost of the computation a third time.
That is, we pay the Toffoli cost of the calculation twice in preparation, and once in inverse preparation.
Thus the amplitude amplification results in triple the cost of computing $\|k_\nu\|^2$ and the multiplication.
This is similar to the tripling of the cost obtained in Ref.~\cite{SuPRXQuantum21} (see below Eq.~(90) of that work).
The other costs are relatively trivial compared to this main cost.

Another subtlety is that we are preparing $k_\nu$ via amplitude amplification for $V$, but by just preparation over the nested boxes for the pseudopotential components of the Hamiltonian.
We can perform these preparations in separate registers, then perform a controlled swap to move $\nu$ into the working register.
This may be performed before the computation of $\|k_\nu\|^2$ in step 8 above, so the value of $\|k_\nu\|^2$ can be used in the implementation of the pseudopotential without recomputing.
The controlled swap has complexity linear in $n$, which can be ignored in this approximate costing.
There are also a number of other small selection costs that we are ignoring here.
For example, the flag on the result of the inequality test is also controlled on the qubits selecting the $V$ term in the Hamiltonian.

\subsection{Implementation of the phase factor}
\label{sec:phasecomp}
The last step is to apply the phase factor of $e^{-ik_{\nu}\cdot R_\ell}$.
For this calculation the vector $k_{\nu}$ is given in terms of the reciprocal lattice vectors as
\begin{equation}
    k_{\nu} = \nu_{x}\boldsymbol{g}^{(1)} + \nu_{y}\boldsymbol{g}^{(2)} + \nu_{z}\boldsymbol{g}^{(3)} \, .
\end{equation}
Therefore, we can compute
\begin{equation}
    R_\ell^{(1)} :=  \boldsymbol{g}^{(1)} \cdot R_\ell \, , \qquad R_\ell^{(2)} :=  \boldsymbol{g}^{(2)} \cdot R_\ell \, , \qquad R_\ell^{(3)} :=  \boldsymbol{g}^{(3)} \cdot R_\ell \, .
\end{equation}
The QROM in Step \ref{step5} can be used to give these quantities, so we can calculate the dot product in the form
\begin{equation}
    k_{\nu}\cdot R_\ell =  \nu_{x} R_\ell^{(1)} + \nu_{y} R_\ell^{(2)} + \nu_{z} R_\ell^{(3)} \, .
\end{equation}
The cost of the QROM in Step \ref{step5} is unchanged by this modification, so the extra complexity to consider here is that from the three multiplications.
The result may be added into a phase gradient state to implement the phase rotation, which is a smaller complexity in comparison.

Using the result in Eq.~(D9) of \cite{Sanders2020}, the complexity of multiplying a $d_B$-qubit integer and a real number is approximately
\begin{equation}
    d_B^2 + 2 d_B b \, ,
\end{equation}
when giving the result to $b$ bits of precision.
The complexity for the three multiplications is then
\begin{equation}
    n_x^2+n_y^2+n_z^2 + 2 (n_x+n_y+n_z) b \, .
\end{equation}

The precision needed for the nuclear positions was previously described for the case without pseudopotentials in \cite{SuPRXQuantum21}.
For the case with pseudopotentials we can bound the error due to the approximation of the nuclear position in the following way.
Suppose that the nuclear positions $R_\ell$ are approximated by $\widetilde{R}_\ell$ such that $\norm{\widetilde{R}_\ell-R_\ell}\leq\delta_R$ for all $\ell$. Then, we have the following bounds on the distance between the corresponding Hamiltonian terms:
\begin{align}
    \norm{\widetilde{U}_{\rm loc}-U_{\rm loc}}
    &=\norm{\sum_{\ell=1}^L \sum_{\nu\in G_0} \locl\!\left(k_\nu\right) \left(e^{-ik_{\nu}\cdot \widetilde{R}_\ell}-e^{-ik_{\nu}\cdot R_\ell} \right)\sum_{j=1}^\eta \sum_{\substack{q\in G\\ (q-\nu) \in G}} \ket{q-\nu}\!\!\bra{q}_{j}}\nn
    &\leq\eta\sum_{\ell=1}^L \sum_{\nu\in G_0} \left|\locl\!\left(k_\nu\right)\right| \left|e^{-ik_{\nu}\cdot \widetilde{R}_\ell}-e^{-ik_{\nu}\cdot R_\ell} \right|
    \leq\eta\sum_{\ell=1}^L \sum_{\nu\in G_0} \left|\locl\!\left(k_\nu\right)\right| \left|k_{\nu}\cdot \left(\widetilde{R}_\ell-R_\ell\right)\right|\nn
    &\leq\eta\sum_{\ell=1}^L \sum_{\nu\in G_0} \left|\locl\!\left(k_\nu\right)\right|\norm{k_\nu}\norm{\widetilde{R}_\ell-R_\ell}
    \leq\eta\delta_R\sum_{\ell=1}^L \sum_{\nu\in G_0} \left|\locl\!\left(k_\nu\right)\right|\norm{k_\nu}
\end{align}
and
\begin{align}
    \norm{\widetilde{U}_{\rm nonloc}-U_{\rm nonloc}}
    &=\norm{\sum_{\ell=1}^L\sum_{\nu\in G_0} \left(e^{-ik_{\nu}\cdot \widetilde{R}_\ell}-e^{-ik_{\nu}\cdot R_\ell}\right) \sum_{j=1}^{\eta}\sum_{\substack{q\in G\\ (q-\nu) \in G}} \nonlocl(k_q,k_{q-\nu})  \ket{q-\nu}\!\!\bra{q}_j}\nn
    &\leq\eta\sum_{\ell=1}^L\sum_{\nu\in G_0} \left|e^{-ik_{\nu}\cdot \widetilde{R}_\ell}-e^{-ik_{\nu}\cdot R_\ell}\right|\max_{\substack{q\in G\\ (q-\nu) \in G}}\left|\nonlocl(k_q,k_{q-\nu})\right|\nn
    &\leq\eta\sum_{\ell=1}^L\sum_{\nu\in G_0} \left|k_{\nu}\cdot \left(\widetilde{R}_\ell-R_\ell\right)\right|\max_{\substack{q\in G\\ (q-\nu) \in G}}\left|\nonlocl(k_q,k_{q-\nu})\right|\nn
    &\leq\eta\sum_{\ell=1}^L\sum_{\nu\in G_0} \norm{k_{\nu}}\norm{\widetilde{R}_\ell-R_\ell}\max_{\substack{q\in G\\ (q-\nu) \in G}}\left|\nonlocl(k_q,k_{q-\nu})\right|\nn
    &\leq\eta\delta_R\sum_{\ell=1}^L\sum_{\nu\in G_0} \norm{k_{\nu}}\max_{\substack{q\in G\\ (q-\nu) \in G}}\left|\nonlocl(k_q,k_{q-\nu})\right| \, .
\end{align}
Because the exponential decay in the local and nonlocal pseudopotentials has factors of $r_{\rm loc}^\alpha$ and $r_l^\alpha$ respectively, the expected values of $\norm{k_\nu}$ should be on the order of $1/r_{\rm loc}^\alpha$ and $1/r_l^\alpha$, which are order 1.
Since the sums omitting $\norm{k_\nu}$ give values of $\lambda$, we can infer that
\begin{align}
    \norm{\widetilde{U}_{\rm loc}-U_{\rm loc}} &\lesssim \delta_R\lambda_{\rm loc}/r_{\rm loc}^\alpha \, , \\
    \norm{\widetilde{U}_{\rm nonloc}-U_{\rm nonloc}} &\lesssim \delta_R\lambda_{\rm nonloc}/r_l^\alpha \, .
\end{align}
That is, $\delta_R$ approximately gives the relative error.
For the purpose of the costing we will assume that the positions are specified with $b$ bits, similarly to other quantities in the coherent arithmetic.

\subsection{The kinetic energy}
\label{sec:kinetic}
In the prior work costing plane-wave simulation \cite{SuPRXQuantum21}, the arithmetic for the kinetic energy was completely avoided by expressing the sum of squares as a linear combination of bitwise products.
That was used to avoid needing to compute $\|q\|^2$.
Here we are using $\|k_q\|^2$, which is a norm computed using the Gramian.
That makes it more difficult to use the approach from \cite{SuPRXQuantum21}, but it is not needed.
Here, we are already computing $\|k_q\|^2$ for use in the nonlocal pseudopotential.
That means we can use that value for the block encoding of $T$ as well.

We can simply perform an inequality test between the value calculated for $\|k_q\|^2$ and a number in an equal superposition in an ancilla register.
This would suggest that we only have the cost of an inequality test, except that the maximum value of $\|k_q\|^2$ will typically not be a power of 2, and we would want the maximum value to match the maximum value in the ancilla register in order to minimize $\lambda_T$.
When the equal superposition state is prepared via Hadamards the maximum value is (one less than) a power of 2.

One approach would be to multiply $\|k_q\|^2$ by a constant in order to provide a matching maximum value.
The drawback is that multiplications have high complexity.
Instead, we can prepare the equal superposition state over a number of basis states that is not a power of 2.
That can be performed in the usual way using inequality testing and amplitude amplification, with complexity $3\lceil\log d\rceil+2b_r -9$ \cite{Lee2020}, where $d$ is the number of basis states to take an equal superposition over.

When the number of basis states to take a superposition over is a multiple of a power of 2, then the complexity can have 3 times that power subtracted from it.
In Ref.~\cite{Lee2020} that power was denoted $\eta$ (distinct from the notation $\eta$ for the number of electrons here), so that the complexity had $3\eta$ subtracted from it.
For our application here we have some freedom in the choice of $d$, because we can take it to be larger than necessary at the cost of a slightly larger value of $\lambda_T$.
For example, if we took $d$ to be 10 bits multiplied by a power of 2, then the cost of the preparation would be about 30 Toffolis, but the increase in $\lambda_T$ due to any imprecision would be less than 1 part in $1000$.
That is negligible for our application here.

\subsection{Total complexity}
Here we summarise the costs, and give the complete additional complexity for a block encoding of the Hamiltonian, beyond that given above for the pseudopotentials in Section \ref{ref:completepseudo}.
That will then need to be multiplied by the value of $\lambda$ discussed in the next section.
In the case without pseudopotentials, the costs are summarised in Table II of Ref.~\cite{SuPRXQuantum21}.
Here we summarise the costs using the same lines as in that Table, because many of the costs are equivalent.
In the following we give the number of bits for each component of momentum as $n$ for simplicity in many cases where the cost is small.
\begin{enumerate}[(i)]
    \item Registers need to be prepared to select between $T$, $V$, $U_{\rm loc}$ and $U_{\rm nonloc}$.
    We will first prepare registers for selecting between $T,V,U$, similarly to Ref.~\cite{SuPRXQuantum21}, then prepare registers selecting between $U_{\rm loc}$ and $U_{\rm nonloc}$ as in Section \ref{sec:stateprep}.
    The costing given in Eq.~(61) of Ref.~\cite{SuPRXQuantum21} assumes a particular relation between the values of $\lambda$ for $V$ and $U$ that does not hold here.
    Nevertheless, this is a small cost compared to the arithmetic in this block encoding, so will be ignored.
    The complexity for the preparation selecting between $U_{\rm loc}$ and $U_{\rm nonloc}$ is part of the costing in Section \ref{ref:completepseudo}.
    \item The preparation of equal superposition states over $i,j$ for selecting electrons will have the same cost as in Ref.~\cite{SuPRXQuantum21}, which is $14n_\eta + 8b_r - 36$.
    Note that it is typically reasonable to take $b_r=7$, which would give cost $14n_\eta + 20$.
    Here, $n_\eta$ is the number of qubits needed to store the electron number.
    \item The preparation of registers needed for $T$ is now changed from that in \cite{SuPRXQuantum21} because we are preparing a superposition over a number of basis states that is not a power of 2.
    The cost is doubled when accounting for inverse preparation.
    The complexity can be taken to be $3\lceil\log d\rceil+2b_r -9$ when preparing a superposition over $d$ basis states, which is negligible compared to the other costs here.
    \item \label{part4} The complexity of swapping the momentum registers for electrons into temporary registers is unchanged at $12\eta n+4\eta -8$.
    In the case with different numbers of bits for the components, this complexity is modified by replacing $3n\to n_x+n_y+n_z$ so it becomes $4\eta (n_x+n_y+n_z)+4\eta -8$.
    \item The select cost for $T$ is now just $b$ for an inequality test.
    \item \label{part6} The cost of preparing the state with $1/\|k_\nu\|$ amplitudes is covered in Section \ref{sec:nuprep} above, and is approximately triple the cost of computing $\|k_\nu\|^2$ plus $3b^2$ for three multiplications.
    In comparison, the leading-order complexity in Ref.~\cite{SuPRXQuantum21} was $3n^2$ for computing $\|k_\nu\|^2$ plus $4n_M n$ for the multiplications (with $n_M$ being the number of qubits for the equal superposition state).
    \item The cost of the QROM for outputting $R_\ell$ was already included in Section \ref{ref:completepseudo}.
    \item The cost of additions and subtractions of $\nu$ into momentum registers is unchanged at $24n$, or $8(n_x+n_y+n_z)$.
    \item \label{part9} The complexity of the phase factor is addressed in Section \ref{sec:phasecomp} above, and is $n_x^2+n_y^2+n_z^2 + 2 (n_x+n_y+n_z) b$ in the more general case.
\end{enumerate}
Of these costs, the leading three are the $1/\|k_\nu\|$ preparation cost in part \ref{part6}, the phase factor in part \ref{part9}, and the cost of swapping the momentum registers from part \ref{part4}.

\section{Computing \texorpdfstring{$\lambda$}{lambda}}
\label{sec:lambdas}
Here we describe the computation of $\lambda$ for the simulation.
We break up $\lambda$ into four parts, $\lambda_T,\lambda_V,\lambda_{\rm loc},\lambda_{\rm nonloc}$ corresponding to the four terms in the Hamiltonian.

\subsection{Kinetic energy \texorpdfstring{$\lambda$}{lambda}}
For comparison, the value of $\lambda_T$ for cubic unit cells was given in Eq.~(71) of \cite{SuPRXQuantum21} as
\begin{equation}
    \lambda_T = \frac{3\eta\pi^2}{2\Omega^{2/3}} 2^{2n_p}.
\end{equation}
In contrast, here we are accounting for general non-orthogonal Bravais vectors.
In our case here, we have
\begin{equation}
    \lambda_T = \frac{\eta}2 \max_q \|k_q\|^2 \, .
\end{equation}
This can be computed as
\begin{equation}\label{eq:lambdaT}
    \lambda_T = \frac{\eta}8 \max \| (N_x-1)\boldsymbol{g}^{(1)} \pm (N_y-1) \boldsymbol{g}^{(2)} \pm (N_z-1) \boldsymbol{g}^{(3)} \|^2 \, ,
\end{equation}
where the maximum is over the two $\pm$ signs, which are taken to be independent.
This expression is used to calculate values in Tables \ref{TAB:lambda_qpe_Costs_LNO} and \ref{TAB:lambda_qpe_Costs}.
We do not need to consider an arbitrary sign on the first term, because an overall sign flip leaves the norm unaffected.

This value of $\lambda_T$ results from using our approach with simply performing an inequality test using the computed value of $\|k_q\|^2$.
If we were to use the approach based on the individual qubits of $q$ similar to \cite{SuPRXQuantum21}, or the proposal for Bravais vectors in \cite{Shokrian}, then it can result in a larger $\lambda_T$.
In that approach one is effectively considering the expansion
\begin{equation}
    \|k_q\|^2 = q_x^2 \|\boldsymbol{g}^{(1)}\|^2 + q_y^2 \|\boldsymbol{g}^{(2)}\|^2 + q_z^2 \|\boldsymbol{g}^{(3)}\|^2 + 2 q_x q_y (\boldsymbol{g}^{(1)})^T \boldsymbol{g}^{(2)} + 2 q_x q_z (\boldsymbol{g}^{(1)})^T \boldsymbol{g}^{(3)} + 2 q_y q_z (\boldsymbol{g}^{(2)})^T \boldsymbol{g}^{(3)} \, ,
\end{equation}
and implementing each term according to a linear combination.
That means the value of $\lambda_T$ comes from taking the sum of the maximum of each of the terms, rather than the maximum of $\|k_q\|^2$.
In cases such as LiNiO$_2$ (C2/m) that gives a larger value, though in other cases it can be the same.

\subsection{Electron-electron potential \texorpdfstring{$\lambda$}{lambda}}
To explain the method to determine $\lambda_V$ for non-orthogonal Bravais vectors, we first summarise the method for cubic unit cells.
That can be be determined from the expression in Eq.\ (25) of \cite{SuPRXQuantum21},
\begin{equation}\label{eq:Su25}
    \lambda_V = \frac{\eta(\eta-1)}{2\pi\Omega^{1/3}} \lambda_\nu , \qquad \lambda_\nu = \sum_{\nu\in G_0} \frac 1{\|\nu\|^2} \, .
\end{equation}
When using $N_x=N_y=N_z$, $\lambda_\nu = \mathcal{O}(N^{1/3})$.
In particular, we obtain
\begin{align}
    \sum_{\nu\in G_0} \frac 1{\|\nu\|^2}  &\le \int_{-(N_x-1/2)}^{N_x-1/2} dx \int_{-(N_y-1/2)}^{N_y-1/2} dy \int_{-(N_z-1/2)}^{N_z-1/2} dz \frac 1{x^2+y^2+z^2} - 
    \int_{-1/2}^{1/2} dx \int_{-1/2}^{1/2} dy \int_{-1/2}^{1/2} dz \frac 1{x^2+y^2+z^2} \nn
    &< \int_{-N_x}^{N_x} dx \int_{-N_y}^{N_y} dy \int_{-N_z}^{N_z} dz \frac 1{x^2+y^2+z^2} \nn
    &= 24 N_x \left[ \frac{\pi}2 \ln\left( 1+\sqrt{2} \right) +  {\rm Ti}_2(3-\sqrt 8) - C \right] \nn
    &\approx 15.3482 N^{1/3} \, ,
\end{align}
where ${\rm Ti}_2$ is the Lewin inverse-tangent integral (which can be evaluated in terms of polylogarithms), and $C$ is Catalan's constant
(see Ref.\ \cite{Bailey2010}, or Eq.\ (29) of \cite{BabbushContinuum}).

The constant obtained in the case of a cube upper bounds that for a rectangular prism by noting the behaviour of the integral as the prism is reshaped to a cube of the same volume.
If, for example, $N_x>N_y$, then reducing the value of $N_x$ by an infinitesimal amount $\delta x$ and increasing the value of $N_y$ by $\delta y$, to maintain the same volume we need
\begin{equation}
    N_x\delta y = N_y\delta x \, .
\end{equation}
The change in the value of the integral over $x,y$ for each $z$ is
\begin{align}
   & -2\delta x \int_{-N_y}^{N_y} dy  \frac 1{N_x^2+y^2+z^2}
    + 2\delta y \int_{-N_x}^{N_x} dx \frac 1{x^2+N_y^2+z^2} \nn
    & = -2\delta x \int_{-N_y}^{N_y} dy  \frac 1{N_x^2+y^2+z^2}
    + 2\delta x \frac{N_y}{N_x} \int_{-N_x}^{N_x} dx \frac 1{x^2+N_y^2+z^2}\nn
   & = -2\delta x \, N_y \int_{-1}^{1} d\tilde y  \frac 1{N_x^2+\tilde y^2+z^2}
    + 2\delta x \, N_y \int_{-1}^{1} d\tilde x \frac 1{\tilde x^2+N_y^2+z^2} \nn
    & > 0 \, ,
\end{align}
where $\tilde x=x/N_x$ and $\tilde y=y/N_y$, and the last inequality is because $N_x>N_y$.
This shows that for any non-cubic (rectangular prism) region, changing the shape towards a cube while preserving the volume can only increase the integral.
The reason for this is that regions of the shape where $1/\|\nu\|$ is smaller are being replaced with regions where it is larger in reshaping the region to a cube.

That is still for orthogonal Bravais vectors.
In the general case with non-orthogonal Bravais vectors, we would use
\begin{align}
     \lambda_V &= \frac{2\pi}{\Omega} \eta(\eta-1) \sum_{\nu\in G_0}\frac 1{\|\nu_x \boldsymbol{g}^{(1)} + \nu_y \boldsymbol{g}^{(2)} + \nu_z \boldsymbol{g}^{(3)}\|^2} \label{eq:lambdaV} \\
     & < \frac{2\pi}{\Omega} \eta(\eta-1)\int_{-N_x}^{N_x} d\nu_x \int_{-N_y}^{N_y} d\nu_y \int_{-N_z}^{N_z} d\nu_z \frac 1{\|\nu_x \boldsymbol{g}^{(1)} + \nu_y \boldsymbol{g}^{(2)} + \nu_z \boldsymbol{g}^{(3)}\|^2} \, .
\end{align}
The expression in the first line is used to calculate $\lambda_V$ in Tables \ref{TAB:lambda_qpe_Costs_LNO} and \ref{TAB:lambda_qpe_Costs}.
Now let us use the change of variables where $x,y,z$ are the components of $\nu_x \boldsymbol{g}^{(1)} + \nu_y \boldsymbol{g}^{(2)} + \nu_z \boldsymbol{g}^{(3)}$.
That is, we have
\begin{equation}\label{eq:xyztrans}
    \begin{pmatrix}
        x \\ y \\ z
    \end{pmatrix} = [\boldsymbol{g}^{(1)} , \boldsymbol{g}^{(2)} , \boldsymbol{g}^{(3)}] \begin{pmatrix}
        \nu_x \\ \nu_y \\ \nu_z
    \end{pmatrix} \, .
\end{equation}

According to the usual rule for changing variables in integrals, we replace the integral over $\nu$ with one over $x,y,z$
\begin{equation}
    \int_{-N_x}^{N_x} d\nu_x \int_{-N_y}^{N_y} d\nu_y \int_{-N_z}^{N_z} d\nu_z \frac 1{\|\nu_x \boldsymbol{g}^{(1)} + \nu_y \boldsymbol{g}^{(2)} + \nu_z \boldsymbol{g}^{(3)}\|^2} = \iiint \det(J)\, dx \, dy\, dz\, \frac 1{x^2+y^2+z^2} \, ,
\end{equation}
where
\begin{equation}
    J = \begin{pmatrix}
        \frac{\partial \nu_x}{\partial x} &  \frac{\partial \nu_x}{\partial y} &  \frac{\partial \nu_x}{\partial z} \\
        \frac{\partial \nu_y}{\partial x} &  \frac{\partial \nu_y}{\partial y} &  \frac{\partial \nu_y}{\partial z} \\
        \frac{\partial \nu_z}{\partial x} &  \frac{\partial \nu_z}{\partial y} &  \frac{\partial \nu_z}{\partial z}
    \end{pmatrix} = [\boldsymbol{g}^{(1)} , \boldsymbol{g}^{(2)} , \boldsymbol{g}^{(3)}]^{-1} .
\end{equation}
Using $[\boldsymbol{g}^{(1)} , \boldsymbol{g}^{(2)} , \boldsymbol{g}^{(3)}] = 2\pi(\boldsymbol{a}^{-1})^T$, we have
\begin{equation}\label{eq:JvalV}
    \det(J) = \det([\boldsymbol{g}^{(1)} , \boldsymbol{g}^{(2)} , \boldsymbol{g}^{(3)}])^{-1} = \frac{\det(\boldsymbol{a})}{(2\pi)^3} = \frac{\Omega}{(2\pi)^3} \, .
\end{equation}

However, this transformation of variables also changes the volume of the region.
The region corresponds to those values of $x,y,z$ obtained from Eq.~\eqref{eq:xyztrans} for $\nu_x\in [-N_x,N_x]$, $\nu_y\in [-N_y,N_y]$, $\nu_z\in [-N_z,N_z]$.
That region is a parallelepiped, where the sides are $2N_x \boldsymbol{g}^{(1)}$, $2N_y \boldsymbol{g}^{(2)}$, $2N_z \boldsymbol{g}^{(3)}$.
The volume of a parallelepiped is given by the triple scalar product of the vectors for its sides, and so here is $8N_xN_yN_z\det([\boldsymbol{g}^{(1)} , \boldsymbol{g}^{(2)} , \boldsymbol{g}^{(3)}])$.
That is, the volume of the region is increased by a factor of $\det([\boldsymbol{g}^{(1)} , \boldsymbol{g}^{(2)} , \boldsymbol{g}^{(3)}])$.
In a similar way as we bounded the integral for the rectangular prism with that of the cube of the same volume,
modifying the parallelepiped to a rectangular prism of the same volume can only increase the integral.
This is because we are again replacing regions where $1/(x^2+y^2+z^2)$ is smaller with those where it is larger in reshaping the region.
Therefore we can use the upper bound for the cube of the same volume.
Since the integral gives a factor of the volume to the power of $1/3$, the change of variables gives a factor of $\det([\boldsymbol{g}^{(1)} , \boldsymbol{g}^{(2)} , \boldsymbol{g}^{(3)}])^{1/3}=2\pi/\Omega^{1/3}$.
Multiplying that by $\Omega/(2\pi)^3$ from the Jacobian, it gives an overall factor of $\Omega^{2/3}/(2\pi)^2$.
This means that exactly the same factor of $\Omega^{2/3}/(2\pi)^2$ is obtained as when we used a cubic region.
For specific examples it is more accurate to compute $\lambda_V$ by explicitly performing the sum, though.

\subsection{Local pseudopotential \texorpdfstring{$\lambda$}{lambda}}
For the local part of the pseudopotential, the value of $\lambda_{\rm loc}$ can be as small as
\begin{equation}
    \lambda_{\rm loc} = \eta \sum_{\alpha} L_\alpha \sum_{\nu\in G_0} \left|\loc\!\left(k_\nu\right) \right| \, .
\end{equation}
This is the value of $\lambda_{\rm loc}$ that would be obtained if we implemented the block encoding via a tight preparation over $\nu$.
We will give the amended expression for the actual implementation below.
This value of $\lambda_{\rm loc}$ is derived by noting that the expression in large round brackets for $U_{\rm loc}$ is approximately unitary.
(The only nonunitarity is from the restriction that $(q-\nu)\in G$, which is implemented by eliminating any transfers outside the state space in the block encoding.)
Here we have switched from the sum over $\ell$ to that over $\alpha$, and used the number of atoms of each species $L_\alpha$.
This is an expression which can be computed via an explicit sum over $\nu$.
An approximate value can be found by approximating the sum by an integral, and integrating over an infinite region.
In the case where $r_{\rm loc}^\alpha$ is significantly smaller than the size of the region, this should be accurate.

First, note that for the cubic case $k_\nu=2\pi\nu/\Omega^{1/3}$, so the integral over $\nu$ gives
\begin{align}\label{eq:loclam}
\int \left|\loc\!\left(k_\nu\right) \right| d^3\nu &= \int \left|- \frac{4 \pi Z_\alpha}{\Omega\|k_\nu\|^2}e^{-\left \|k_\nu \right\|^2 (r_{\rm loc}^\alpha)^2/2} + \sqrt{8 \pi^3}\frac{(r_{\rm loc}^\alpha)^3}{\Omega} \sum_{j=1}^4 C_j^\alpha \widetilde F_0^j \! \left( r_{\rm loc}^\alpha \|k_\nu \| \right) \right| d^3\nu  \nn
 &= 2 \int_0^{\infty} \left|-\frac{Z_{\alpha}}{\pi r_{\rm loc}^\alpha}e^{-r^2/2} + \frac{r^2}{\sqrt{2\pi}} \sum_{j=1}^4 C_j^\alpha \widetilde F_0^j \! \left( r \right) \right| dr \, ,
\end{align}
where $r=\|k_\nu\|r_{\rm loc}^\alpha$, so
\begin{equation}\label{eq:cov}
    d^3 \nu = d^3 \vec r \frac{\Omega}{(2\pi r_{\rm loc}^\alpha)^3} \mapsto dr \frac{4\pi r^2 \Omega}{(2\pi r_{\rm loc}^\alpha)^3} = dr \frac{r^2 \Omega}{2\pi^2(r_{\rm loc}^\alpha)^3}.
\end{equation}
Here, $\vec r=k_\nu r_{\rm loc}^\alpha$ and `$\mapsto$' indicates integration over the angular coordinates.
The form in Eq.~\eqref{eq:loclam} gives an approximation of $\lambda_{\rm loc}$ depending only on the parameters of the pseudopotential for the nucleus, and not on $N$ or $\Omega$.

When using non-orthogonal Bravais vectors, we can use the change of variables
\begin{equation}
    \vec r = k_\nu r_{\rm loc}^\alpha = r_{\rm loc}^\alpha [\boldsymbol{g}^{(1)} , \boldsymbol{g}^{(2)} , \boldsymbol{g}^{(3)}] \begin{pmatrix}
        \nu_x \\ \nu_y \\ \nu_z
    \end{pmatrix} ,
\end{equation}
so $r=\|(x,y,z)\|=\|k_\nu\|r_{\rm loc}^\alpha$.
This just differs from the change of variables used to bound $\lambda_V$ above by the factor of $r_{\rm loc}^\alpha$, so instead of Eq.~\eqref{eq:JvalV} we obtain
\begin{equation}
    \det(J) = \frac{\Omega}{(2\pi r_{\rm loc}^\alpha)^3} \, .
\end{equation}
This means that we have the same rule for the change of variables as in Eq.~\eqref{eq:cov}, and exactly the same derivation as in Eq.~\eqref{eq:loclam} holds.

Hence the same expression is obtained regardless of whether a cubic region or non-orthogonal Bravais vectors are used.
Note also that there is a factor of $1/r_{\rm loc}^\alpha$ on the first term in Eq.~\eqref{eq:loclam}, and no factors of $1/\Delta$.
In contrast, for $\lambda_V$ and $\lambda_T$ there are factors of $1/\Delta$ and $1/\Delta^2$, respectively, where $\Delta=(\Omega/N)^{1/3}$ is the spacing of the plane-wave reciprocal lattice.
It can therefore be expected that the value of $\lambda_{\rm loc}$ is small compared to $\lambda_V$ when the lattice spacing is smaller than $r_{\rm loc}^\alpha$ or $1$, which is the typical situation that should be expected.

\begin{table}[!htpb]
    \begin{tabular}{|c|c|c|}
    \hline\hline
    atom & Eq.\ \eqref{eq:loclam} & Eq.\ \eqref{eq:aprx2b} \\
    \hline
    Li & 1.80296 & 3.43499 \\
    C  & 9.92985 & 19.9358 \\
    N  & 14.6176 & 29.2986 \\
    O  & 20.1626 & 40.3460 \\
    F  & 26.3031 & 52.5513 \\
    Al & 7.38817 & 13.8106 \\
    Mn & 8.72686 & 8.72686\\
    Ni & 14.2479 & 14.2479 \\
    Pt & 23.9801 & 23.9801 \\
    Pd & 18.5970 & 18.5970 \\
    Rh & 16.9535 & 16.9535 \\ 
    \hline\hline
        \end{tabular}
    \caption{The integral approximation $\int \left|\loc\!\left(k_\nu\right) \right| d^3\nu$ from Eq.\ \eqref{eq:loclam}, and the corresponding approximation obtained when separating the terms given in the brackets in Eq.\ \eqref{eq:aprx2b}.
    \label{TAB:lamvals}}
\end{table}

In practice, our block encoding will work by separately implementing the terms in the sum for $U_{\rm loc}$.
That means that $\lambda_{\rm loc}$ should be given by
\begin{align}
    \lambda_{\rm loc} &\coloneqq \eta \sum_{\alpha} L_\alpha \sum_{\nu\in G_0}  \left[ \frac{4 \pi Z_\alpha}{\Omega\|k_\nu\|^2}e^{-(r_{\rm loc}^\alpha\left \|k_\nu \right\|)^2/2} + \sqrt{8 \pi^3}\frac{(r_{\rm loc}^\alpha)^3}{\Omega} \sum_{j=1}^4 \left|C_j^\alpha \widetilde F_0^j \! \left( r_{\rm loc}^\alpha \|k_\nu \| \right) \right| \right] \label{eq:aprx2} \\
    & \approx \eta \sum_{\alpha} L_\alpha
\left( \sqrt{\frac{2}{\pi}} \frac{Z_\alpha}{r_{\rm loc}^\alpha} + |C_1^\alpha| + 1.85016|C_2^\alpha| + 7.00306|C_3^\alpha| + 40.9966 |C_4^\alpha| \right).\label{eq:aprx2b}
\end{align}
The expression on the first line is used to calculate $\lambda_{\rm loc}$ in Tables \ref{TAB:lamvals1}, \ref{TAB:lambda_qpe_Costs_LNO}, and \ref{TAB:lambda_qpe_Costs}.
In the second line we have used the integral approximation again.
This integral approximation and that without separating components are shown in
Table \ref{TAB:lamvals}.
It can be seen that separately implementing the terms in the sum gives about a factor of 2 increase in many cases, but it is still relatively small.

When we are performing the preparation, we are only choosing the amplitudes according to the box that $\nu$ is in, which increases the effective value of $\lambda$.
For the local pseudopotential, we are able to choose to implement a different component of the Hamiltonian (for example $T$) in the case of the failure of the inequality test.
This is similar to the approach used in \cite{SuPRXQuantum21} to account for the failures in state preparation for $V$.
Since the value of $\lambda$ for the local pseudopotential is small, it can be expected that we can always apply $V$ in the failure cases, so we can still use this expression for $\lambda_{\rm loc}$.

The expression in Eq.~\eqref{eq:aprx2} can slightly overestimate the values for real materials, since it is using an upper bound.
For the example of C (diamond), if the simulation cell consists of one primitive cell then we obtain $17.1594$, which is slightly less.
For diamond it would be appropriate to use more primitive cells within a simulation cell.
Using 2 in each direction increases the value to $18.5678$, 4 in each direction gives $19.2545$, and 8 gives $19.5956$, which is closer to the value found from Eq.\ \eqref{eq:aprx2}.
There are many examples of crystal structures listed in Table \ref{TAB:LATTICE_VECTORS}.
In Table \ref{TAB:lamvals1} we give the corresponding contributions to $\lambda_{\rm loc}$ for each of the nuclei in these structures (omitting the factor of $\eta$). 
In each case we find that the value is smaller than that given by Eq.~\eqref{eq:aprx2}, but only by a small amount.

\begin{table}[!htpb]
\begin{tabular}{|c|c|c|c|c|c|c|c|c|c|c|c|}
\hline\hline
structure            & Li & C  & N  & O  & F  & Al & Mn & Ni & Pt & Pd & Rh \\
\hline
LiNiO$_2$ (C2/m)     & 3.1169 &    &    & 38.545 &    &    &    & 11.278 &    &    &    \\
LiNiO$_2$ (P2$_1$/c) & 3.1215 &    &    & 38.586 &    &    &    & 11.345 &    &    &    \\
LiNiO$_2$ (P2/c)     & 3.1208 &    &    & 38.582 &    &    &    & 11.338 &    &    &    \\
Pd       &    &    &    &    &    &    &    &    &    & 17.265 &    \\
Pt $2\times 2$       &    &    &    &    &    &    &    &    & 22.679 &    &    \\
Pt $3\times 3$       &    &    &    &    &    &    &    &    & 22.518 &    &    \\
Pt $4\times 4$       &    &    &    &    &    &    &    &    & 22.653 &    &    \\
Rh      &    &    &    &    &    &    &    &    &    &    & 15.743   \\
Li$_{0.5}$MnO$_3$    & 3.3533 &    &    & 35.890 &    &    & 8.1836 &    &    &    &    \\
Li$_{0.75}$[Li$_{0.17}$Ni$_{0.25}$Mn$_{0.58}$]O$_2$ & 3.3756 &    &    & 34.885 &    &    & 8.3326 & 13.681 &    &    &    \\
Li$_{0.75}$MnO$_2$F  & 3.2840 &    &    & 38.125 & 48.452 &    & 7.6931 &    &    &    &    \\
C (diamond)          &    & 19.026 &    &    &    &    &    &    &    &    &    \\
AlN (wurzite)        &    &    & 27.611 &    &    & 13.408 &    &    &    &    &    \\
\hline\hline
\end{tabular}
    \caption{The values of $\sum_{\nu\in G_0} \left|\loc\!\left(k_\nu\right) \right|$ using the explicit sum for various nuclei and the Bravais vectors for structures given in Table \ref{TAB:LATTICE_VECTORS} with $n=6$, using Eq.~\eqref{eq:aprx2}.
    The supercell size is given for Pt (but not the others) in order to distinguish the cases.
    \label{TAB:lamvals1}}
\end{table}

It is also possible to compute the error due to interpolation of the exponential via a very similar calculation.
That is, in the calculation where we would otherwise be estimating $\lambda$, replace each value of the exponential by the interpolation error in the exponential.
The results of performing this calculation for a quadratic interpolation with 64 points are given in Table~\ref{TAB:lamvals1e}.
This choice of interpolation can be expected to give a \emph{maximum} relative error in the interpolation below $10^8$, but it will be smaller on average.
It can be expected that the average error is more relevant here, as the sum involves the evaluation of the exponential at a large number of possible arguments.
In each case we find that the error is about 4 parts per billion relative to $\lambda$.

\begin{table}[!htpb]
\begin{tabular}{|c|c|c|c|c|c|c|c|c|c|c|c|}
\hline\hline
structure            & Li & C  & N  & O  & F  & Al & Mn & Ni & Pt & Pd & Rh \\
\hline
LiNiO$_2$ (C2/m)     & 11.657 &    &    & 160.80 &    &    &    & 44.851 &    &    &    \\
LiNiO$_2$ (P2$_1$/c) & 13.788 &    &    & 155.15 &    &    &    & 46.761 &    &    &    \\
LiNiO$_2$ (P2/c)     & 12.472 &    &    & 156.15 &    &    &    & 55.780 &    &    &    \\
Pd       &    &    &    &    &    &    &    &    &    & 74.971 &    \\
Pt $2\times 2$       &    &    &    &    &    &    &    &    & 95.523 &    &    \\
Pt $3\times 3$       &    &    &    &    &    &    &    &    & 91.906 &    &    \\
Pt $4\times 4$       &    &    &    &    &    &    &    &    & 98.291 &    &    \\
Rh      &    &    &    &    &    &    &    &    &    &    & 61.602   \\
Li$_{0.5}$MnO$_3$    & 13.577 &    &    & 147.04 &    &    & 34.882 &    &    &    &    \\
Li$_{0.75}$[Li$_{0.17}$Ni$_{0.25}$Mn$_{0.58}$]O$_2$ & 14.257 &    &    & 143.57 &    &    & 33.348 & 55.218 &    &    &    \\
Li$_{0.75}$MnO$_2$F  & 13.906 &    &    & 157.15 & 199.95 &    & 29.801 &    &    &    &    \\
C (diamond)          &    & 80.333 &    &    &    &    &    &    &    &    &    \\
AlN (wurzite)        &    &    & 117.24 &    &    & 55.722 &    &    &    &    &    \\
\hline\hline
\end{tabular}
    \caption{The errors in implementing the local component of the pseudopotential using a quadratic interpolation and 64 interpolation points.
    All values are multiplied by $10^9$, so are in billionths of a Hartree.
    These are contributions per nucleus and electron, similar to Table \ref{TAB:lamvals1}.}
    \label{TAB:lamvals1e}
\end{table}

\subsection{Nonlocal pseudopotential \texorpdfstring{$\lambda$}{lambda}}

Next, based on the definition of $U_{\rm nonloc}$, the value of $\lambda_{\rm nonloc}$ can be as small as
\begin{align}\label{eq:nonlocsum}
    \lambda_{\rm nonloc} = \eta \sum_{\alpha} L_\alpha \sum_{\nu\in G_d} \max_{\substack{q\in G\\ (q-\nu) \in G}} \left| \nonloc(k_q,k_{q-\nu})\right| .
\end{align}
When block encoding the nonlocal part by separating the $l,i,j$ components, the value of $\lambda$ becomes
\begin{align}\label{eq:nonloclam}
\lambda_{\rm nonloc} \coloneqq \frac{\eta}{\Omega} \sum_{\alpha} L_\alpha \sum_{\nu\in G_d}
\sum_{l = 0}^{l_{\text{max}}}
\frac{\left(2 l + 1\right)}{4\pi}
\sum_{i = 1}^3
\sum_{j = 1}^3 |E_{l\alpha}^{ij}| C_{li}^{\alpha} C_{lj}^{\alpha}  \upper_{\alpha,\nu,l,i,j} \, ,
\end{align}
where
\begin{equation}
\upper_{\alpha,\nu,l,i,j} \coloneqq   \frac 1{(r_l^\alpha)^{2l}} \max_{q}
\left| P_l \left(\frac{k_{q+\nu} \cdot k_q}{\left\|k_{q+\nu}\right\| \left\|k_q\right\|}\right) \widetilde F_{l}^i\!\left(r_l^\alpha\|k_{q+\nu}\|\right) \widetilde F_{l}^j\!\left(r_l^\alpha\|k_q\|\right) \right| \, .
\end{equation}
This is the formula used to calculate the $\lambda_{\rm nonloc}$ values in Tables \ref{TAB:lamvals4}, \ref{TAB:lambda_qpe_Costs_LNO}, and \ref{TAB:lambda_qpe_Costs}.
Recall that $C_{lj}^{\alpha}\ge 0$ so the absolute value is not required.
Then we can approximate $\lambda_{\rm nonloc}$ via the integral,
\begin{align}
\lambda_{\rm nonloc} &\approx \frac{\eta}{\Omega} \sum_{\alpha} L_\alpha 
\sum_{l = 0}^{l_{\text{max}}}
\frac{\left(2 l + 1\right)}{4\pi} 
\sum_{i = 1}^3
\sum_{j = 1}^3
|E_{l\alpha}^{ij}| C_{li}^{\alpha} C_{lj}^{\alpha} \int
\upper_{\alpha,\nu,l,i,j} d^3 \nu \, .
\end{align}

Now we consider the change of variables using the general form for non-orthogonal Bravais vectors, and the case for a cubic region can be considered to be a special case.
For each $l$ we make the change of variables
\begin{equation}
    \begin{pmatrix}
        x \\ y \\ z
    \end{pmatrix} = r_l^\alpha [\boldsymbol{g}^{(1)} , \boldsymbol{g}^{(2)} , \boldsymbol{g}^{(3)}] \begin{pmatrix}
        \nu_x \\ \nu_y \\ \nu_z
    \end{pmatrix} ,
\end{equation}
so the change of variables gives the factor
\begin{equation}
    \det(J) = \frac{\Omega}{(2\pi r_l^\alpha)^3} \, .
\end{equation}
If we also upper bound the Legendre polynomial by 1, that gives
\begin{align}
\lambda_{\rm nonloc} &\approx \frac{\eta}{\Omega} \sum_{\alpha} L_\alpha 
\sum_{l = 0}^{l_{\text{max}}}
\frac{\left(2 l + 1\right)}{4\pi} 
\sum_{i = 1}^3
\sum_{j = 1}^3
|E_{l\alpha}^{ij}| C_{li}^{\alpha} C_{lj}^{\alpha} \int_0^\infty
\frac 1{(r_l^\alpha)^{2l}} \widetilde\upper_{r,l,i,j} 4\pi r^2 \frac{\Omega}{(2\pi r_l^\alpha)^3} dr .
\end{align}
where
\begin{equation}
\widetilde \upper_{r,l,i,j} \coloneqq   \max_{r_p\ge 0,r_q\ge 0,|r_p-r_q|\le r}
\left| \widetilde F_{l}^i\!\left(r_p\right) \widetilde F_{l}^j\!\left(r_q\right) \right| .
\end{equation}
If we further define $\widetilde C_{lj}=C_{lj}^\alpha/(r_l^\alpha)^{l+3/2}$ (so they are independent of $\alpha$), then we obtain
\begin{align}\label{eq:lamnonloc}
\lambda_{\rm nonloc} &\approx  \eta \sum_{\alpha} L_\alpha
\left(\sum_{l = 0}^{l_{\text{max}}}
\frac{\left(2 l + 1\right)}{8\pi^3} 
\sum_{i = 1}^3
\sum_{j = 1}^3
|E_{l\alpha}^{ij}| \widetilde C_{li} \widetilde C_{lj} \int_0^\infty
r^2 \widetilde\upper_{r,l,i,j} dr\right)
\, .
\end{align}
This again gives us an expression that only depends on the parameters of the pseudopotential, and not independently on $\Omega$ and $N$.
In fact, the way it is written is now independent of $r_l^\alpha$, and depends on the nucleus only through $E_{l\alpha}^{ij}$.
It is also independent of the geometry of the non-orthogonal Bravais vectors.
The values computed for a range of atoms are given in Table \ref{TAB:lamvals2}.
For the atoms with $l_{\max}=0$, for example B, the value of $\lambda_{\rm nonloc}$ is just $8E_0^{00}$.

\begin{table}[!htpb]
    \begin{tabular}{|l|c|}
    \hline\hline
    atom & $\lambda_{\rm nonloc}$  \\
    \hline
    Li & 15.0167 \\
    C  & 76.1827 \\
    N  & 108.4179 \\
    O  & 146.1353 \\
    F  & 188.6795 \\
    Al & 141.3821 \\
    Mn & 539.4224 \\
    Ni & 740.9611 \\
    Pt & 538.7159 \\
    Pd & 354.0983 \\
    Rh & 1113.8664 \\
    \hline\hline
        \end{tabular}
    \caption{The sum in round brackets in Eq.~\eqref{eq:lamnonloc}, corresponding to the    
    approximate value of $\lambda_{\rm nonloc}$ per electron and nucleus.}
    \label{TAB:lamvals2}
\end{table}

\begin{table}[!htpb]
\begin{tabular}{|c|c|c|c|c|c|c|c|c|c|c|c|}
\hline\hline
structure            & Li & C  & N  & O  & F  & Al & Mn & Ni & Pt & Pd & Rh \\
\hline
LiNiO$_2$ (C2/m)     & 14.146 &    &    & 145.19 &    &    &    & 686.17 &    &    &    \\
LiNiO$_2$ (P2$_1$/c) & 14.034 &    &    & 145.07 &    &    &    & 684.33 &    &    &    \\
LiNiO$_2$ (P2/c)     & 14.041 &    &    & 145.07 &    &    &    & 684.49 &    &    &    \\
Pd       &    &    &    &    &    &    &    &    &    & 330.23 &    \\
Pt $2\times 2$       &    &    &    &    &    &    &    &    & 487.16 &    &    \\
Pt $3\times 3$       &    &    &    &    &    &    &    &    & 499.80 &    &    \\
Pt $4\times 4$       &    &    &    &    &    &    &    &    & 504.28 &    &    \\
Rh      &    &    &    &    &    &    &    &    &    &    & 1022.7   \\
Li$_{0.5}$MnO$_3$    & 14.546 &    &    & 136.76 &    &    & 488.99 &    &    &    &    \\
Li$_{0.75}$[Li$_{0.17}$Ni$_{0.25}$Mn$_{0.58}$]O$_2$ & 14.568 &    &    & 131.53 &    &    & 474.38 & 612.34 &    &    &    \\
Li$_{0.75}$MnO$_2$F  & 14.684 &    &    & 144.45 & 184.11 &    & 507.87 &    &    &    &    \\
C (diamond)          &    & 75.673 &    &    &    &    &    &    &    &    &    \\
AlN (wurzite)        &    &    & 106.88 &    &    & 138.36 &    &    &    &    &    \\
\hline\hline
\end{tabular}
    \caption{The values of $\lambda_{\rm nonloc}$ for various nuclei computed using the explicit sum over $\nu$ with maximization as in Eq.~\eqref{eq:nonlocsum}, and $n=6$.}
    \label{TAB:lamvals3}
\end{table}

\begin{table}[!htpb]
\begin{tabular}{|c|c|c|c|c|c|c|c|c|c|c|c|}
\hline\hline
structure            & Li & C  & N  & O  & F  & Al & Mn & Ni & Pt & Pd & Rh \\
\hline
LiNiO$_2$ (C2/m)     & 123.41 &    &    & 1281.3 &    &    &    & 5978.0 &    &    &    \\
LiNiO$_2$ (P2$_1$/c) & 127.12 &    &    & 1289.9 &    &    &    & 6043.0 &    &    &    \\
LiNiO$_2$ (P2/c)     & 124.13 &    &    & 1260.3 &    &    &    & 5876.1 &    &    &    \\
Pd       &    &    &    &    &    &    &    &    &    & 2433.9 &    \\
Pt $2\times 2$       &    &    &    &    &    &    &    &    & 5574.5 &    &    \\
Pt $3\times 3$       &    &    &    &    &    &    &    &    & 4339.3 &    &    \\
Pt $4\times 4$       &    &    &    &    &    &    &    &    & 5412.3 &    &    \\
Rh      &    &    &    &    &    &    &    &    &    &    & 6571.9   \\
Li$_{0.5}$MnO$_3$    & 110.62 &    &    & 597.21 &    &    & 2768.6 &    &    &    &    \\
Li$_{0.75}$[Li$_{0.17}$Ni$_{0.25}$Mn$_{0.58}$]O$_2$ & 159.07 &    &    & 728.06 &    &    & 3452.3 & 3713.4 &    &    &    \\
Li$_{0.75}$MnO$_2$F  & 115.80 &    &    & 780.40 & 883.04 &    & 3290.6 &    &    &    &    \\
C (diamond)          &    & 639.83 &    &    &    &    &    &    &    &    &    \\
AlN (wurzite)        &    &    & 644.16 &    &    & 1197.0 &    &    &    &    &    \\
\hline\hline
\end{tabular}
    \caption{The values of $\lambda_{\rm nonloc}$ for various nuclei computed using the explicit sum over $\nu$ with maximization as in Eq.~\eqref{eq:nonloclam}, and $n=6$. For this we are using the maxima over nested cubes which increases the effective $\lambda$.
    We use the same $\delta$ shifts for the cubes as given in Table \ref{TAB:sucprobs}.}
    \label{TAB:lamvals4}
\end{table}

For both the local and nonlocal pseudopotential, we are only choosing the amplitudes according to the box that $\nu$ is in, which increases the effective value of $\lambda$.
For the local pseudopotential we were able to choose to implement a different component of the Hamiltonian for failure of the inequality test.
That isn't possible for the \emph{nonlocal} pseudopotential, because the amplitude for success of the inequality test depends on the state as well as $\nu$.
Typically, determining a $\nu$-dependent upper bound (as in $\upper$ above) is complicated, and would require further function interpolation complexity to implement.
For this reason we do not apply any other components of the Hamiltonian in the case of failure of the inequality test, and there will be an increase to the effective value of $\lambda_{\rm nonloc}$.

To give an explicit formula for this effective value of $\lambda_{\rm nonloc}$, we adopt the notation used in Appendix \ref{app:nested}.
We use $B_\mu$ for box $\mu$, and $\mu(\nu)$ for the minimum $\mu$ such that $\nu\in B_\mu$.
We can then use 
\begin{equation}\label{eq:alephbox}
    \upper_{\alpha,\nu,l,i,j}^{\rm box} \coloneqq \max_{\nu'\in B_{\mu(\nu)}\backslash B_{\mu(\nu)-1}}
    \upper_{\alpha,\nu',l,i,j} \, .
\end{equation}
That is, the effective $\lambda$ is obtained by using the maximum value in the current box excluding the inner box; see Appendix \ref{app:nested} for further discussion.
We can then calculate the effective $\lambda_{\rm nonloc}$ by using Eq.~\eqref{eq:nonloclam} with $\upper_{\alpha,\nu,l,i,j}^{\rm box}$.

\section{Cost estimates for quantum simulating interesting materials}
\label{sec:examples}

In this section we describe the proposed use case of CO adsorption on transition metals; its
motivation, the details of DFT calculations and their results which are used to estimate inputs
for the proposed quantum algorithm.
CO adsorption on transition metals finds application in many heterogeneous catalytic systems
including methanation~\cite{sehested2005methanation}, steam reforming of fossil fuels~\cite{jones2008first}, methanol synthesis~\cite{grabow2011mechanism, behrens2012active}, water-gas-shift
and its reverse~\cite{grabow2008mechanism}, as well as the electrochemical reduction of $\mathrm{CO_{2}}$. CO is also responsible for
deactivation of catalysts through coking (e.g.\ steam reforming) or poisoning (e.g.\ if present in
feeds of low temperature fuel cells~\cite{takenaka2004complete}). Thus, it is important to provide reliable and accurate
estimates of the binding (i.e.\ adsorption) energies, as well as adsorption sites of CO on
transition metal surfaces to accurately describe reaction energetics, deactivation mechanisms,
and realistic active sites for the reactions involved.

However, generalized gradient
approximation (GGA) family of exchange correlation functionals in the context of density functional
theory (DFT-GGA) – commonly used within the communities of heterogeneous catalysis and
surface science – routinely overestimate the binding energies and wrongly assign the binding
sites of CO on transition metals, relative to experiments~\cite{PhysRevB.68.073401, PhysRevB.100.035442}. This is commonly known in
literature as the `CO/Pt(111) puzzle'~\cite{feibelman2001co, olsen2003co}, though the issue is manifested still for other metal
surfaces that form a strong chemisorption bond with CO. The fundamental reason behind this
problem is attributed to the self-interaction error associated with GGA density functionals
(among other functionals), which lead to the overestimation of the back donation of electrons
from the surface to adsorbed CO. Essentially, these functionals incorrectly place the highest
occupied and lowest unoccupied molecular orbitals of CO relative to the Fermi level of the
metal~\cite{PhysRevB.100.035442}. These quantitative (i.e.\ overestimation of binding energies) and qualitative (i.e.\ wrong
adsorption sites) errors can be mitigated by reducing density-driven errors via using higher
level functionals on Jacob’s ladder~\cite{perdew2013climbing}. However, these approaches are not universal as they
often rely on (semi)empirical parameters to arrive at the correct HOMO-LUMO gap, and
often require impractical computational expense. For these reasons, we believe this scientific
topic with wide fundamental interest and industrial applicability is an impactful use case for
quantum computing applications/algorithms which could circumvent the shortcomings of
DFT.

\subsection{Classical Computational Methodology}
All periodic DFT calculations are performed with the Perdew-Burke-Ernzerhof exchange correlation functional~\cite{PhysRevLett.77.3865} as implemented in the Quantum
Espresso code~\cite{giannozzi2009quantum}. We calculate CO adsorption on the (111) facet of three strong-adsorbing
transition metals: Pt, Pd, and Rh. All surfaces are modeled with a three-layered slab, with the
bottom layer fixed at the bulk positions and the upper two, together with the adsorbed CO
fully relaxed. A vacuum of 10 Å is applied in the direction normal to the surface to avoid
spurious interactions among vertical images. We used two sets of pseudopotentials to
implement the frozen core approximation: the projector augmented wave (PAW)
pseudopotentials and the GTH pseudopotentials. The latter are
implemented in our quantum computing algorithm, while we present the former for
comparison as they are widely used in the computational catalysis literature. The number of
valence electrons for Pt, Pd, and Rh are 16, 18, 17 for the Vanderbilt pseudopotentials and
10, 10, and 9 with the GTH pseudopotentials, respectively. We provide comprehensive
convergence studies for three different parameters: kinetic energy cutoff of the plane-wave
basis set through which the Kohn-Sham one-electron valence states are expanded, the
Monkhorst-Pack \textit{k}-point mesh density and the unit cell size for Pt(111) surface. The
converged parameters were then adopted for the respective Pd and Rh surfaces. The (111)
facet of these FCC metals exhibit four unique adsorption sites: top, bridge, hollow fcc, and
hollow hpc. Thus, we calculated adsorption on all four sites, but for brevity, results and
convergence tests were reported on only the most favorable (e.g.\ CO adsorption on fcc
hollow site of Pt(111) surface).
The binding energy (BE) of CO on a given surface is defined as:
\begin{equation}\label{eq:binding_energy_co}
\mathrm{BE} = \mathrm{E}|\mathrm{CO}^{*} - \mathrm{E|Clean}  - \mathrm{E|CO(g)}
\end{equation}
where $\mathrm{E|CO^*}$ is the total electronic energy of CO adsorbed on the surface, $\mathrm{E|Clean}$ is the total
electronic energy of the clean metal slab, and $\mathrm{E|CO(g)}$ is the total electronic energy of CO in
the gas phase. The more negative the binding energy, the more favorable adsorption is.

\subsection{Computational Assessment}
\begin{figure}[tbh]
    \centering
    \includegraphics[width=6.55cm]{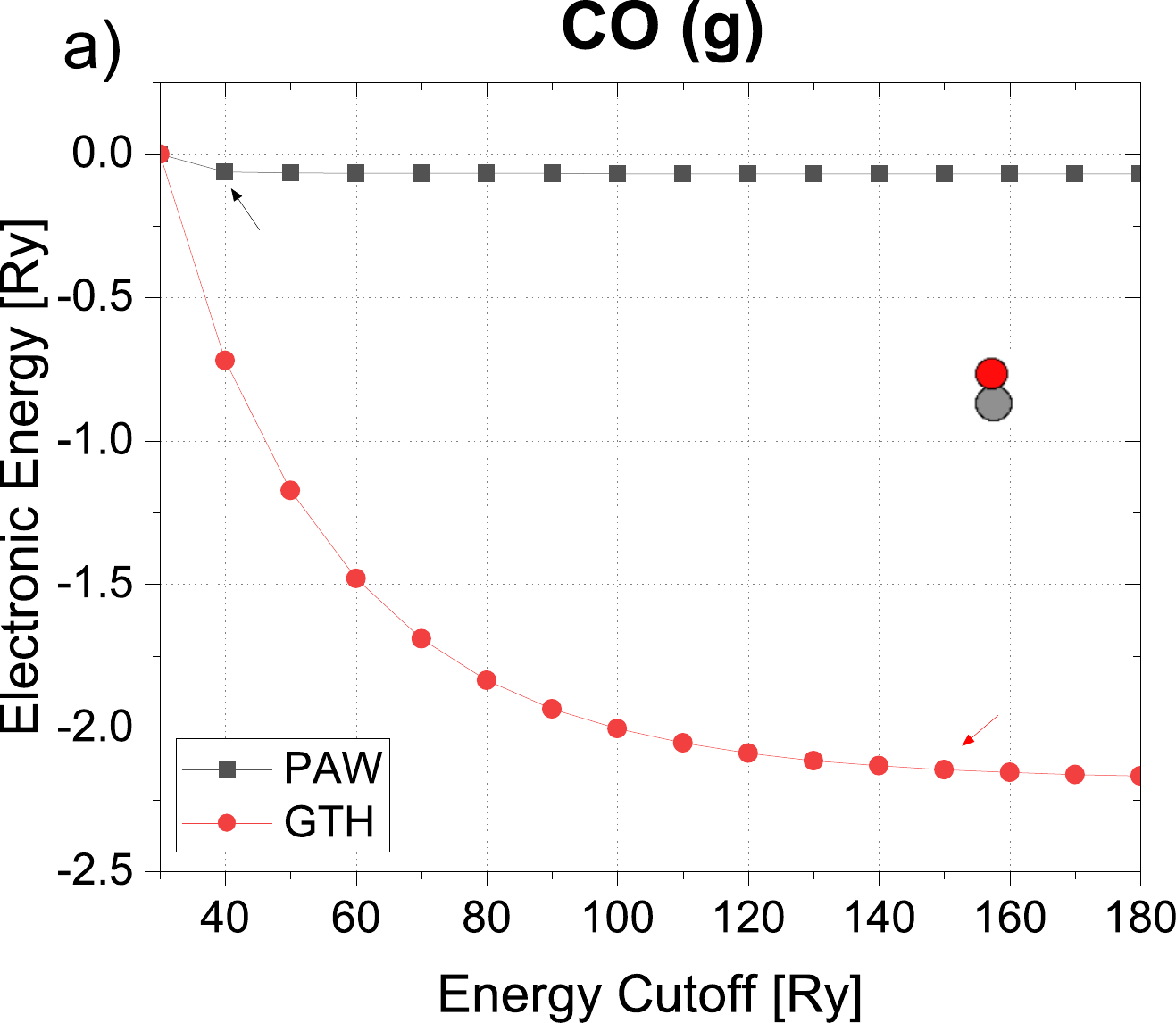}\qquad
    \includegraphics[width=6.55cm]{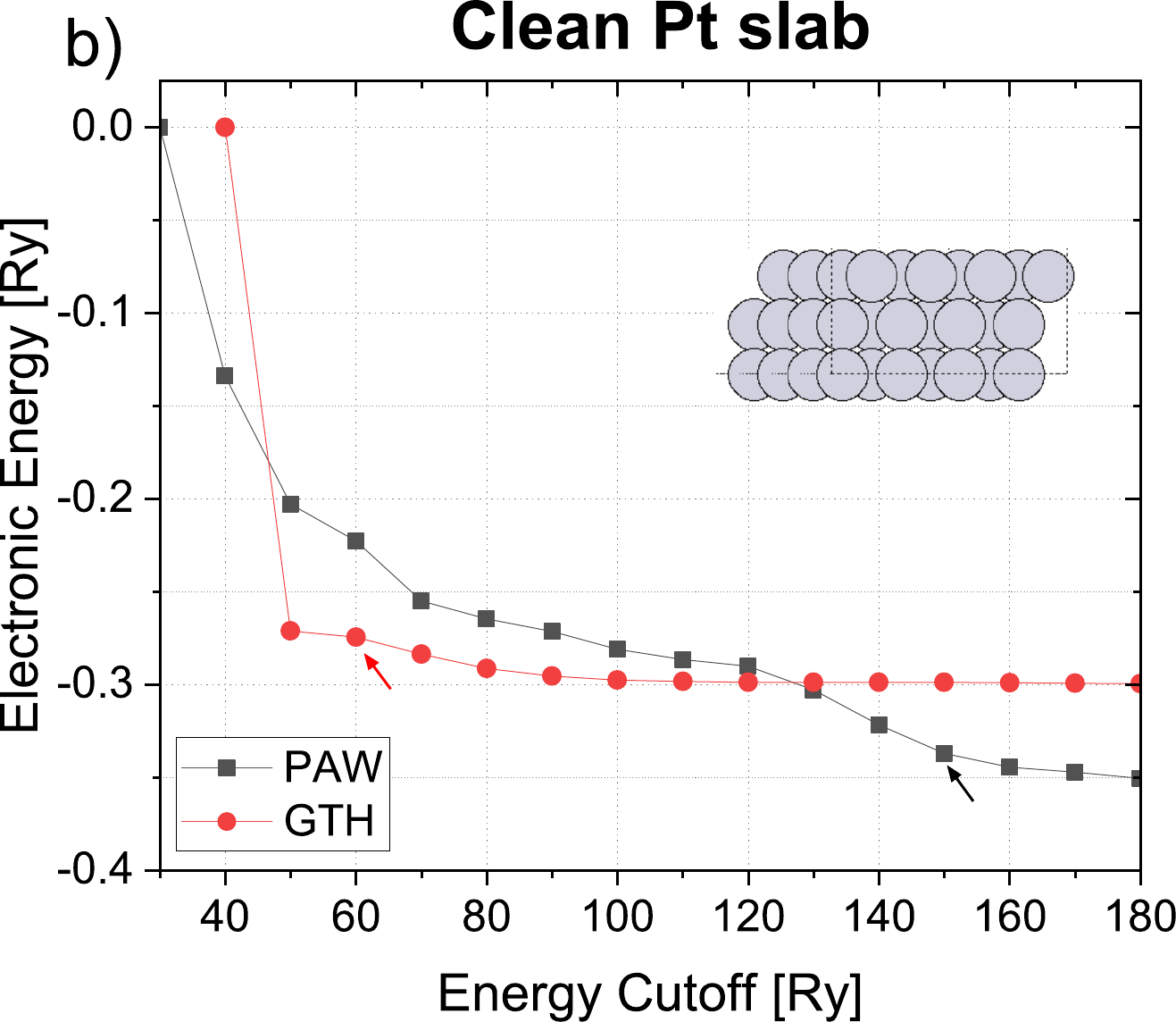}\\
    \includegraphics[width=6.55cm]{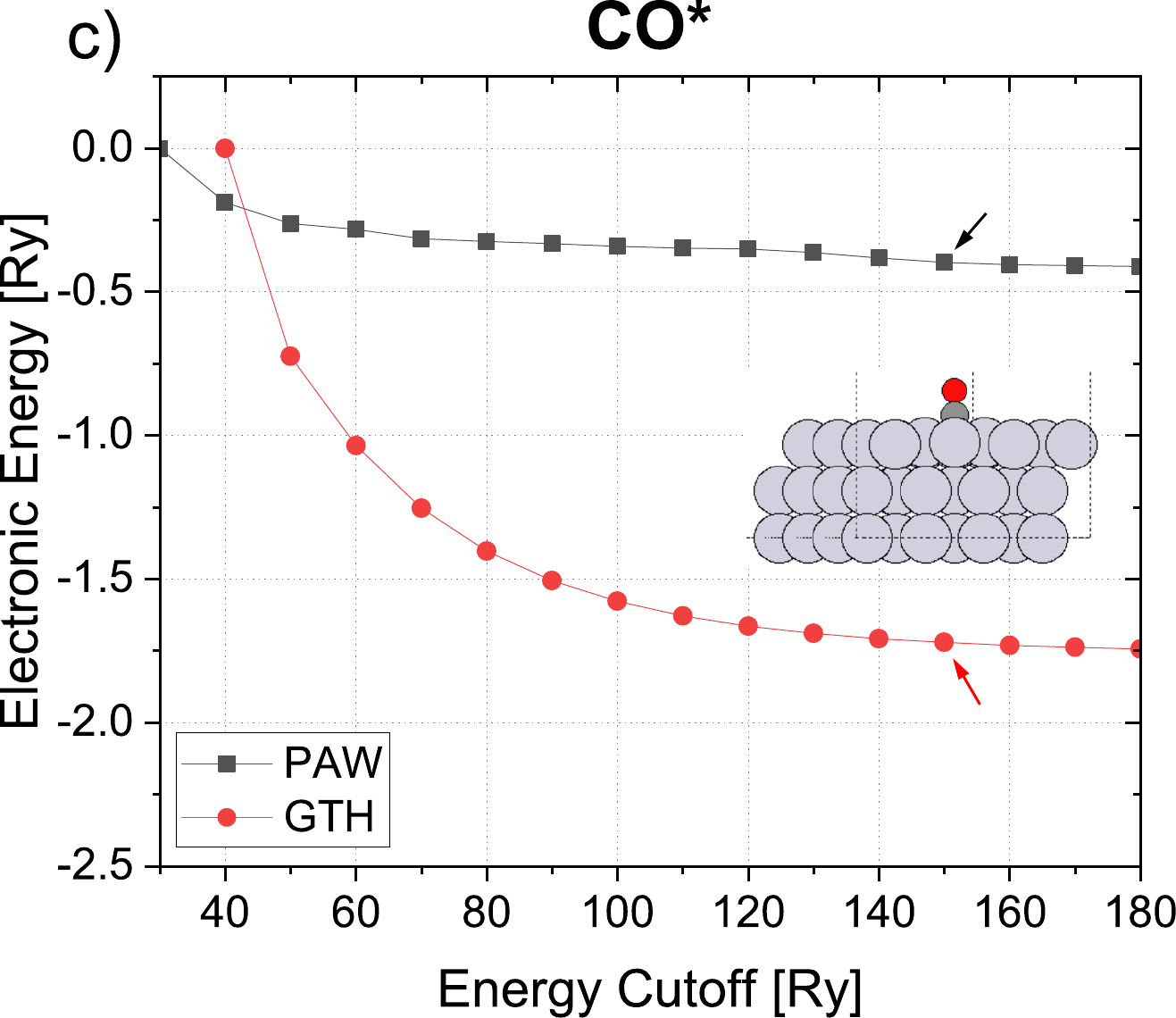}
    \caption{The convergence of electronic energies (in Ry) versus the plane-wave energy cutoff (in Ry) using the PAW (black lines) and GTH (red lines) pseudopotentials of the three components making up the binding energies: a) isolated CO molecule in vacuum, b) clean Pt slab with no adsorbates, and c) the Pt slab with a CO molecule adsorbed on its surface. Arrows point out the energy cutoff at which convergence of electronic energies has been achieved to within 0.01 Ry. In panels b) and c) the data point for GTH at 30 Ry energy cutoff is omitted because of too large a difference in the corresponding electronic energies. Notice the different scales for the vertical axis in all three panels. The data is generated from a ($4\times 4$) Pt(111) slab, with the Brillouin zone sampled at a $4\times 4\times 1$ Monkhorst-Pack \textit{k}-point mesh. The insets in each panel depict the atomistic models on which convergence was studied.}
\label{fig:covestro_pw_convergence_1}
\end{figure}

\begin{figure}
    \centering
    \includegraphics[width=8.08cm]{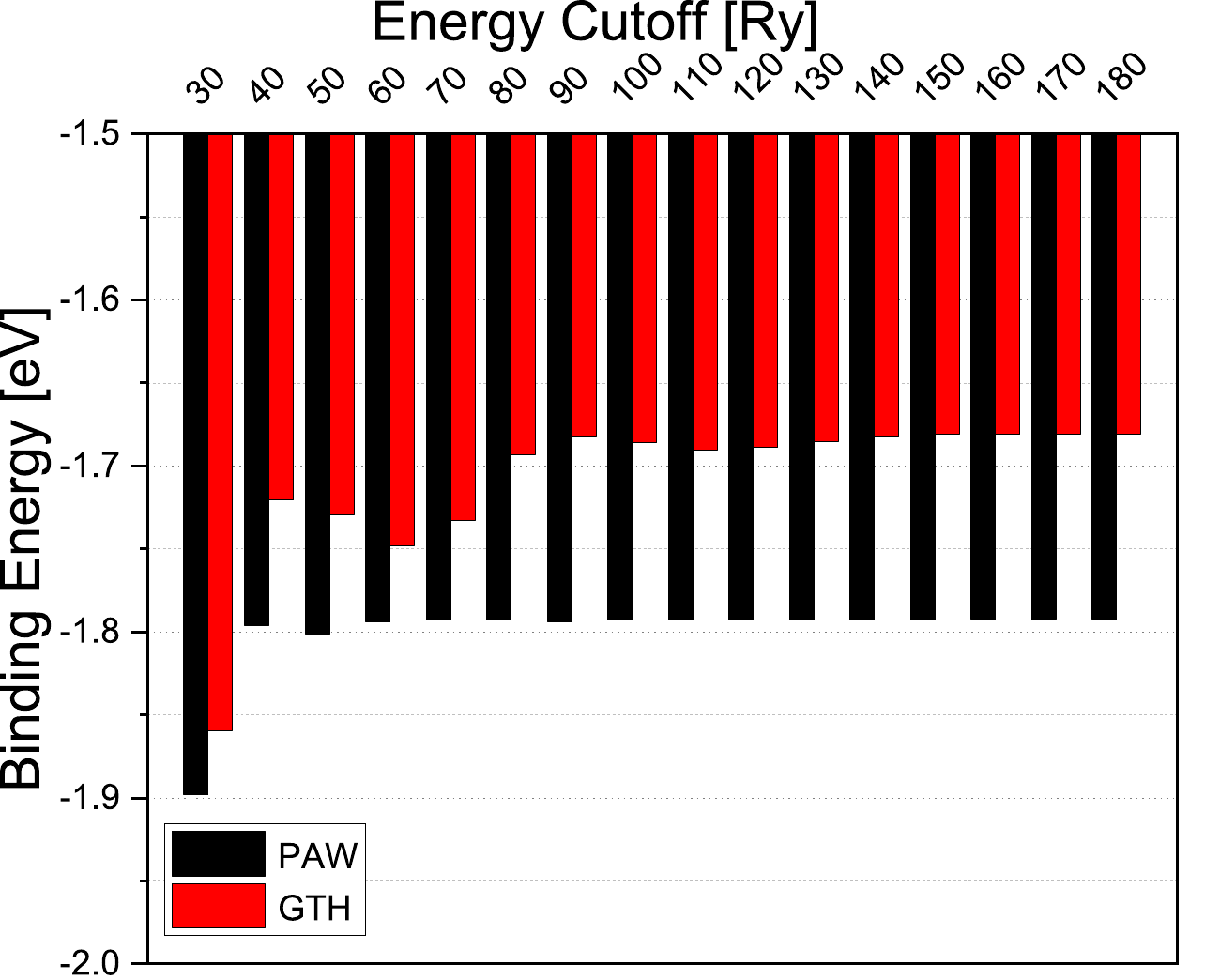}
    \caption{The convergence of the binding energy (in eV) of CO on the fcc hollow site on Pt(111) with respect to the energy cutoff (in Ry) used to calculate each of the three terms of Eq.~\eqref{eq:binding_energy_co}, for both pseudopotential: PAW (blue columns) and GTH (orange columns). We present the binding energies in electronvolts (eV) because it is a more common unit than Ry in applications of computational catalysis. For either pseudopotential, the binding energies converge to within 0.10 eV at 40 Ry energy cutoff.}
    \label{fig:cutoff_gth_vs_paw}
\end{figure}

To get reliable resource estimates for conducting quantum chemical simulations on a
quantum computer, respective DFT simulations of CO adsorption on (111) FCC metal
surfaces have to be analyzed. Since -- in this paper -- the focus lies on the usage of GTH
pseudopotentials for periodic systems, the determination of optimal simulation parameters for GTH
pseudopotentials is crucial. Therefore, in this section we are focusing on systematic
convergence studies for DFT calculation parameters for GTH pseudopotentials, and for direct
comparison PAW pseudopotentials.
One of the first parameters to analyze is the energy cutoff value, which is used to truncate the
plane waves, for both pseudopotentials studied: GTH and PAW. Thereby, we studied the
convergence for each of the three binding energy terms on Pt(111) with the convergence
threshold of 0.01 Rydberg (Ry), as shown in Figure~\ref{fig:covestro_pw_convergence_1}.
From the calculation results it emerges, that with the exception of CO(g) calculated with PAW and the clean Pt slab calculated with GTH, electronic energies generally require high energy cutoffs of 150 Ry to converge. By contrast, binding energies converge at much lower energy cutoffs, as shown in Figure~\ref{fig:covestro_pw_convergence_1}. This is intelligible due to the cancellation of errors when calculating relative energies like binding energies. Figure~\ref{fig:cutoff_gth_vs_paw} reports adsorption on the preferred hollow fcc site. We deem the binding energies converged when differences across consecutive energy cutoff values are less than 0.10 eV (1 Ry = 13.61 eV). The binding energies calculated with either PAW or GTH reach convergence at a minimum of just 40 Ry of energy cutoff.

Having established the convergence of binding energies with respect to energy cutoff, we then studied the convergence of the binding energies with respect to the density of the Monkhorst-Pack \textit{k}-point mesh and the size of the unit cell of the Pt(111) slab. Denser \textit{k}-point settings barely have any effect on the calculated binding energies. On the other hand, as we make the unit cell smaller the binding energies weaken a little: the binding energies on the ($4\times 4$), ($3\times 3$), and ($2\times 2$) unit cells are -1.79, -1.77, and -1.72 eV calculated with PAW/Vanderbilt pseudopotentials, and -1.69, -1.67, and -1.62 eV calculated with GTH, respectively. This is unsurprising, given that smaller unit cells offer higher coverages of adsorbates: the coverage of adsorbed CO on ($4\times 4$), ($3\times 3$), and ($2\times 2$) unit cells are 1/16, 1/9, and 1/4 ML, respectively. The differences are more pronounced going from the 1/9 ML coverage on the ($3\times 3$) unit cell to the 1/4 ML coverage on the ($2\times 2$) unit cell than at the higher end of unit cell size, even though all differences in binding energies are smaller than the 0.10 eV convergence threshold. The hollow site was the most stable adsorption site for both ($4\times 4$) and ($3\times 3$) unit cells as commonly predicted by DFT, but this preference switches to the bridge site on the ($2\times 2$) unit cell~\cite{PhysRevB.100.035442, ford2005atomic}. Therefore, for our further analysis, we choose the ($3\times 3$) unit cell; the smallest size that captures the DFT trend of favorable adsorption on fcc hollow site.
Using those parameters ($4\times 4\times 1$ Monkhorst-Pack \textit{k}-point mesh and ($3\times 3$) unit cell), together with 80 Ry for energy cutoff, we report the adsorption energy of CO on Pd(111) and Rh(111) in Table~\ref{tab:covestro_table}. The energy cutoff of 80 Ry is higher than suggested by the convergence tests; nevertheless we err on the side of accuracy to capture small differences in energetics of the different adsorption sites.

\begin{table}[H]
    \centering
    \begin{tabular}{|c|c|c|c|}
    \hline
    \hline
         & PAW & GTH & Experiment~\cite{PhysRevB.100.035442} \\
         \hline
       Pt  & -1.77 (fcc)  & -1.67 (fcc) & -1.37 (top)~\cite{steininger1982adsorption} \\
       Pd  &  -1.88 (fcc) & -1.83 (fcc) & -1.48 (fcc)~\cite{szanyi1993co}\\
       Rh  &  -1.91 (hcp) & -1.83 (hcp) & -1.45 (top)~\cite{wei1997desorption} \\
       \hline \hline
    \end{tabular}
    \caption{Comparison of calculated and experimental adsorption energies (in eV); adsorption sites are provided in parentheses. Experimental methods include low-energy electron diffraction (LEED) and electron energy loss spectroscopy (EELS).}
    \label{tab:covestro_table}
\end{table}
Table~\ref{tab:covestro_table} details the CO adsorption puzzle on transition metals: the CO adsorption energy is overestimated by at least 0.4 eV by DFT compared to experiments, and DFT often fails to predict the correct adsorption site (the exception here is Pd).

\subsection{Resource estimates}
We now compile a series of resource estimates using the plane-wave cutoff estimates from the aforementioned DFT calculations.  For all metal slabs we determine the maximum number of bits for each Miller index direction determined from an 80 Rydberg (Ry) spherical energy cutoff; double what is needed for energy difference convergences.  The number of grid points in each direction is determined from the maximum Miller index of the spherical cutoff. The total number of electrons is computed using the large-core LDA pseudopotential parameters listed in Table~\ref{TAB:GTH_LOCAL_LDA_PARAMETERS}. The nonlocal pseudopotential parameters used for the resource estimate are described in Table~\ref{TAB:GTH_NONLOCAL_LDA_PARAMETERS}. We also investigated Diamond in a $3\times 3 \times 3$ simulation cell and Wurzite (AlN) both with an 80 Ry spherical energy cutoff. The number of bits, total number of electrons, number of atoms of each species, and the block encoding costs are listed in Table~\ref{TAB:BE_Costs}. We find that for systems with $\eta = 100-500$ the block encoding for the entire Hamiltonian is on the order of $10^{4}$ Toffoli gates. In some systems we have also considered a doubling of the grid size in each direction (corresponding to a doubling in the spherical energy cutoff) and observe modest increases in the Toffoli complexity. 

We also compare the block encoding costs using our protocol against the resources reported in Ref.~\cite{Shokrian} for three cathode structures.
We find that our protocol provides a factor of 30-40 improvement in the Toffoli cost of block encoding while faithfully simulating the entire nonlocal pseudopotential instead of just the diagonal angular momentum projectors proposed in Ref.~\cite{Shokrian}. While 30-40 may seem modest, in terms of orders of magnitude improvement, we note that 
this is an improvement over a method that incurred an approximation error on the order of hundreds of milliHartree \textit{per atom}. 
These estimates from Ref.~\cite{Shokrian} are for a number of plane waves calculated from a spherical cutoff uniformly distributed in each Miller index direction. Due to the non-cubic nature of the simulation cells this provides a non-uniform grid spacing.
In our Toffoli costs the majority of the cost originates from the controlled swaps accessing the electronic registers in order to apply the \textsc{SELECT} operator.  The coherent swaps (of which there are two sets to move electrons $\eli$ and $\elj$ into a work register) is large in part because the number of electrons is large. Our Toffoli costs for the pseudopotential are about a factor of 100 times lower than those in Ref.~\cite{Shokrian}. It can be expected that the approach of Ref.~\cite{Shokrian} could also be extended to compute non-diagonal angular momentum projectors in the nonlocal pseudopotental, but it would require terms with $k_p\cdot k_q$ and $(k_p\cdot k_q)^2$.
That would require a large number of additional states that would need to be prepared using QROM in that approach, and could increase the cost by a further order of magnitude.  
\begin{table*}[!htpb]
    \begin{tabular}{|c|c|c|c|c|c|c|c|}
    \hline
    \hline
    system & super cell size &  $\eta$ &  atoms & $n_{x}$  &  $n_{y}$ & $n_{z}$ & C$_{\mathrm{B.E.}}$\\
    \hline
Pd & $3\times 3$ & 270 & Pd 27 & 6 & 6 & 7 & 32931 \\ 
Pd$_{\mathrm{CO}}$ & $3\times 3$ & 280 & Pd 27, C 1, O 1 & 6 & 6 & 7 & 34067\\ 
Pt & $2\times 2$ & 120 & Pt 12 & 5 & 5 & 7 & 19627 \\
Pt$_{\mathrm{CO}}$ & $2\times 2$ & 130 & Pt 12, C 1, O 1 & 5 & 5 & 7 &  20697\\
Pt & $2\times 2$ & 120 & Pt 12 & 6 & 6 & 7 & 20903\\
Pt$_{\mathrm{CO}}$ & $2\times 2$ & 130 & Pt 12, C 1, O 1 & 6 & 6 & 7 &  22053 \\
Pt & $3\times 3$ & 270 & Pt 27 & 6 & 6 & 7 &  32931 \\
Pt$_{\mathrm{CO}}$ & $3\times 3$ & 280 & Pt 27, C 1, O 1  & 6 & 6 & 7 & 34067 \\
Pt & $4\times 4$ & 480 & Pt 48 & 6 & 6 & 7 & 49731 \\
Pt$_{\mathrm{CO}}$ & $4\times 4$ & 490 & Pt 48, C 1, O 1 &  6 & 6 & 7 &  50867 \\
Rh & $3 \times 3$ & 243 & Rh 27 & 6 & 6 & 7  & 30757\\
Rh$_{\mathrm{CO}}$ & $3 \times 3$ & 253 & Rh 27, C 1, O 1& 6 & 6 & 7 & 31893\\
Li$_{0.5}$MnO$_3$ & $2\times 2\times 1$ & 408 & Li 8, Mn 16, 48 O & 6 & 7 & 5 &  45428 \\
Li$_{0.5}$MnO$_3$ & $2\times 2\times 1$ & 408 & Li 8, Mn 16, 48 O & 7 & 7 & 6 & 49868 \\
Li$_{0.75}$[Li$_{0.17}$Ni$_{0.25}$Mn$_{0.58}$]O$_2$ & $2\times 3 \times 2$ & 468 & Li 22, Mn 14, Ni 6, O 48 & 5 & 6 & 7 & 50450 \\
Li$_{0.75}$MnO$_2$F & $3\times 2\times 2$ & 428 & Li 12, Mn 16, F 16, O 32 &  6 & 6 & 6 & 42814 \\
Li$_{0.75}$MnO$_2$F & $3\times 2\times 2$ & 428 & Li 12, Mn 16, F 16, O 32 &  7 & 6 & 6 & 45221 \\
C (diamond) & $3\times 3\times 3$ & 216 & C 54 & 6 & 6 & 6 & 23576\\
AlN (wurzite) & $3\times 3\times 3$ &  432 & Al 54, N 54 &  6 & 6 & 7 &  45249 \\
LiNiO$_2$ (C2/m) & $2\times 2 \times 1$ &  92 & Li 4, Ni 4, O 8 & 5 & 5 & 5 & 18569\\
 LiNiO$_2$ (C2/m) & $2\times 2 \times 1$ &  92 & Li 4, Ni 4, O 8 & 6 & 6 & 6 & 21498\\
LiNiO$_2$ (P2$_1$/c) & $1\times 2 \times 1$ &  92 & Li 4, Ni 4, O 8 & 5 & 5 & 5 & 18419\\ 
LiNiO$_2$ (P2$_1$/c) & $1\times 2 \times 1$ &  92 & Li 4, Ni 4, O 8 & 6 & 6 & 6 & 21282 \\ 
LiNiO$_2$ (P2/c) & $1\times 1\times 1$ &  92 & Li 4, Ni 4, O 8  
 & 5  & 5 & 5 & 18419\\
LiNiO$_2$ (P2/c) & $1\times 1\times 1$ &  92 & Li 4, Ni 4, O 8  
 & 6  & 6 & 6 &  21282 \\
% CaTiO$_3$ & $3 \times 3 \times 3$  &  & \\
    \hline
    \hline
    \end{tabular}
    \caption{Block encoding costs C$_{\mathrm{B.E.}}$, electron count $\eta$,  cutoff for slabs calculated with 80 Rydberg cutoff ($n_{xyz}$), and atom counts. The number of bits for the systems registers are derived from the energy cutoffs for each cathode structure detailed in Ref.~\cite{Shokrian}.  For each system we set the bits of precision for coherent arithmetic to be 20. We use first order interpolation with 256 points. The block encoding cost for Li$_{0.5}$MnO$_3$ from Ref.~\cite{Shokrian} is 2,106,011 Toffolis which is a factor of 42 higher than our estimates while simulating the incomplete Hamiltonian. For the other two cathode structures taken from Ref.~\cite{Shokrian} our block encoding costs are 27-38 times lower.\label{TAB:BE_Costs}}
\end{table*}

An important feature of the block encoding is the value of $\lambda$ that scales the block-encoded Hamiltonian.
In many algorithms $\lambda$ plays an important role in quantifying the total costs. In phase estimation $\lambda$ scales the number of queries to the block encoding for $\epsilon$ precision. Table~\ref{TAB:lambda_qpe_Costs} tabulates each $\lambda$ for the components of the electronic structure LCU determined using Eq.~\eqref{eq:nonloclam} with $\upper_{\alpha,\nu,l,i,j}^{\rm box}$ as given in Eq.~\eqref{eq:alephbox}.  The $\lambda$ component corresponding to the normalization for the nonlocal pseudopotential term is now the dominant cost over the two-electron integral $\lambda_{V}$.  For all systems we observe both $\lambda_{\rm nonloc}$ and $\lambda_{V}$ are on the order of $10^{7}$. Combined with an $\epsilon = 1.0\times 10^{-3}$ Hartree and the $10^{4}$ block-encoding costs the $10^{15}$ Toffoli complexity is not surprising. All pseudopotentials documented in this work used the LDA large-core GTH potentials. The small core forms (more electrons) include more electrons per atom (in some cases doubling the number of electrons) which if used, would be more accurate at substantial cost.

\subsubsection{Comparing to symmetry-adapted second quantization simulation}
\begin{figure}[tbh]
    \centering
    \includegraphics[width=8.5cm]{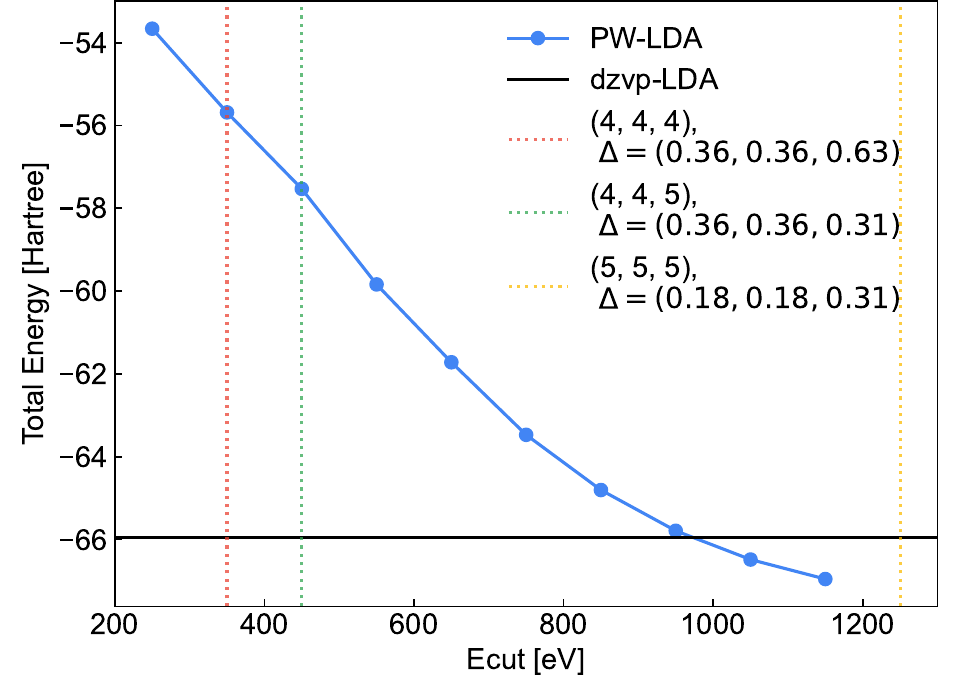}
    \caption{A comparison of the convergence of the plane-wave basis towards the MOLOPT-DZVP basis for the primitive cell of the LNO-C2m structure from Ref.~\cite{PRXQuantum.4.040303} using LDA-DFT.  $\Delta$ is the grid resolution in Bohr. The primitive cell of C2m has one formula unit. Thus in the simulation cell we use a supercell of $2\times 2\times 1$ corresponding to $4$ formula units consistent with the $\mathrm{P2_{1}c}$ structure. For the supercell we need a grid of $n=[5, 5, 5]$ qubits to achieve a similar resolution of less than $0.36$ Bohr in each direction.  Thus we use a $n=[5, 5, 5]$ for each of the LNO structures. We note that in Ref.~\cite{PRXQuantum.4.040303} the GTH-HF-REV pseudopotential was used which is a small core pseudopotential. Thus there were more electrons in that simulation (132 electrons for small-core versus 92 electrons for large-core).}
    \label{fig:C2m_convergence}
\end{figure}

A benefit of developing a simulation protocol for pseudopotentials using a first quantized plane-wave representation is that we can now directly compare to the most recent quantum algorithms for simulating materials in second quantization. In order to make this comparison, we first estimate the plane wave grid resolution needed to reproduce a localized orbital simulation accuracy. We take as an assumption that convergence to the thermodynamic limit is independent of reaching the complete basis set limit in terms of bands. Thus we can use a single primitive cell to verify the number of plane waves (and thus momentum space grid resolution) needed to reproduce the DFT energy of a localized orbital calculation. Shown in Figure~\ref{fig:C2m_convergence} is the total energy convergence of plane-wave DFT using an LDA functional to the MOLOPT-DZVP basis. The plane-wave LDA-DFT calculations are run using qcpanop~\cite{qcpanop} which relies on pseudopotentials implemented in PySCF~\cite{sun2020recent}. Dashed vertical lines in Figure~\ref{fig:C2m_convergence} are the plane-wave cutoffs where the grid resolution requires another bit. The grid resolutions ($\Delta$ values) are determined from the number of bits needed to represent the Miller index grid using Eq.~\eqref{eq:bits_to_Nx}.

\begin{figure}
    \centering
    \includegraphics[width=8.5cm]{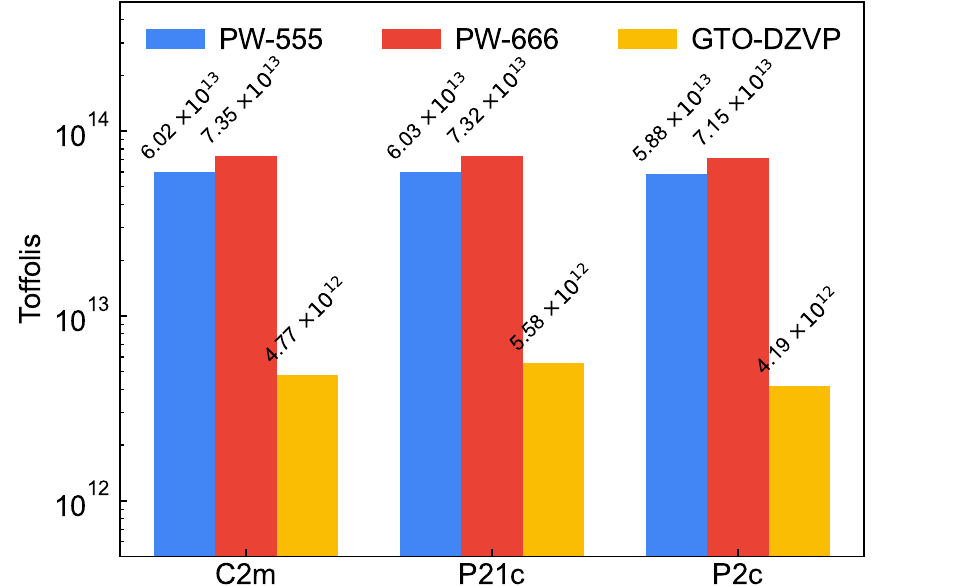}
    \caption{Toffoli resources (listed above each bar) needed to perform phase estimation to $\epsilon=1.6\times 10^{-3}$ Hartree precision using the large-core GTH-LDA pseudopotential in first and second quantization.  Each system (C2m, $\mathrm{P2_{1}c}$, P2c) contains four formula units.  The first quantized plane-wave simulations are compared against symmetry adapted second-quantized simulation using the MOLOPT-DZVP basis. The second quantized costs differ from Ref.~\cite{PRXQuantum.4.040303} due to using a larger pseudopotential (large-core GTH-LDA versus GTH-HF-REV) and not normalizing by the number of formula units (which results in $\lambda$ being divided by four).}
    \label{fig:LNO-barchart}
\end{figure}

Table~\ref{TAB:lambda_qpe_Costs_LNO} tabulates the first-quantized plane wave $\lambda$ for each component and total Toffoli complexity needed to perform phase estimation to precision $\epsilon = 1.6\times 10^{-3}$ Hartree. The LNO systems are slightly smaller than the metal surfaces and cathode structures previously considered with the smallest simulation cell involving 92 electrons.  Using the plane-wave DFT calculations we determine that for these 92 electron simulations a qubit register of $[5, 5, 5]$ for each electron in each of the Miller index directions would be required to reproduce the MOLOPT-DZVP basis quality using a large-core GTH-LDA pseudopotential. In Figure~\ref{fig:LNO-barchart} we compare the first quantized plane wave resource estimates to the second quantized symmetry-adapted localized orbital resource requirements described in Ref.~\cite{PRXQuantum.4.040303}. The second quantized simulations are about an order of magnitude lower for the same system in terms of Toffoli complexity. It is important to note though, that due to the extensive use of the QROM primitive to output the Hamiltonian the second quantized simulations of LNO require approximately seventy-five thousand logical qubits.

For the pseudopotential Toffoli costs given in Table~\ref{TAB:lambda_qpe_Costs_LNO} we have used a large number of ancilla qubits (around 8000) to minimize the Toffoli cost of uncomputing the pseudopotential.
Without using those ancilla qubits, the Toffoli cost is increased by only about 60\%, so $9.6\times 10^{13}$ Toffolis for the first line in Table~\ref{TAB:lambda_qpe_Costs_LNO}.
Then
the dominant space complexity for the first quantized simulations is the system register. For LNO the system register requires 1380 logical qubits.
Thus the total spacetime volume of the second quantized calculation is higher than the first quantized simulation.

\addtolength{\tabcolsep}{5pt}
\begin{table*}[!htpb]
    \begin{tabular}{|c|c|c|c|c|r|c|}
    \hline
    \hline
    {system} & $[n_{x},n_{y},n_{z}]$ & $\lambda_{\rm{loc}}$ &  $\lambda_{\rm{nonloc}}$ & \multicolumn{1}{c|}{$\lambda_{T}$}  &  \multicolumn{1}{c|}{$\lambda_{V}$} & QPE-Toffolis\\
    \hline
LNO-C2m  & [5,5,5] &  \num{33666.64455676} &  \num{3188341.80268461} &  \num{20230.41590738} &  \num{64010.31713000} &    \num{6.0273}$\times 10^{13}$\\ 
LNO-C2m  & [6,6,6] &  \num{33667.39896147} &  \num{3235914.70016500} &  \num{86406.35416442} &  \num{130303.2568274} &    \num{7.3580}$\times 10^{13}$\\
LNO-P21c & [5,5,5] &  \num{33722.69607491} &  \num{3219974.95007803} &  \num{17205.56062173} &  \num{65160.56366929} &    \num{6.0325}$\times 10^{13}$\\
LNO-P21c & [6,6,6] &  \num{33723.09015272} &  \num{3268231.20691254} &  \num{73486.86114439} &  \num{132593.6942023} &    \num{7.3295}$\times 10^{13}$\\
LNO-P2c  & [5,5,5] &  \num{33717.58378667} &  \num{3135639.28824505} &  \num{17656.05304654} &  \num{65094.73316146} &    \num{5.8807}$\times 10^{13}$\\
LNO-P2c  & [6,6,6] &  \num{33717.93057476} &  \num{3180792.38962238} &  \num{75410.96434546} &  \num{132463.6753791} &    \num{7.1505}$\times 10^{13}$\\
    \hline
    \hline
    \end{tabular}
    \caption{$\lambda$ calculated for select systems and the total Toffoli cost for QPE derived by multiplying the block encoding costs from Table~\ref{TAB:BE_Costs} by $\lceil \lambda \pi / (2\epsilon)\rceil$ with $\epsilon = 1.6\times 10^{-3}$ Hartree. 
    The value of $\lambda_{\rm loc}$ is calculated with Eq.~\eqref{eq:aprx2},
    $\lambda_{\rm nonloc}$ using Eq.~\eqref{eq:nonloclam},
    $\lambda_T$ using Eq.~\eqref{eq:lambdaT}, and
    $\lambda_V$ using Eq.~\eqref{eq:lambdaV}. \label{TAB:lambda_qpe_Costs_LNO}}
\end{table*}
\begin{table*}[!htpb]
    \begin{tabular}{|c|c|r|r|r|r|c|}
    \hline
    \hline
    system & \multicolumn{1}{c|}{$[n_{x},n_{y},n_{z}]$} &  \multicolumn{1}{c|}{$\lambda_{\rm{loc}}$} &  \multicolumn{1}{c|}{$\lambda_{\rm{nonloc}}$} & \multicolumn{1}{c|}{$\lambda_{T}$}  &  \multicolumn{1}{c|}{$\lambda_{V}$} & QPE-Toffolis\\
    \hline
Pd 3x3                                              & [6,6,7] &  \num{125859.76451407}  &  \num{19478925.97031171} &  \num{109324.52178850} &  \num{764307.66470876}  &   \num{6.6206}$\times 10^{14}$ \\ 
Pd$_{\mathrm{CO}}$ 3x3                              & [6,6,7] &  \num{147026.51503251}  &  \num{20720435.42687289} &  \num{113373.57815104} &  \num{822080.61086395}  &   \num{7.2920}$\times 10^{14}$ \\
Pt 2x2                                              & [5,5,7] &  \num{32657.94174834 }  &  \num{8163337.03339567 } &  \num{34182.04952875 } &  \num{124240.26674333}  &   \num{1.6097}$\times 10^{14}$ \\
Pt$_{\mathrm{CO}}$ 2x2                              & [5,5,7] &  \num{43047.30639906 }  &  \num{9125932.07458512 } &  \num{37030.55365615 } &  \num{145904.01073429}  &   \num{1.9002}$\times 10^{14}$ \\
Pt 2x2                                              & [6,6,7] &  \num{32657.94174834 }  &  \num{8165031.78348551 } &  \num{96382.38984147 } &  \num{202881.28200785}  &   \num{1.7436}$\times 10^{14}$ \\
Pt$_{\mathrm{CO}}$ 2x2                              & [6,6,7] &  \num{43047.37284347 }  &  \num{9130475.86796979 } &  \num{104414.25566159} &  \num{238257.64000502}  &   \num{2.0603}$\times 10^{14}$ \\
Pt 3x3                                              & [6,6,7] &  \num{164153.88992528}  &  \num{34165988.76537946} &  \num{115341.04245517} &  \num{786940.75265860}  &   \num{1.1390}$\times 10^{15}$ \\
Pt$_{\mathrm{CO}}$ 3x3                              & [6,6,7] &  \num{186703.56278689}  &  \num{36007219.05931250} &  \num{119612.93291647} &  \num{846424.50223998}  &   \num{1.2428}$\times 10^{15}$ \\
Pt 4x4                                              & [6,6,7] &  \num{521918.99157296}  &  \num{32364841.31270695} &  \num{143432.96342983} &  \num{2017576.3630485}  &   \num{1.7111}$\times 10^{15}$ \\
Pt$_{\mathrm{CO}}$ 4x4                              & [6,6,7] &  \num{561680.21026363}  &  \num{34094066.46384626} &  \num{146421.15016795} &  \num{2102607.3084119}  &   \num{1.8429}$\times 10^{15}$ \\
Rh 3x3                                              & [6,6,7] &  \num{103292.74918883}  &  \num{52147804.81382003} &  \num{103226.53423748} &  \num{633088.12243109}  &   \num{1.5999}$\times 10^{15}$ \\
Rh$_{\mathrm{CO}}$ 3x3                              & [6,6,7] &  \num{122454.10711166}  &  \num{54754297.45165874} &  \num{107474.53976165} &  \num{686378.36842698}  &   \num{1.7430}$\times 10^{15}$ \\
Li$_{0.5}$MnO$_3$                                   & [6,7,5] &  \num{845321.91175207}  &  \num{50684598.44613531} &  \num{91546.69807120 } &  \num{1427981.0356096}  &   \num{2.3659}$\times 10^{15}$ \\
Li$_{0.5}$MnO$_3$                                   & [7,7,6] &  \num{845322.75290329}  &  \num{50748285.06070714} &  \num{288881.56286013} &  \num{2203184.3586764}  &   \num{2.6479}$\times 10^{15}$ \\
Li$_{0.75}$[Li$_{0.17}$Ni$_{0.25}$Mn$_{0.58}$]O$_2$ & [5,6,7] &  \num{1026445.8969298}1 &  \num{94845755.77831738} &  \num{128488.88427527} &  \num{1852815.2342955}  &   \num{4.8466}$\times 10^{15}$ \\
Li$_{0.75}$MnO$_2$F                                 & [6,6,6] &  \num{962591.34009697}  &  \num{60519816.73228221} &  \num{80284.67885412 } &  \num{1531888.9435455}  &   \num{2.6520}$\times 10^{15}$ \\
Li$_{0.75}$MnO$_2$F                                 & [7,6,6] &  \num{962715.67858416}  &  \num{61226716.65323664} &  \num{125975.02393736} &  \num{1954408.1672054}  &   \num{2.8532}$\times 10^{15}$ \\
C (diamond)                                         & [6,6,6] &  \num{221922.98589511}  &  \num{7528714.20023193 } &  \num{110218.14675463} &  \num{541362.75157725}  &   \num{1.9447}$\times 10^{14}$ \\
AlN (wurzite)                                       & [6,6,7] &  \num{981514.12499348}  &  \num{52310984.70082929} &  \num{147908.92399308} &  \num{1806111.1774864}  &   \num{2.4542}$\times 10^{15}$ \\
    \hline
    \hline
    \end{tabular}
    \caption{$\lambda$ calculated for select systems and the total Toffoli cost for QPE derived by multiplying the block encoding costs from Table~\ref{TAB:BE_Costs} by $\lceil \lambda \pi / (2\epsilon)\rceil$ with $\epsilon = 1.6\times 10^{-3}$ Hartree.
    The value of $\lambda_{\rm loc}$ is calculated with Eq.~\eqref{eq:aprx2},
    $\lambda_{\rm nonloc}$ using Eq.~\eqref{eq:nonloclam},
    $\lambda_T$ using Eq.~\eqref{eq:lambdaT}, and
    $\lambda_V$ using Eq.~\eqref{eq:lambdaV}. \label{TAB:lambda_qpe_Costs}}
\end{table*}
\addtolength{\tabcolsep}{-5pt}

\section{Conclusions}
\label{sec:conclude}
Quantum simulations of electronic structure using plane waves can have high cost due to the large number of plane waves needed to resolve the core orbitals, as well as the total number of electrons. Pseudopotentials address both these issues by using a smoothly-varying effective potential to model the behavior of the valence electrons. The difficulty is now that the description of these effective potentials requires more classical data to be input to the quantum algorithm, as compared to non-pseudized potential interactions which can be simply calculated on the fly as in prior plane wave approaches. As seen in second quantized simulations, inputting classical data into quantum algorithms can be a major contributor to the cost and should be minimized.

Prior work in Ref.~\cite{Shokrian} used a very data-intensive approach to implement the pseudopotentials where the entire functional variation needs to be input as data.
That negates the advantage of the functional form for pseudopotentials by linking the cost of block encoding to something that is scaling with the total basis set size.
Moreover, they used an incomplete form of the pseudopotentials, corresponding to errors on the order of thousands of times the desired accuracy of the calculation. We demonstrate the magnitude of the error in Section~\ref{sec:omiterr} and further emphasize that this error will grow as a function of system size or more sophisticated GTH-type pseudopotentials are used. While it is possible to extend the QROM data-intensive approach to include inner products of basis states and higher powers of inner products of basis states, that classical data would incur a substantial quantum resource overhead.

In contrast, we have fully taken advantage of the functional form of the pseudopotentials to provide a simulation method that minimizes the use of the classical data. As well as providing the complete pesudopotential for accurate simulations, the Toffoli count for a single block encoding of the Hamiltonian is far lower than that in \cite{Shokrian} and recovers the expected $\widetilde{\mathcal{O}}(\eta)$ scaling of block encoding in first quantization. We also provide a set of routines for handling non-cubic simulation cells that maintains uniform grid resolutions.  This involves modifying basis state inner products to include reciprocal cell geometry via the coefficients of the reciprocal cell Gramian. Allowing different bits in each Miller index direction necessitates revisiting some of the block encoding primitives such as the $\nu$ state preparation.

The improvement in performance by using the pseudopotentials in the quantum algorithm reflects the improvement obtained by using pseudopotentials in classical algorithms.
The accuracy is improved for a given basis size, enabling a smaller basis to be used with lower-energy plane waves.
In contrast to classical algorithms, the improvement in the quantum algorithm comes about because the $\lambda$-value is reduced.
In the classical literature, the improvements provided by pseudopotentials are not described in a complexity-theoretic way, so our improvement is not given in big O notation.
However, our improvement over the pseudopotential algorithm of Ref.~\cite{Shokrian} can be so described.
That algorithm is polynomial in the basis size $N$, whereas ours is logarithmic.
More specifically it scales as $\mathcal{O}(\log^2 N)$ due to the need to perform multiplications to calculate norms.

A drawback to our approach is that we have simplified the implementation of the nonlocal pseudopotential in a way that results in the $\lambda$-value being increased. The overall simulation complexity scales as the product of $\lambda$ and the block-encoding complexity. The $\lambda$-value is increased modestly by breaking the pseudopotential into the sum to reduce the arithmetic needed. A more serious increase in the value of $\lambda$ comes from the state preparation over $\nu$ for the nonlocal pseudopotential. In contrast, Ref.~\cite{Shokrian} used a representation that reduces $\lambda$. That being said, the large $\lambda$ values are inherent to the plane-wave basis we have selected.

Using the new algorithmic primitives for handling non-cubic unit cells and pseudopotentials we assess the fault-tolerant resources needed to simulate a much broader class of materials simulation problems than prior work that was isolated to cubic unit cells. To provide canonical costs, we estimate the quantum resources needed to resolve the CO-puzzle in heterogenous catalysis. While the costs are substantial it is important to note that these systems would have resulted in unrealistically high Toffoli complexity in a similar all-electron calculation. Another large benefit of developing the algorithmic primitives needed to accurately simulate in a non-cubic unit cell and with pseudopotentials is that we can now, for the first time, directly compare to second quantized simulation of materials recently introduced in Ref.~\cite{PRXQuantum.4.040303}. In this work we have compared the flagship Lithium-Nickel-Oxide battery cathode simulation problem outlined in~\cite{PRXQuantum.4.040303} and estimated that the space-time volume of our new first quantized simulation approach provides substantial value.

Another benefit of the full compilation of non-local pseudopotentails is that it does provide a direction for further improvements. For example, when block encoding the nonlocal pseudopotential, a strategy to reduce $\lambda$ would be to use a state preparation step with more detailed dependence on $\nu$ rather than just the nested boxes we have used here. Alternatively it may be possible to provide a different form of block encoding to reduce $\lambda$ in a similar way as in Ref.~\cite{Shokrian}, though that seems to make it difficult to take advantage of the functional form to reduce the complexity.

\subsection*{Acknowledgements}

The authors thank Yuan Su, Nathan Wiebe, Sam Pallister, and Burak Şahinoğlu for helpful discussions and analysis. DWB worked on this project under a sponsored research agreement with Google Quantum AI. DWB is also supported by Australian Research Council Discovery Projects DP190102633, DP210101367 and DP220101602.

%%%%%%%%%%%%%%%%%%%%%%%%%%%%%%%%%%%%%%%%%%%%%%%%%%%%%%%%%%%%%%%%%%%%%%%%%%%%%%

\bibliographystyle{apsrev4-2}

\appendix
\section{Large-core LDA Pseudopotential Parameters}\label{app:lda_pp_params}
Here we document the large-core LDA pseudopotential parameters used in this work and in Ref.~\cite{Shokrian}.
\begin{table}[H]
\centering
    \begin{tabular}{|l|r|r|rrrr|}
    \hline
    \hline
    atom & $Z_\alpha$ & $r_{\rm loc}^\alpha$ & $C_1^\alpha$ & $C_2^\alpha$ & $C_3^\alpha$ & $C_4^\alpha$ \\
    \hline
    %LDA:
    % in Pyscf: pseudo = GTH-LDA-q1
    Li & 1 & 0.78755305 & -1.89261247 & 0.28605968 & 0.00000000 & 0.00000000 \\
    %B  & 3 & 0.43392956 & -5.57864173 & 0.80425145 & 0.00000000 & 0.00000000 \\
    C  & 4 & 0.34883045 & -8.51377110 & 1.22843203 & 0.00000000 & 0.00000000 \\
    N  & 5 & 0.28917923 & -12.23481988 & 1.76640728 & 0.00000000 & 0.00000000 \\
    O  & 6 & 0.24762086 & -16.58031797 & 2.39570092 & 0.00000000 & 0.00000000 \\
    F  & 7 & 0.21852465 & -21.30736112 & 3.07286942 & 0.00000000 & 0.00000000 \\
    Al & 3 & 0.45000000 & -8.49135116 & 0.00000000 & 0.00000000 & 0.00000000 \\
    %Si & 4 & 0.44000000 & -7.33610297 & 0.00000000 & 0.00000000 & 0.00000000 \\
    %Cl & 7 & 0.41000000 & -6.86475431 & 0.00000000 & 0.00000000 & 0.00000000 \\
    Mn  & 7 & 0.64000000 & 0.00000000 & 0.00000000 & 0.00000000 & 0.00000000 \\
    Ni & 10 & 0.56000000 & 0.00000000 & 0.00000000 & 0.00000000 & 0.00000000 \\
    % pseudo = 'gth-lda-q10'
    Pt & 10 & 0.61600000 & 11.02741707 & 0.00000000 & 0.00000000 & 0.00000000 \\
    % pseudo = 'gth-lda-q10'
    Pd & 10 & 0.59600000 & 5.20966476 & 0.00000000 & 0.00000000 & 0.00000000 \\
    % pseudo = 'gth-lda-q9'
    Rh & 9 & 0.62142857 & 5.39796233 & 0.00000000 & 0.00000000 & 0.00000000 \\
    % pseudo = 'gth-lda-q2'
    Ca & 2 & 0.80000000 & 0.00000000 & 0.00000000 & 0.00000000 & 0.00000000 \\
    % pseudo = 'gth-lda-q4'
    Ti & 4 & 0.72000000 & 0.00000000 & 0.00000000 & 0.00000000 & 0.00000000 \\
    \hline
    \hline
        \end{tabular}
    \caption{GTH local pseudopotential parameters for LDA, given in atomic units.  The Li, Mn, Ni, and O correspond to LDA optimized parameters with `large-core' potentials. Parameters for `small-core' and other functionals can be found in Ref.~\cite{kuhne2020cp2k}. }
    \label{TAB:GTH_LOCAL_LDA_PARAMETERS}
\end{table}

\begin{table}[H]
\centering
    
    \begin{tabular}{|c|c|c|ccc|}
    \hline\hline
        &     &          & & {$c_{x,li}$} & \\
        \cline{4-6}
        &     &          & & {$x$} & \\
        \cline{4-6}
    $~~l~~$ & $~~i~~$ & {$C_{li}^\alpha$} & 0 & 1 & 2 \\
    \hline
    0   &  1  & $4\sqrt{2}                          \pi^{5/4}(r_l^\alpha)^{ 3/2}$   &   1 &     &   \\
    0   &  2  & $8\sqrt{\frac{2}{15}}               \pi^{5/4}(r_l^\alpha)^{ 3/2}$   &   3 &  -1 &   \\
    0   &  3  & $\frac{16}{3}\sqrt{\frac{2}{105}}   \pi^{5/4}(r_l^\alpha)^{ 3/2}$   &  15 & -10 & 1 \\
    1   &  1  & $8\sqrt{\frac{1}{3}}                \pi^{5/4}(r_l^\alpha)^{ 5/2}$   &   1 &     &   \\
    1   &  2  & $16\sqrt{\frac{1}{105}}             \pi^{5/4}(r_l^\alpha)^{ 5/2}$   &   5 &  -1 &   \\
    1   &  3  & $\frac{32}{3}\sqrt{\frac{1}{1155}}  \pi^{5/4}(r_l^\alpha)^{ 5/2}$   &  35 & -14 & 1 \\
    2   &  1  & $8\sqrt{\frac{2}{15}}               \pi^{5/4}(r_l^\alpha)^{ 7/2}$   &   1 &     &   \\
    2   &  2  & $\frac{16}{3}\sqrt{\frac{2}{105}}   \pi^{5/4}(r_l^\alpha)^{ 7/2}$   &   7 &  -1 &   \\
    2   &  3  & $\frac{32}{3}\sqrt{\frac{2}{15015}} \pi^{5/4}(r_l^\alpha)^{ 7/2}$   &  63 & -18 & 1 \\
\hline\hline
\end{tabular}
    \caption{GTH nonlocal pseudopotential projector parameters.
    \label{TAB:PROJECTORS}
    }
\end{table}

\clearpage
\begin{table}[H]
\centering
    \setlength{\tabcolsep}{0.5em}
    \begin{tabular*}{0.6\textwidth}{@{\extracolsep{\fill}}|c|c|c|ccc|}
    \hline
    atom & $l$ & $r_l^\alpha$ &  & {$E^{ij}_{l\alpha}$} & \\
    \hline
    %LDA:
    Li        & 0 & 0.66637518 & 1.85881111 & & \\
              & 1 & 1.07930561 & -0.00589504 & & \\  \hline
    %B    & 0 & 0.37384326 & 6.23392822 \\
    C    & 0 & 0.30455321 & 9.52284179 & & \\  \hline
    N    & 0 & 0.25660487 & 13.55224272 & & \\  \hline
    O    & 0 & 0.22178614 & 18.26691718 & & \\  \hline
    F    & 0 & 0.19556721 & 23.58494211 & & \\  \hline
    Al   & 0 & 0.46010427 & 5.08833953 & -1.03784325 & \\
         &   &            & -1.03784325 & 2.67969975 & \\
         & 1 & 0.53674439 & 2.19343827 & & \\  \hline
    %Si   & 0 & 0.42273813 & 5.90692831 & -1.26189397 \\
    %     &   &            & -1.26189397 & 3.25819622 \\
    %     & 1 & 0.48427842 & 2.72701346 \\
    %Cl   & 0 & 0.33820832 & 9.06223968 & -1.96193036 \\
    %     &   &            & -1.96193036 & 5.06568240 \\
    %     & 1 & 0.37613709 & 4.46587640 \\
    Mn      & 0 & 0.48124608 & 2.79903057 & -0.96286281 & 0.62594958 \\
            &   &            & -0.96286281 & 2.48610107 & -1.61619487 \\
            &   &            & 0.62594958 & -1.61619487 & 2.56562982 \\
            & 1 & 0.66930432 & 1.36877564 & -0.13385695 & \\
            &   &            & -0.13385695 & 0.31676337 & \\
            & 2 & 0.32776314 & -7.99541784 & & \\  \hline
    Ni      & 0 & 0.42539870 & 3.61965071 & -1.19635099 & 0.74622158 \\
            &   &            & -1.19635099 & 3.08896496 & -1.92673583 \\
            &   &            & 0.74622158 & -1.92673583 & 3.05859831 \\
            & 1 & 0.58408076 & 1.74222007 & -0.16325873 & \\
            &   &            & -0.16325873 & 0.38634067 & \\
            & 2 & 0.27811348 & -11.60842823 & & \\  \hline
    Pt      & 0 & 0.52013211 & 2.44743006 & -1.02260695 & \\
            &   &            & -1.02260695 & 2.64035978 & \\
            & 1 & 0.65897566 & 0.40845297 & -0.69628711 & \\
            &   &            & -0.69628711 & 1.64771604 & \\
            & 2 & 0.45124318 & -4.55229454 & 0.92706936 & \\
            &   &            & 0.92706936 & -2.10239568 & \\  \hline
    Pd      & 0 & 0.58220422 & 2.41107608 & -0.89811395 & \\
            &   &            & -0.89811395 & 2.31892024 & \\
            & 1 & 0.68878713 & 1.22725330 & -0.32032214 & \\
            &   &            & -0.32032214 & 0.75802053 & \\
            & 2 & 0.44283523 & -4.37713124 & -0.18223540 & \\
            &   &            & -0.18223540 & 0.41327105 & \\  \hline
    Rh      & 0 & 0.59807881 & 2.24211131 & -0.83330981 & \\
            &   &            & -0.83330981 & 2.15159669 & \\
            & 1 & 0.70958567 & 1.15527830 & -0.29785195 & \\
            &   &            & -0.29785195 & 0.70484635 & \\
            & 2 & 0.36920687 & -1.05305819 & 4.81701344 & \\
            &   &            & 4.81701344 & -10.92395969 & \\  \hline
    Ca      & 0 & 0.66973721 & 1.64501442 & -0.59004546 & 0.07221571 \\
            &   &            & -0.59004546 & 1.52349082 & -0.18646017 \\
            &   &            & 0.07221571 & -0.18646017 & 0.29599634 \\
            & 1 & 0.94647402 & 0.58547893 & -0.05338365 & \\
            &   &            & -0.05338365 & 0.12632878 & \\
            & 2 & 0.52654998 & -3.03232095 & & \\  \hline
    Ti      & 0 & 0.52841076 & 1.86661330 & -0.55779994 & 0.89250253 \\
            &   &            & -0.55779994 & 1.44023325 & -2.30443161 \\
            &   &            & 0.89250253 & -2.30443161 & 3.65817177 \\
            & 1 & 0.79114554 & 0.96791577 & -0.11016023 & \\
            &   &            & -0.11016023 & 0.26068669 & \\
            & 2 & 0.40871178 & -4.82645635 & & \\
    \hline
    \hline
        \end{tabular*}
    \caption{GTH nonlocal pseudopotential parameters for LDA, given in atomic units.
    }
    \label{TAB:GTH_NONLOCAL_LDA_PARAMETERS}
\end{table}

\section{Preparing via nested boxes}
\label{app:nested}
The procedure for preparing states using nested boxes was previously explained in Ref.~\cite{SuPRXQuantum21}.
Here, we give a more detailed explanation to also take into account the preparation of states with more general amplitudes (as compared to $1/\|\nu\|$ in Ref.~\cite{SuPRXQuantum21}).
The procedure for the nested boxes may be described in detail as follows.
In the most general case, say we are aiming to produce a state of the form
\begin{equation}
    \sum_{\nu_x=-(2^{n_x}-1)}^{2^{n_x}-1} \sum_{\nu_y=-(2^{n_y}-1)}^{2^{n_y}-1}
    \sum_{\nu_z=-(2^{n_z}-1)}^{2^{n_z}-1} \sqrt{u(\nu)} \ket{\nu_x} \ket{\nu_y} \ket{\nu_z},
\end{equation}
where we are allowing for different dimensions in each direction.
For our application here, $u(\nu)=1/\|k_\nu\|$ for $V$, and there we use amplitude amplification in order to prepare this state with high probability for success.
We also treat the pseudopotential using the nested boxes, but do not use amplitude amplification.

The quantities $n_x,n_y,n_z$ are the numbers of bits excluding the sign bits for $\nu_x,\nu_y,\nu_z$, respectively.
We are allowing sums up to $2^{n_x}-1$ (and similarly for $y$ and $z$) rather than $2^{n_x}-2$, which is the strict bound of the region $G_d$ for $\nu$.
That is for simplicity of the explanation, and it is also possible to prepare over just the region $G_d$.
The implementation with this form is correct, because it just contributes to parts of the Hamiltonian where $q-\nu\notin G$, which are eliminated.
The more strict preparation over $G_d$ would just give a (very small) improvement in the $\lambda$-value for the block-encoding.

First, we prepare a state of the form
\begin{equation}
    \sum_{\mu=2}^{\mu_{\max}} \psi_\mu \ket{\mu},
\end{equation}
where the weighting $\psi_\mu$ is $\sqrt{2^\mu}$ when we are preparing a state with amplitudes $1/\|\nu\|$, but has a more complicated dependence for $1/\|k_\nu\|$ with non-cubic unit cells.
The principle is that $\mu$ indexes the boxes so that the size (width) of the boxes goes up as $2^\mu$ corresponding to larger values of $\|\nu\|$.
The initial amplitude increases with $\mu$, rather than decreasing with $\mu$ as might be expected (since larger $\mu$ corresponds to smaller $1/\|\nu\|$).
The reason for the initial amplitude in the superposition is that it needs to account for \emph{all} amplitudes in the box, rather than just one.
As the size of the box is increased the number of values of $\nu$ in the box scales as $2^{3\mu}$, which gives an extra factor of $2^{3\mu/2}$ for each value of $\mu$, meaning that the amplitude needed in the state preparation goes like $\psi_\mu\propto\sqrt{2^\mu}$.

In this work we have $\|k_\nu\|$ not proportional to $\|\nu\|$, as well as different numbers of bits in each direction, requiring a more general treatment. 
The form of the boxes we will take here is
\begin{align}
    B_\mu &=\{\nu\ |\ (|\nu_x|<2^{\mu_x-1})\land(|\nu_y|<2^{\mu_y-1})\land(|\nu_z|<2^{\mu_z-1})\land (|\nu_x|<2^{n_x})\land(|\nu_y|<2^{n_y})\land(|\nu_z|<2^{n_z})\} \, ,
\end{align}
where we have shifted $\mu$ according to $\mu_x=\mu-\delta_x$, $\mu_y=\mu-\delta_y$, $\mu_z=\mu-\delta_z$.
This general form of the boxes is convenient to take account of non-orthogonal Bravais vectors and unequal $n_x,n_y,n_z$.
For this set of boxes we do not eliminate inner boxes, so they are overlapping.
This approach boosts the probability of success, as well as simplifying the preparation.

The key part of this definition is the first three inequalities.
These restrict the dimension of the box in the three directions.
The other inequalities (four to six) are just the conditions that are required for all values of $\nu$.
We include these because the generality of the definition allows for cases where (for example) $\mu_x-1>n_x$.
Without loss of generality, we can take $\mu$ to be the maximum of $\mu_x,\mu_y,\mu_z$, which corresponds to taking $\delta_x,\delta_y,\delta_z\ge 0$, and the minimum of them zero.

Now consider the example where $\mu_x$ is larger than $\mu_y$ and $\mu_z$, so $\mu=\mu_x$.
(Other cases with one largest value of $\mu_{x,y,z}$ are equivalent from symmetry.)
This inequality relation (between $\mu_x$ and $\mu_y,\mu_z$) would then hold for all boxes $B_\mu$, because the relative values of $\mu_{x,y,z}$ are given by the offsets $\delta_{x,y,z}$.
Then, for $\mu=\mu_x=2$ we need $|\nu_x|<2$, but $\nu_y=\nu_z=0$ from the first three inequalities.
Another situation is where there are two equal largest values of $\mu_{x,y,z}$, for example $\mu_x=\mu_y>\mu_z$.
Then the box for $\mu=2$ will require $|\nu_x|<2$ and $|\nu_y|<2$, but $\nu_z=0$.
In the case with equal $\mu_x=\mu_y=\mu_z$ the cube with $\mu=2$ is that where $|\nu_{x,y,z}|<2$.

In each case, the box with $\mu=1$ includes only the point $\nu=0$.
That point is excluded for $V$, as well as the local pseudopotential.
For those cases $\nu=0$ would give a term proportional to the identity.
That means the minimum value of $\mu$ would be 2.
In the case of the nonlocal pseudopotential we need $\nu=0$, so can take $\mu=1$ to be the smallest allowed value.

For the maximum of $\mu$, note that if $\mu_x-1> n_x$, $\mu_y-1> n_y$, and $\mu_z-1> n_z$, the size of the box is being constrained by inequalities four to six.
Moreover, we could decrease $\mu$ by 1, and the box would be unchanged.
Because we are avoiding identical boxes, we therefore take the maximum $\mu$ to be that where at least one of $\mu_x> n_x+1$, $\mu_y> n_y+1$, $\mu_z> n_z+1$ is violated.
This corresponds to the maximum value of $\mu$ being
\begin{equation}\label{eq:maxmu}
    \mu_{\max}=\max(n_x+\delta_x,n_y+\delta_y,n_z+\delta_z)+1.
\end{equation}

After preparing a state with a superposition over $\mu$, we would prepare a superposition of $\nu$ values within $B_\mu$, giving
\begin{equation}\label{eq:genprep}
	\sum_{\mu=1}^{\mu_{\max}} \psi_\mu \frac{1}{\sqrt{|B_\mu|}}\sum_{\nu\in B_\mu}\ket{\mu}\ket{\nu_x}\ket{\nu_y}\ket{\nu_z} \, .
\end{equation}
After this, we may use the QROM interpolation to output the value of $u(\nu)$ in a register and perform an inequality test with a register in superposition in order to obtain the desired amplitude.

For our definition, the sizes of the boxes $|B_\mu|$ are given as
\begin{equation}
    |B_\mu| = \left( 2 \min(\lceil 2^{\mu_x-1}\rceil,2^{n_x})-1\right)\left( 2 \min(\lceil 2^{\mu_y-1}\rceil,2^{n_y})-1\right)\left( 2 \min(\lceil2^{\mu_z-1}\rceil,2^{n_z})-1\right)\, .
\end{equation}
The ceiling is to take account of cases where, for example, $2^{\mu_x-1}<1$, which would mean the only acceptable value of $\nu_x$ is 0.
In the state preparation, we would actually generate positive and negative zeros in the signed integers.
The negative zeros need to be eliminated, but the net effect is that we have a number
\begin{equation}
    |B'_\mu| = 8 \min(\lceil 2^{\mu_x-1}\rceil,2^{n_x})  \min(\lceil 2^{\mu_y-1}\rceil,2^{n_y})  \min(\lceil 2^{\mu_z-1}\rceil,2^{n_z}) \, .
\end{equation}
This effective set $B'_\mu$ with negative zeros is used to account for the negative zeros generated in the state preparation.

We need to account for the fact that each $\nu$ will not just be an element of one box, but if it is an element of one box $B_\mu$ for a minimum value of $\mu$, then it will also be an element of boxes for $B_{\mu'}$ for $\mu'\ge \mu$.
To account for this, let us write $\mu(\nu)$ as the minimum $\mu$ such that $\nu\in B_\mu$.
Then the state from \eq{genprep} can be rewritten as
\begin{equation}
    \sum_\nu \sum_{\mu'=\mu(\nu)}^{\mu_{\max}} \psi_{\mu'} \frac{1}{\sqrt{|B'_{\mu'}|}}\ket{\mu'}\ket{\nu_x}\ket{\nu_y}\ket{\nu_z},
\end{equation}
Note that we are using $B'_\mu$ to account for the production of negative zeros.
So that the correct amplitude can be obtained by inequality testing, the obvious requirement would be (we require $u(\nu)\ge 0$)
\begin{equation}
    \sum_{\mu'=\mu}^{\mu_{\max}}\frac{\psi_{\mu'}^2}{|B'_{\mu'}|} \propto \max_{\nu\in B_{\mu}\backslash B_{\mu-1}} u(\nu) \, .
\end{equation}
This definition could cause problems if $|u(\nu)|$ can be larger for larger boxes, so we modify it slightly to
\begin{equation}\label{eq:psipdef}
    \sum_{\mu'=\mu}^{\mu_{\max}}\frac{\psi_{\mu'}^2}{|B'_{\mu'}|} \propto \max_{\nu\in B_{\mu_{\max}}\backslash B_{\mu-1}} u(\nu) \, .
\end{equation}
With this definition, the total amplitude for $\nu$ prior to inequality testing is proportional to $\max_{\nu'\in B_{\mu_{\max}}\backslash B_{\mu(\nu)-1}} |u(\nu')|$.

To be more specific about how we compute $\psi_\mu$, let $\tilde \psi_\mu$ be an unnormalised form satisfying
\begin{equation}
    \sum_{\mu'=\mu}^{\mu_{\max}} \frac{\tilde\psi_{\mu'}^2}{|B'_{\mu'}|} = \max_{\nu\in B_{\mu_{\max}}\backslash B_{\mu-1}} u(\nu) \, ,
\end{equation}
so that
\begin{equation}\label{eq:psinorm}
    \psi_\mu = \left[\sum_{\mu'=1}^{\mu_{\max}} \tilde\psi_{\mu'}^2 \right]^{-1/2} \tilde\psi_\mu \, .
\end{equation}
Here we have taken the sum over $\mu$ from 1, but for cases where the minimum $\mu$ is 2 the value of $\tilde\psi_{1}$ can be taken to be zero.
To obtain $\tilde\psi_\mu$, we can use the formula
\begin{equation}
    \tilde\psi_{\mu_{\max}} = \sqrt{|B'_{\mu_{\max}}|} \max_{\nu\in B_{\mu_{\max}}\backslash B_{\mu_{\max}-1}} \sqrt{u(\nu)} \, ,
\end{equation}
and for $\mu< \mu_{\max}$,
\begin{equation}
     \tilde\psi_{\mu}^2 = |B'_{\mu}| \left\{ \max_{\nu\in B_{\mu_{\max}}\backslash B_{\mu-1}} u(\nu) - \max_{\nu\in B_{\mu_{\max}}\backslash B_{\mu}} u(\nu) \right\} .
\end{equation}
Because we are taking the maximum over $B_{\mu_{\max}}\backslash B_{\mu-1}$, this expression cannot give negative values.
(That is the problem that can be encountered if we were to use $B_\mu\backslash B_{\mu-1}$.)
To check this formula, it gives
\begin{align}
    \sum_{\mu'=\mu}^{\mu_{\max}} \frac{\tilde\psi_{\mu'}^2}{|B'_{\mu'}|} &= \sum_{\mu'=\mu}^{\mu_{\max}-1} \left\{ \max_{\nu\in B_{\mu_{\max}}\backslash B_{\mu'-1}} u(\nu) - \max_{\nu\in B_{\mu_{\max}}\backslash B_{\mu'}} u(\nu) \right\} + \max_{\nu\in B_{\mu_{\max}}\backslash B_{\mu_{\max-1}}} u(\nu) \nn
    &= \sum_{\mu'=\mu}^{\mu_{\max}-1} \max_{\nu\in B_{\mu_{\max}}\backslash B_{\mu'-1}} u(\nu) - \sum_{\mu'=\mu+1}^{\mu_{\max}}\max_{\nu\in B_{\mu_{\max}}\backslash B_{\mu'-1}} u(\nu)  + \max_{\nu\in B_{\mu_{\max}}\backslash B_{\mu_{\max-1}}} u(\nu) \nn
    &= \max_{\nu\in B_{\mu_{\max}}\backslash B_{\mu-1}} u(\nu) \, ,
\end{align}
as required.

We can use this approach to compute the probability for success of the state preparation via inequality testing, as
\begin{equation}
    \frac{\sum_{\nu} u(\nu)}{\sum_{\mu=1}^{\mu_{\max}} \tilde\psi_{\mu}^2}.
\end{equation}
To show this expression, the prepared state is
\begin{align}\label{eq:prepderv}
    \sum_{\mu=1}^{\mu_{\max}} \psi_\mu \ket{\mu} &= \left[\sum_{\mu'=1}^{\mu_{\max}} \tilde\psi_{\mu'}^2 \right]^{-1/2} \sum_{\mu=1}^{\mu_{\max}} \tilde\psi_\mu \ket{\mu} \nn
    &\mapsto \left[\sum_{\mu'=1}^{\mu_{\max}} \tilde\psi_{\mu'}^2 \right]^{-1/2} \sum_\nu \sum_{\mu'=\mu(\nu)}^{\mu_{\max}} \frac{\tilde\psi_{\mu'}}{\sqrt{|B'_{\mu'}|}}\ket{\mu'}\ket{\nu_x}\ket{\nu_y}\ket{\nu_z} \nn
    &\equiv \left[\sum_{\mu'=1}^{\mu_{\max}} \tilde\psi_{\mu'}^2 \right]^{-1/2} \sum_\nu \sqrt{\sum_{\mu'=\mu(\nu)}^{\mu_{\max}} \frac{\tilde\psi_{\mu'}^2}{|B'_{\mu'}|}} \ket{\nu_x}\ket{\nu_y}\ket{\nu_z} \nn
    &= \left[\sum_{\mu'=1}^{\mu_{\max}} \tilde\psi_{\mu'}^2 \right]^{-1/2} \sum_\nu \max_{\nu'\in B_{\mu_{\max}}\backslash B_{\mu(\nu)-1}} \sqrt{u(\nu')} \ket{\nu_x}\ket{\nu_y}\ket{\nu_z} \nn
    &\mapsto \left[\sum_{\mu'=1}^{\mu_{\max}} \tilde\psi_{\mu'}^2 \right]^{-1/2} \sum_\nu \sqrt{u(\nu)} \ket{\nu_x}\ket{\nu_y}\ket{\nu_z} .
\end{align}
The first line is just from the definition of $\psi_\mu$ as the normalised form of $\tilde\psi_\mu$.
The second line is the state resulting from the preparation of superpositions for each $\mu$.
The third line is the equivalent state where we ignore the $\ket{\mu'}$ register (which is allowed for in the state preparation for linear combinations of unitaries).
The fourth line is from the definition of $\tilde\psi_\mu$.
The final line is that obtained from an inequality test.
This subnormalised state gives the probability of success as its norm.

For preparation of $1/\|k_\nu\|$, we perform amplitude amplification boosting the probability for success.
We can then adjust $\delta_x,\delta_y,\delta_z$ to obtain probabilities of success after the amplitude amplification that are close to 1.
Appropriate choices of $\delta_x,\delta_y,\delta_z$ and success probabilities are given in Table \ref{TAB:sucprobs}, where we use the lattice vectors given in Table \ref{TAB:LATTICE_VECTORS}.
In all cases the initial probability is larger than or close to $1/4$, so the success probability after amplitude amplification is close to 1.

\begin{table*}[!htpb]
    \begin{tabular}{|c|ccc|ccc|}
    \hline
    \hline
    system & $\delta_x$ & $\delta_y$ & $\delta_z$ & $n=5$ & $n=6$ & $n=7$  \\
    \hline
LiNiO$_2$ (C2/m)  &   0 &   0 &   0 &   0.22940 &   0.21898 &   0.21309 \\
LiNiO$_2$ (P2$_1$/c) &   0 &   0 &   0 &   0.23059 &   0.22130 &   0.21592 \\
LiNiO$_2$ (P2/c)   &   0 &   0 &  0 &   0.23392 &   0.22450 &   0.21904 \\
Pd ($3\times 3$) &  1 &   1 &   0 &   0.26917 &  0.25529 &   0.24751 \\
Pt ($2\times 2$) &  1 &   1 &   0 &   0.24901 &  0.23553 &   0.22793 \\
Pt ($3\times 3$) &  1 &   1 &   0 &   0.25469 &  0.24157 &   0.23422 \\
Pt ($4\times 4$) &  0 &   0 &   0 &   0.22110 &  0.21222 &   0.20706 \\
Rh ($3\times 3$) &  1 &   1 &   0 &   0.27182 &  0.25791 &   0.25010 \\
Li$_{0.5}$MnO$_3$ &  1 &   0 &   2 &   0.30576 &   0.28594 &   0.27495 \\
Li$_{0.75}$[Li$_{0.17}$Ni$_{0.25}$Mn$_{0.58}$]O$_2$ &  2 &   1 &   0 &   0.23836 &   0.22193 &   0.21298 \\
Li$_{0.75}$MnO$_2$F &  0 &   1 &   1 &   0.27760 &   0.26426 &   0.25665 \\
C (diamond)         &   0 &   0 &   0 &  0.23334 &   0.22339 &   0.21771 \\
AlN (wurzite) &   1 &   1 &   0 &   0.25770 &   0.24442 &   0.23698 \\
CaTiO$_3$ &   1 &   1 &   0 &   0.26494 &  0.25221 &  0.24494 \\
    \hline
    \hline
        \end{tabular}
    \caption{Choices of $\delta_x,\delta_y,\delta_z$ and success probabilities for preparation of $1/\|k_\nu\|$ states.}
    \label{TAB:sucprobs}
\end{table*}

For the pseudopotentials it is convenient to use a state preparation that is as simple as possible, then apply the appropriate amplitudes as part of the controlled operations in the block encoding.
If we use the state preparation with amplitudes $\psi_\mu$ as above, then there is an increase in the effective $\lambda$ due to the negative zeros and the possibly larger values of $u(\nu)$ for the larger boxes.
It is therefore convenient to use the similar amplitude amplification as above but only for eliminating the inner boxes and negative zeros.
This will be used to prepare a state of the form
\begin{equation}
	\sum_{\mu=1}^{\mu_{\max}} \frac{\psi'_\mu}{\sqrt{|B_\mu\backslash B_{\mu-1}|}}\sum_{\nu\in B_\mu\backslash B_{\mu-1}}\ket{\mu}\ket{\nu_x}\ket{\nu_y}\ket{\nu_z} \, ,
\end{equation}
where
\begin{equation}
     (\widetilde\psi_{\mu}')^2 = |B_\mu\backslash B_{\mu-1}| \max_{\nu\in B_{\mu}\backslash B_{\mu-1}} u(\nu) \, ,
\end{equation}
and $\psi'_{\mu}$ is obtained from $\widetilde\psi'_{\mu}$ via normalisation.
This state has a large amplitude for preparation because it is mainly eliminating the negative zeros.
Moreover, it is giving an effective $\lambda$ where in each box excluding the inner box $B_\mu\backslash B_{\mu-1}$, the value of $u(\nu)$ is replaced with its maximum over this region.

\section{Complexity of norm with Bravais vectors}
\label{app:bravaisnorm}
In the simple case of orthogonal Bravais vectors of equal length, computing $\|\nu\|^2$ requires the sum of three squares.
In the case with $n_x+n_y+n_z=n$, the sum of squares may be computed using $3n^2-n-1$ Toffolis using the approach of Lemma 8 of \cite{SuPRXQuantum21}.
More generally, we would need to compute $\|\nu\|^2$ for general Bravais vectors, and $n_x,n_y,n_z$ may be different from each other.
Here we discuss the complexity for the case of general vectors, but in specific examples there may be simplifications.
The general form we need to calculate is
\begin{equation}
    g_{11} |\nu_x|^2 + g_{22} |\nu_y|^2 + g_{33} |\nu_z|^2 + 2 g_{12} \nu_x \nu_y + 2 g_{23} \nu_y \nu_z + 2 g_{13} \nu_x \nu_z \, .
\end{equation}
The arithmetic needed is then as follows, where we give only leading-order complexities.
\begin{enumerate}
    \item Squaring of $\nu_x,\nu_y,\nu_z$, with complexity $n_x^2,n_y^2,n_z^2$, respectively.
    \item Three multiplications of pairs of different $\nu_x,\nu_y,\nu_z$, with complexities $2n_x n_y,2n_y n_z,2n_z n_x$.
    The leading-order complexity of this and the previous step would then be $(n_x+n_y+n_z)^2$.
    \item Six multiplications by real numbers.
    \item Five additions, with complexity that may be ignored if we consider only leading order complexities.
\end{enumerate}

For simplicity of the estimation of the complexity, we will give the number of bits to be used for real numbers as $b$.
This will be used for the number of bits for the Bravais vectors as well as the precision of the result.

Using the result in Eq.~(D9) of \cite{Sanders2020}, the complexity of multiplying a $d_B$-qubit integer and a real number is given by
\begin{equation}
    d_B^2 + (2d_B-1)\lceil \log(d_B/\epsilon) \rceil - d_B \, ,
\end{equation}
where $\epsilon$ is the required precision in the resulting product, so $\log(1/\epsilon)$ would be equivalent to $b$ here.
That complexity is for the real number in a quantum register, and the complexity is approximately halved when the real number is given as a classical constant, as it is in the case of computing the Bravais vectors.
To leading order the complexity would then be
\begin{equation}
    d_B^2/2 + d_B b.
\end{equation}
The values of $d_B$ would be $2n_x$, $2n_y$, $2n_z$, $n_x+n_y$, $n_y+n_z$, $n_x+n_z$, so the complexity in the worst case of six multiplications is
\begin{align}
    & [4n_x^2+4n_y^2+4n_z^2+(n_x+n_y)^2+(n_y+n_z)^2+(n_x+n_z)^2]/2+4b(n_x+n_y+n_z) \nn
    &= (5/2)(n_x^2+n_y^2+n_z^2) + (n_x+n_y+n_z)^2 + 4b(n_x+n_y+n_z) \, .
\end{align}
An assumption for this complexity is that $b$ is larger than the numbers of bits in the integers.
Adding to that the complexity of the squaring and multiplications of the integers is, in the worst case, $(n_x+n_y+n_z)^2$, which would give a worst-case complexity
\begin{align} \label{eq:normcomplex}
 (5/2)(n_x^2+n_y^2+n_z^2) + 2(n_x+n_y+n_z)^2 + 4b(n_x+n_y+n_z) \, .
\end{align}

There are specific examples of Bravais vectors given in Table~\ref{TAB:LATTICE_VECTORS}.
For these examples the complexity can be reduced as follows.
In the following we are ignoring costs linear in the numbers of bits.
That means that when we add (for example) $\nu_x$ and $\nu_y$ we approximate the number of bits as $\max(n_x,n_y)$ for estimation of the complexity.
\begin{enumerate}
\item For LiNO$_2$ (C2/m) there are no zeros in the vectors, suggesting that the full complexity would be needed.
However, we first note that we may remove any overall multiplicative factor in this calculation.
This is because the norm $\|k_\nu\|^2$ is always multiplied by a further constant in the algorithm, and any overall multiplying factor may be absorbed into that constant.
It is found that $g_{11}=g_{22}$ and $g_{23}=-g_{13}$, so we may use a choice of multiplying constant so that (for example) $-g_{23}=g_{13}=1$.
Then we can rewrite the expression to calculate as
\begin{equation}\label{eq:lino1}
    g_{11} (\nu_x + \nu_y)^2 + g_{33}\nu_z^2 + 2(g_{12}-g_{11})\nu_x\nu_y + 2(\nu_x-\nu_y)\nu_z \, .
\end{equation}
First, only two squares and two products are needed, reducing the arithmetic complexity of that part to
    \begin{equation}\label{eq:lincmp1}
        \max(n_x,n_y)^2 + n_z^2 + 2n_x n_y + 2\max(n_y,n_x)n_z \, .
    \end{equation}
Then only two multiplications by the squares and one multiplication by a product are needed.
That means the complexity of that part is
\begin{align}
   2\max(n_x,n_y)[\max(n_x,n_y)+b] + 2n_z(n_z+b) + (n_x+n_y)^2/2+(n_x+n_y)b \, .
\end{align}

\item For LiNO$_2$ (P2$_1$/c), LiNO$_2$ (P2/c), or Li$_{0.5}$MnO$_3$ there are no equalities between terms, but $g_{23}=g_{12}=0$.
Similarly, for Li$_{0.75}$[Li$_{0.17}$Ni$_{0.25}$Mn$_{0.58}$]O$_2$ there are no equalities but $g_{23}=g_{13}=0$.
Then we need to compute all three squares, but the only product is $\nu_x\nu_z$, or $\nu_x\nu_y$ for Li$_{0.75}$[Li$_{0.17}$Ni$_{0.25}$Mn$_{0.58}$]O$_2$.
That gives complexities
\begin{align}
    (n_x+n_z)^2+n_y^2, \qquad (n_x+n_y)^2+n_z^2 \, ,
\end{align}
for the two cases.
If we choose a scaling such that $g_{13}=1$, or $g_{12}=1$ for Li$_{0.75}$[Li$_{0.17}$Ni$_{0.25}$Mn$_{0.58}$]O$_2$, then we only need multiplications by the squares of the components, giving complexity
    \begin{equation}
        2(n_x^2+n_y^2+n_z^2) + 2b(n_x + n_y + n_z) \, .
    \end{equation}  
\item For Pd ($3\times 3$), Pt ($2\times 2$), Pt ($3\times 3$), or Rh ($3\times 3$) we have $g_{11}=g_{22}=-2g_{12}$, so we may scale such that all these are equal to 1.
Similarly, for Pt ($4\times 4$) or AlN (wurzite) we have $g_{11}=g_{22}=2g_{12}$, and the same scaling may be used.
In both cases $g_{23}=g_{13}=0$.
The expression to calculate can then be given as
\begin{equation}
    (\nu_x-\nu_y)^2 + g_{33}\nu_z^2 +\nu_x\nu_y \, , \qquad (\nu_x+\nu_y)^2 + g_{33}\nu_z^2 -\nu_x\nu_y \, ,
\end{equation}
in the two cases.
Then we only have two squarings and one product, with complexity
\begin{equation}
    \max(n_x,n_y)^2 + n_z^2 + 2n_x n_y \, .
\end{equation}
Then there is the additional complexity for the multiplication by $g_{33}$, which is $2n_z (n_z + b)$.

\item For Li$_{0.75}$MnO$_2$F there is $g_{22}=g_{33}$, but $g_{12}=g_{23}=g_{13}$ are all zero.
The complexity of the three squarings is $n_x^2+n_y^2+n_z^2$.
We can choose a constant factor such that $g_{22}=g_{33}=1$, and the only multiplication needed is for the square of the $x$-component, giving complexity $2n_x (n_x + b)$.

\item For CaTiO$_3$ we have no equalities but $g_{12}=g_{23}=g_{13}$ are all zero.
Again the complexity of the three squarings is $n_x^2+n_y^2+n_z^2$.
We can scale such that, for example, $g_{33}=1$, in which case we only have complexity for multiplication for the $x$- and $y$- components giving
    \begin{equation}
        2n_x (n_x + b) + 2n_y (n_y + b) \, .
    \end{equation}  

    \item For Diamond all the products are needed, but $\|k_\nu\|^2$ can be rewritten as proportional to
    \begin{equation}
        (\nu_x+\nu_y-\nu_z)^2+(\nu_x-\nu_y+\nu_z)^2+(-\nu_x+\nu_y+\nu_z)^2  .
    \end{equation}
    This means that the complexity is just that for \emph{three} squarings.
    That is, the complexity would be $3\max(n_x,n_y,n_z)^2$.
    Then there is no further complexity for multiplications by real numbers.
\end{enumerate}

In the case of an inner product between different vectors using the Gramian, the costs of multiplication by the $g_{ij}$ constants are unchanged, but there are different costs for the products of integers.
In the worst case there are 9 products between integers needed, rather than three products and three squarings.
The (leading) overall complexity is multiplied by $2$, because all the squarings are replaced with products of different numbers, and all products of different components are replaced with two products of different components.
That gives a complexity $2(n_x+n_y+n_z)^2$, so the combination with the worst-case complexity of the multiplications above gives complexity
\begin{align} \label{eq:dotcomplex}
 (5/2)(n_x^2+n_y^2+n_z^2) + 3(n_x+n_y+n_z)^2 + 4b(n_x+n_y+n_z).
\end{align}

For the individual examples listed above, the costs are again multiplied by $2$ for the same reason as for the worst-case complexity.
To be more specific, the analysis for the first case is as follows.
The dot product can be written as, instead of Eq.~\eqref{eq:lincmp1},
    \begin{equation}
    g_{11} (\nu_x^a + \nu_y^a)(\nu_x^b + \nu_y^b) + g_{33}\nu_z^a\nu_z^b + (g_{12}-g_{11})(\nu_x^a\nu_y^b+\nu_x^b\nu_y^a) + (\nu_x^a-\nu_y^a)\nu_z^b+(\nu_x^b-\nu_y^b)\nu_z^a \, .
\end{equation}
Thus the squares are replaced with products, and the products are replaced with two products, doubling the complexity to twice that in Eq.~\eqref{eq:lincmp1}.
The analysis for all other examples is identical.

\clearpage
\section{Lattice vectors and reciprocal lattice vectors}
\label{app:vectors}

\begin{table*}[!htpb]
    \begin{tabular}{|c|c|ccc|ccc|}
    \hline
    \hline
    system & super cell size &  $\vec{a}_1$ & $\vec{a}_2$ & $\vec{a}_3$ & $\vec{g}_1$ & $\vec{g}_2$ & $\vec{g}_3$ \\
    \hline
LiNiO$_2$ (C2/m)$^a$ & $2\times 2 \times 1$ &  10.86498265 &  -5.95793520 &  -4.24456632 &   0.55174545 &   0.01376719 &  -0.08728916 \\
         &   &   0.13075864 &   9.08669104 &   2.25965848 &   0.38793206 &   0.67433284 &   0.04646969 \\
         &   &   1.33714130 &  -1.33714130 &   8.24103381 &   0.17780854 &  -0.17780854 &   0.70472651 \\
 \hline
LiNiO$_2$ (P2$_1$/c)$^a$ & $1\times 2 \times 1$ &  9.17881463 &   0.00000000 &  -3.81775899 &   0.64911019 &   0.00000000 &  -0.08516076 \\
         &   &   0.00000000 &  10.96132916 &   0.00000000 &   0.00000000 &   0.57321382 &   0.00000000 \\
         &   &   1.09817823 &   0.00000000 &   8.37050671 &   0.29605690 &   0.00000000 &   0.71179228 \\
 \hline
LiNiO$_2$ (P2/c)$^a$ & $1\times 1\times 1$ &   9.22368214 &   0.00000000 &  -1.65511564 &   0.70228065 &   0.00000000 &   0.11747103 \\
         &   &   0.00000000 &  10.88085803 &   0.00000000 &   0.00000000 &   0.57745311 &   0.00000000 \\
         &   &  -1.52693970 &   0.00000000 &   9.12855033 &   0.12733190 &   0.00000000 &   0.70959936 \\
 \hline
Pd & $3\times 3$ &  15.86273910 &   7.93136955 &   0.00000000 &   0.39609712 &   0.00000000 &   0.00000000 \\
                &  &   0.00000000 &  13.73753504 &   0.00000000 &  -0.22868678 &   0.45737356 &   0.00000000 \\
                &  &   0.00000000 &   0.00000000 &  27.53179798 &   0.00000000 &   0.00000000 &   0.22821558 \\
\hline
Pt & $2\times 2$ &  10.58834339 &   5.29417170 &   0.00000000 &   0.59340589 &   0.00000000 &   0.00000000 \\
                &  &   0.00000000 &   9.16977436 &   0.00000000 &  -0.34260305 &   0.68520610 &   0.00000000 \\
                &  &   0.00000000 &   0.00000000 &  24.89695283 &   0.00000000 &   0.00000000 &   0.25236764 \\
\hline
Pt & $3\times 3$ &  15.88252454 &   7.94126227 &   0.00000000 &   0.39560369 &   0.00000000 &   0.00000000 \\
                &  &   0.00000000 &  13.75466973 &   0.00000000 &  -0.22840190 &   0.45680379 &   0.00000000 \\
                &  &   0.00000000 &   0.00000000 &  24.89695283 &   0.00000000 &   0.00000000 &   0.25236764 \\
\hline
Pt & $4\times 4$ &  21.18893221 & -10.59446611 &   0.00000000 &   0.29653147 &   0.00000000 &   0.00000000 \\
                &  &   0.00000000 &  18.35015357 &   0.00000000 &   0.17120253 &   0.34240505 &   0.00000000 \\
                &  &   0.00000000 &   0.00000000 &  24.56643973 &   0.00000000 &   0.00000000 &   0.25576296 \\
\hline
Rh & $3 \times 3$ &  15.40909145 &   7.70454572 &   0.00000000 &   0.40775832 &   0.00000000 &   0.00000000 \\
              &    &   0.00000000 &  13.34466464 &   0.00000000 &  -0.23541938 &   0.47083876 &   0.00000000 \\
              &    &   0.00000000 &   0.00000000 &  27.28499974 &   0.00000000 &   0.00000000 &   0.23027984 \\
\hline
Li$_{0.5}$MnO$_3$$~^b$ & $2\times 2\times 1$ &  18.93505585 &   0.00000000 &  -3.20289682 &   0.33182819 &   0.00000000 &   0.00000000 \\
                     &   &   0.00000000 &  32.73005662 &   0.00000000 &   0.00000000 &   0.19196989 &   0.00000000 \\
                     &   &   0.00000000 &   0.00000000 &   9.06974057 &   0.11718212 &   0.00000000 &   0.69276351 \\
\hline
Li$_{0.75}$[Li$_{0.17}$Ni$_{0.25}$Mn$_{0.58}$]O$_2$~$^b$ & $2\times 3 \times 2$ &  10.78674574 &  -8.09010655 &   0.00000000 &   0.58249128 &   0.00000000 &   0.00000000 \\
                                                        &  &   0.00000000 &  14.01250825 &   0.00000000 &   0.33630071 &   0.44839833 &   0.00000000 \\
                                                       &   &   0.00000000 &   0.00000000 &  37.09853652 &   0.00000000 &   0.00000000 &   0.16936478 \\
\hline
Li$_{0.75}$MnO$_2$F$~^c$ & $3\times 2\times 2$ &  23.12230001 &  -0.02683881 &  -0.02005775 &   0.27173758 &   0.00030702 &   0.00021386\\
&&  -0.01813051 &  16.04617832 &   0.00146633 &   0.00045450 &   0.39156947 &  -0.00000841\\
&&  -0.01224671 &   0.00034827 &  15.56081563 &   0.00035022 &  -0.00003650 &   0.40378279\\
\hline
C (diamond)$~^e$ & $3\times 3\times 3$ &   0.00000000 &  10.11097963 &  10.11097963 &  -0.31071101 &   0.31071101 &   0.31071101 \\
          & &  10.11097963 &   0.00000000 &  10.11097963 &   0.31071101 &  -0.31071101 &   0.31071101 \\
          & &  10.11097963 &  10.11097963 &   0.00000000 &   0.31071101 &   0.31071101 &  -0.31071101 \\
\hline
AlN (wurzite)$~^e$& $3\times 3\times 3$ &  17.63114474 &  -8.81557237 &   0.00000000 &   0.35636854 &   0.00000000 &   0.00000000 \\
          & &   0.00000000 &  15.26901924 &   0.00000000 &   0.20574947 &   0.41149894 &   0.00000000 \\
          & &   0.00000000 &   0.00000000 &  28.23250830 &   0.00000000 &   0.00000000 &   0.22255144 \\               
    \hline
    \hline
        \end{tabular}
    \caption{Lattice vectors ($a_0$) and reciprocal lattice vectors of the simulation cell.}
    \label{TAB:LATTICE_VECTORS}

    {\footnotesize 
    $^a$~See Figure~12 of Ref.~\cite{PRXQuantum.4.040303}. \\
    $^b$~Lattice vectors, taken from Ref.~\cite{Shokrian}, correspond to structural models built according to Refs.~\cite{Cho15_7066,Kang20_419}. \\
    $^c$~Ref.~\cite{Islam22_5275}.
    $^d$~Ref.~\cite{Yang12_7346}.
    $^e$~Ref.~\cite{Scuseria05_174101}.
    $^f$~Ref.~\cite{Schulze87_1668}.
    }
\end{table*}

\end{document}